\documentclass[aps,prx,twocolumn,superscriptaddress]{revtex4-2}
\usepackage{times}
\usepackage{graphicx}
\usepackage{amsmath}
\usepackage{amssymb}
\usepackage{euscript}
\usepackage{verbatim}
\usepackage{fancyhdr}
\usepackage{wrapfig}
\usepackage{setspace}
\usepackage{xcolor}
\usepackage{amsfonts}
\usepackage{subfigure}
\usepackage{upgreek}
\usepackage {multirow}
\usepackage{tabularx}
\usepackage{makecell}
\usepackage[hypertexnames=false]{hyperref}

\begin{document}

%\preprint{}

\title{Interference measurements of non-Abelian $e/4$ \& Abelian $e/2$ quasiparticle braiding}


\author{R.~L.~Willett}
\email[]{robert.willett@nokia.com}
\thanks{Correspondence to: R.~L.~Willett, Nokia Bell Labs, Room 1d-217, Murray Hill, NJ 07974, USA, phone number 908-370-4395}
\affiliation{Nokia Bell Labs, Murray Hill, NJ 07974, USA}
\author{K.~Shtengel}
\affiliation{Department of Physics \& Astronomy, University of California,
Riverside, CA 92521, USA}
\author{C.~Nayak}
\affiliation{Microsoft Quantum, Elings Hall, University of California, Santa Barbara, CA  93106, USA}
\affiliation{Department of Physics, University of California,
Santa Barbara, CA 93106, USA}
\author{L.~N.~Pfeiffer}
\affiliation{Department of Electrical Engineering, Princeton University, Princeton, NJ 08544, USA}
\author{Y.~J.~Chung}
\affiliation{Department of Electrical Engineering, Princeton University, Princeton, NJ 08544, USA}
\author{M.~L.~Peabody}
\affiliation{Nokia Bell Labs, Murray Hill, NJ 079784, USA}
\author{K.~W.~Baldwin}
\affiliation{Department of Electrical Engineering, Princeton University, Princeton, NJ 08544, USA}
\author{K.~W.~West}
\affiliation{Department of Electrical Engineering, Princeton University, Princeton, NJ 08544, USA}


\date{\today}

\begin{abstract}
The quantum Hall states at filling factors $\nu=5/2$ and $7/2$  are expected to have Abelian charge $e/2$  quasiparticles and non-Abelian charge $e/4$ quasiparticles. The non-Abelian statistics of the latter has been predicted to display a striking interferometric signature, the even-odd effect.  By measuring resistance oscillations as a function of magnetic field in Fabry–P\'{e}rot interferometers using new high purity heterostructures, we for the first time report experimental evidence for the non-Abelian nature of excitations at $\nu=7/2$. At both $\nu=5/2$ and $7/2$ we also examine, for the first time, the fermion parity, a topological quantum number of an even number of non-Abelian quasiparticles.  The phase of observed $e/4$ oscillations is reproducible and stable over long times (hours) near both filling factors, indicating stability of the fermion parity. At both fractions, when phase fluctuations are observed, they are predominantly $\pi$ phase flips, consistent with either fermion parity change or change in the number of the enclosed $e/4$ quasiparticles. We also examine lower-frequency oscillations attributable to Abelian interference processes in both states.  Taken together, these results constitute new evidence for the non-Abelian nature of $e/4$ quasiparticles; the observed life-time of their combined fermion parity further strengthens the case for their utility for topological quantum computation.
\end{abstract}

%\keywords{}

\maketitle

%\section{}
%% Put \label in argument of \section for cross-referencing
%%\section{\label{}}
%\subsection{}
%\subsubsection{}

\section{Introduction}
\label{sec:intro}
The idea of fault-tolerant topological quantum computation is premised on both the existence of non-Abelian anyons and our ability to manipulate them~\cite{Nayak2008}. Fractional quantum Hall (FQH) states are the best-established examples of topological phases with FQH states at filling fractions $\nu=5/2$ and $7/2$ being arguably the strongest candidates for non-Abelian phases. They are predicted to have non-Abelian charge-e/4 excitations if their ground states are in either the Pfaffian / Moore--Read ~\cite{Moore1991}, anti-Pfaffian~\cite{Lee2007a,Levin2007a} or particle-hole-symmetric Pfaffian (‘PH-Pfaffian’) ~\cite{Bonderson2013b,Chen2014,Son2015,Zucker2016} universality class. In addition to their electrical charge, these excitations also carry the non-Abelian topological charge of Ising anyons ~\cite{Nayak1996c,Nayak2008}, which can be understood as the presence of a Majorana zero mode~\cite{Read2000}. A combined quantum state of a pair of such anyons (known as a \textit{fusion channel}) can be viewed as an Abelian charge-e/2 excitation (which is a ``conventional'' Laughlin quasiparticle, i.e. a fractionally charged quasiparticle corresponding to an insertion of one additional flux quantum), either with or without a neutral fermion~\cite{Milovanovic1996} -- see Figure~\ref{fig:braiding}. The presence or absence of the neutral mode determines the fermion parity of the state.  The nature of these excitations—both their charge and statistics—can be probed by interferometry experiments~\cite{Fradkin1998,Stern2006a,Bonderson2006a,Bonderson2008a,Bishara2009a,Halperin2011a}.  The non-Abelian properties of the $e/4$ quasiparticles should manifest themselves in the even-odd effect, whereby the interference between two different paths for an $e/4$ quasiparticle is switched on or off whenever the difference between the paths encircles, respectively, an even or odd number of $e/4$ quasiparticles, see Figure~\ref{fig:braiding}. Meantime, the Abelian $e/2$ quasiparticle should show interference regardless of the number of encircled quasiparticles, with a pattern similar to other Laughlin quasiparticles. In a realistic Fabry--P\'{e}rot interferometer, the interference pattern should consist of oscillations due to all types of charged quasiparticles present in the system. It is the goal of this study to experimentally determine the full set of observed oscillation frequencies at $\nu=5/2$ and $7/2$ and to compare the observed frequencies with theoretical predictions based on the braiding properties of the $e/2$ and $e/4$ quasiparticles.

Previous interferometry experiments at $\nu=5/2$~\cite{Willett2007,Willett2009a,Willett2010a,Willett2013a,Willett2013b} have observed resistance oscillations consistent with charge $e/4$ and charge $e/2$  excitations displaying, respectively, non-Abelian braiding and Abelian braiding statistics.  Meanwhile, tunneling~\cite{Radu2008a} and charge sensing measurements~\cite{Venkatachalam2011} at $\nu=5/2$ have found signatures of $e/4$ quasiparticles but no indication of  $e/2$ quasiparticles. At the same time shot noise measurements~\cite{Dolev2008a} found evidence for $e/4$ quasiparticles and indicated a crossover from $e/4$ to $e/2$-dominated behavior in different temperature and voltage ranges~\cite{Dolev2010,Carrega2011,Feldman2017}. These measurements did not probe the braiding statistics of the excitations, only their electrical charge.  By contrast, interferometry measurements can provide information about both the charge and the braiding statistics of the quasiparticles. A recent measurement of the thermal Hall conductivity~\cite{Banerjee2018} is an indirect probe of the topological order of the bulk and, therefore, an indirect measure, at best, of the presence of quasiparticles with non-Abelian braiding statistics in the bulk.

This study presents two important new findings: (i) interferometric signatures consistent with the non-Abelian even-odd effect at filling factor $\nu=7/2$, and (ii) stable (over hours or even days) interference oscillations consistent with the even-odd effect at both $\nu=5/2$ and $7/2$, where at both filling factors sporadic interruptions of this stability take the form of phase jumps by $\pi$ indicative of either the fermion parity change or change in the number of the enclosed $e/4$ quasiparticles.

The key experimental advance underpinning the study presented here is the
development of a new class of ultra-high mobility AlGaAs heterostructures in which the Al alloy layers are purified to the extreme, promoting stronger electron-electron correlations, and thus more robust quantum Hall states than previously attained. This improvement in the material purification also results in a substantially larger amplitude of resistance oscillations observed in new interferometer devices. Specifically, in addition to providing more solid evidence for the non-Abelian nature of the $\nu=5/2$ state, we report the first experimental evidence in support of the similar nature of the $\nu=7/2$ state. Furthermore, oscillations consistent with the even-odd effect associated with transport by non-Abelian charge $e/4$ quaiparticles are stable in the time scales of hours or even days.  When an instability occurs, it takes the form of a $\pi$ phase shift consistent with the change of either the fusion channel of the enclosed non-Abelian anyons or their number, thus providing further evidence for the non-Abelian nature of the states. Irrespective of the mechanism for these phase shifts, they are observed
to occur infrequently. This strengthens the case for using such FQH systems as a platform for topological quantum computation.

Both heterostructure and interferometer designs used in this study allow us to address another potential issue that has been plaguing earlier interference studies. Specifically, resistance oscillations of a mesoscopic quantum Hall island can be due to some combination of the Aharonov–Bohm (AB) and Coulomb blockade effects. The latter are expected to dominate in smaller devices~\cite{Bhattacharyya2019}, and some early results~\cite{Mcclure2012a,Camino2005b,Camino2007a} are consistent with this. However, several aspects of our heterostructure design work to suppress Coulomb effects.  Most importantly, we use special heterostructures that contain additional conducting layers able to screen the long-range Coulomb interactions (see Section~\ref{sec:S2} of Supplementary Materials). This layering promotes AB oscillations even in relatively small quantum Hall interferometers at $\nu=5/2$ and $7/3$~\cite{Willett2007,Willett2009a, Willett2010a,Willett2013a,Willett2013b}; it is akin to employing surface top gates used for the same purpose in previous Fabry--P\'{e}rot interferometry studies~\cite{Zhang2009a,Ofek2010}. In spite of these additional conducting
layers, our devices allow illumination of the samples to achieve the necessary high quality  needed to observe $\nu=5/2$ and $\nu=7/2$ states. The case for AB oscillations aided by parallel conductors is supported by recent measurements~\cite{Nakamura2019} at $\nu=1/3$. A second important design feature of our devices is their high electron densities, $4\times10^{11}\,\text{cm}^{-2}$, which also suppresses Coulomb effects. In addition, our measurements employ large aperture interferometers that contribute to further suppression of Coulomb blockade effects.

The manuscript is organized as follows. In Section~\ref{sec:interference} we present an extensive review of quantum Hall interferometry, emphasizing the expected periodicities of the AB oscillations for both Abelian and non-Abelian quasiparticles expected at $\nu=5/2$ and $7/2$. For non-Abelian quasiparticles, we focus on the consequences of the existence
of two fusion channels corresponding to the two possible fermion parities
that may be contained within the loop (displayed schematically in Figure~\ref{fig:braiding}). In Section~\ref{sec:methods} we turn to the description of the devices  and experimental methods used to probe these predictions. We present evidence in support of the Aharonov--Bohm mechanism as the dominant mechanism behind the interference oscillations.  Section~\ref{sec:results} is dedicated to our principal experimental results – stable oscillations observed at $\nu=5/2$ and $7/2$,  which demonstrate occasional $\pi$ phase shifts.  In Section~\ref{sec:results_oscillations} we present full power spectra of these oscillations at $\nu=7/2$ and identify spectral peak positions observed at these filling fractions. In Section~\ref{sec:results_stability} we present evidence of the temporal stability of non-Abelian quasiparticle fusion. We observe occasional $\pi$ phase shifts that we attribute to changes in the fusion channel (fermionic parity) or the parity of the enclosed non-Abelian quasiparticles. In Section~\ref{sec:results_five_halves} we present similar findings for $\nu=5/2$. In addition, in Section~\ref{sec:results_abelian} we focus on the signatures of Abelian braiding processes and demonstrate the ability to control a specific component of the interference spectrum attributable to different braiding processes. We conclude that, taken together, these data significantly strengthen the case for both the non-Abelian nature of the FQH states at $\nu=5/2$ and $7/2$ (while providing the first such experimental evidence for the latter state) and for their potential applications for quantum information processing.

\section{Introduction to fractional quantum Hall interferometry}
\label{sec:interference}

\begin{figure}[htb]
\centering
    \includegraphics[width=0.9\columnwidth]{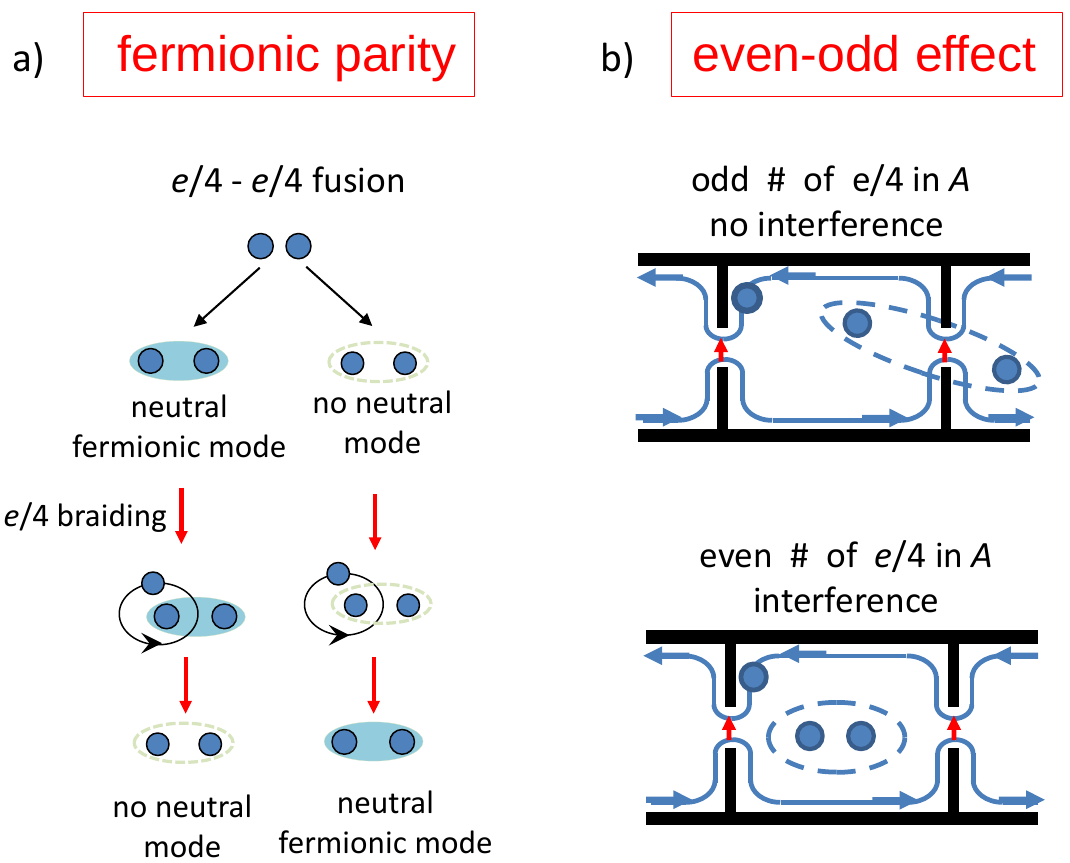}
    \caption{Fermion parity and the non-Abelian even-odd effect.\\
(a) The fusion of two non-Abelian e/4  quasiparticles has two possible outcomes, one with and one without a neutral fermionic mode.  The fermion parity is a quantum number associated with an even number of such quasiparticles. It is changed when one constituent e/4  quasiparticle is braided by an external  e/4 quasiparticle.\\
(b) When $e/4$ quasiparticles backscatter from the lower to the upper edge of the Hall bar at the two constrictions, the outcome depends dramatically on the parity of $e/4$ quasiparticles inside the interferometer. If their number is even, quasiparticles propagating along two possible paths interfere akin to the familiar double slit interference. If this number is odd, the interference disappears since with each tunneling quasiparticle the fermion parity is either switched or not, depending on which path is taken. The final quantum states are therefore orthogonal to one another, which precludes interference.
}
    \label{fig:braiding}
\end{figure}

In this section, we will describe various types of resistance oscillations
that can be observed in quantum Hall interferometers. We review the oscillations expected to be observed in magnetic field sweeps in Abelian quantum Hall states with contributions from both the Aharonov--Bohm and statistical phases.  We explain how this picture is altered if the state in question is non-Abelian and focus on the role the fermionic parity plays in this case.
We also mention an alternative mechanism for resistance oscillations originating
from breathing of Coulomb-dominated electron droplets.
As we discuss, these different phenomena can be distinguished
by their resistance oscillation spectra.

Fabry--P\'{e}rot interferometry in the two-dimensional electron gas (2DEG) in the quantum Hall regime is due to interference between two different paths by which electrical current can flow from source to drain along edge states and across constrictions (see Figure~\ref{fig:interferometer}). The interference pattern is thus determined by the total phase difference accumulated along the two paths, which in turn consists of both the Aharonov--Bohm phase, determined by the charge of the propagating quasiparticles and the enclosed flux, and the statistical contribution, determined by the statistics of the quasiparticles and the number and type of the quasiparticles enclosed between the two paths. This overall phase difference can be changed experimentally by changing either the enclosed flux or the number of quasiparticles within the interferometer loop. As is discussed in more detail in Section~\ref{sec:S1} of Supplementary Materials, for Abelian quasiparticles this change is given by
\begin{equation}
\Delta \gamma_{e^*}  =2 \pi\left(\frac{\Delta\Phi}{\Phi_0} \right)\left(\frac{e^*}{e}\right)+2\theta_{e^*}\Delta N_{e^*}
\label{eq:AB phase}
\end{equation}	
In this expression, $\Delta\Phi$ is the change in the encircled flux, $\Phi_0=hc/e \approx 41\,\text{G}\,\upmu \text{m}^2$ is the flux quantum, and $\Delta N_{e^*}$ is the change in number of the enclosed quasiparticles of charge $e^*$. Their braiding statistics is described by statistical angle $\theta_{e^*}$ – a phase acquired by the wavefunction upon counter-clockwise exchange of two identical quasiparticles.

A key difference between the Abelian interference described by Eq.~(\ref{eq:AB phase}) and non-Abelian interference is that in the former case both the encircled flux and the number of enclosed quasiparticles play a similar role: they simply contribute to the phase difference between the two interfering paths. However, if the interfering quasiparticles are \textit{non-Abelian}, the number of enclosed quasiparticles does not just contribute to the phase difference, it can change the \textit{amplitude} of the interference term~\cite{Bonderson2006b}.
The most dramatic manifestation of this property is the even-odd effect predicted for the interference of $e/4$ quasiparticles in non-Abelian $\nu=5/2$ or $7/2$ QH states. Specifically, the interference is only possible when the number of encircled $e/4$ quasiparticles is even whereas no Aharonov--Bohm oscillations should be observed if the number is odd -- see Figure~\ref{fig:braiding}.

According to Eq.~(\ref{eq:AB phase}), two parameters can potentially be varied in interferometric studies: the encircled flux and the number of enclosed quasiparticles. While varying them independently may seem experimentally hard, the variation of different combinations of them is achieved by: (i) varying the side gate ($V_s$) at fixed magnetic field~\cite{Willett2009a,Willett2010a,Willett2013a}, and (ii) varying the magnetic field at fixed gate voltage~\cite{Willett2013b}.  The active area $A$ in an interferometer -- the area encircled by the current paths -- is defined by surface gates. Varying the applied voltages on these gates changes this area; consequently, it changes both the enclosed flux and the number of $e/4$ quasiparticles randomly localized within the area. This method has shown signatures of both $e/2$ and $e/4$ quasiparticles and also demonstrated a pattern of oscillations consistent with the non-Abelian nature of the latter, specifically the aforementioned even-odd effect~\cite{Stern2006a,Bonderson2006a} whereby the Aharonov--Bohm oscillations associated with electrical transport by $e/4$ quasiparticles  are observed only when an even number of $e/4$ quasiparticles is localized within the interferometer loop. Meantime Aharonov--Bohm oscillations associated with electrical transport by $e/2$ quasiparticles are always present.

When the magnetic field is varied with fixed gate voltage, the enclosed magnetic flux number and the enclosed quasiparticle number change in tandem (see Section~\ref{sec:S1} of Supplementary Materials). Specifically, in the simplest model we assume that the active area of the interferometer
is independent of the magnetic field (thus discounting the possibility of so-called Coulomb domination ~\cite{Halperin2011a,Keyserlingk2015} -- see section~\ref{sec:S1} of Supplementary Materials for more details).
In this case, the change in the number of bulk quasiparticles in response to the change in flux $\Delta\Phi$ is given by $\Delta N_{e^*}=-(\Delta\Phi/\Phi_0 )(\nu e/e^* )$, resulting in
\begin{equation}
\Delta \gamma_{e^*}=\left(\frac{\Delta\Phi}{\Phi_0}\right)\left[2 \pi\left(\frac{e^*}{e}\right)-2\theta_{e^*}  \left(\frac{\nu e}{e^*}\right)\right].
\label{eq:AB_phase2}
\end{equation}

The putative non-Abelian nature of $e/4$ quasiparticles should  result in specific small-period oscillations centered near $5f_0$ for the $\nu=5/2$ state (and  $7f_0$ for the $\nu=7/2$ state), with $f_0= 1/\Phi_0$ being the oscillation frequency corresponding to the period of one flux quantum.  These small period oscillations are a manifestation of the even-odd effect (illustrated schematically in Figure~\ref{fig:braiding}; see also schematic for fermion parity) driven by the systematic variation of the $e/4$ quasiparticle number as the magnetic field is changed ($B$-field sweep): the interference of $e/4$ quasiparticles traversing the interferometer should be on/off if there is an even/odd number
of $e/4$ quasiparticles within the loop. Repeated switching between these regimes results in additional resistance oscillations. The corresponding period is determined by the flux needed to increase the number of $e/4$ quasiparticles within the active area of the interferometer by two, namely a period of  $2\Phi_0 e^*/\nu e=\Phi_0/5$ at $\nu=5/2$ and $\Phi_0/7$ at $\nu=7/2$. Such high-frequency peaks should be an unmistakable signature of non-Abelian statistics and they were first reported in the earlier study~\cite{Willett2013b} at 5/2 filling.

However, a more complicated picture emerges when all quasiparticle types are considered.  Specifically, if both $e/4$ and $e/2$ excitations are present, one should observe oscillations due to all permutations of interfering and enclosed quasiparticles. A straightforward generalization of Eqs.~(\ref{eq:AB phase})--(\ref{eq:AB_phase2}) for the Abelian phase acquired by interfering quasiparticles of type $a$ encircling bulk quasiparticles of type $b$ is given by:
\begin{multline}
\Delta \gamma_{ab}=2 \pi\left(\frac{\Delta\Phi}{\Phi_0} \right)\left(\frac{e_a}{e}\right)+2\theta_{ab}\Delta N_b
\\
=\left(\frac{\Delta\Phi}{\Phi_0} \right)\left[2 \pi\left(\frac{e_a}{e}\right)-2\theta_{ab} \left(\frac{\nu e}{e_b}\right)\right].
\label{eq:interferenece_general}
\end{multline}
Direct application of this expression results in oscillation periods of $\Phi_0$ for $e/4$ quasiparticles interfering around $e/2$ quasiparticles and $\Phi_0/2$ for the $e/2$ quasiparticles interfering around either $e/4$ or $e/2$ quasiparticles.  Finally, the interference of $e/4$ around $e/4$ quasiparticles would naïvely result in the period of $4/9\,\Phi_0$ for the Moore--Read Pfaffian state and $4/11\,\Phi_0$ for the anti-Pfaffian state (see Section~\ref{sec:S1} of Supplementary Materials for more detail). However, this is not the case since in both states the $e/4$ excitations are actually non-Abelian. Therefore the interference turns on and off with each shift $\Delta N_{e/4}=\pm 1$, resulting in the aforementioned small period of $\Delta\Phi=\Phi_0/5$.

When the number of bulk $e/4$ excitations is even, the interference is not simply governed by $\theta_{e/4}$; it also depends on the fusion channel of the enclosed quasiparticles. There are three basic possibilities: (i) the fusion channel, which determines the fermion parity, is fixed by the energetics and remains largely stable during the magnetic field sweep across the $\nu=5/2$ plateau, (ii) the fusion channel is random but its autocorrelation time is longer or comparable to the time it takes to change the flux by one flux quantum, and (iii) the fusion channel fluctuates rapidly on the time scale of changing the flux by $\Delta\Phi=\Phi_0$. Focusing on the first scenario, let us assume that the net fusion channel of the bulk quasiparticles is always trivial. Physically this means that from the point of view of interference, the bulk is equivalent to a collection of $e/2$ Laughlin quasiparticles, which would result in the aforementioned Abelian factor in the interference pattern, with period of $\Phi_0$ irrespective of the exact nature of the $\nu=5/2$ state. The net result would be a convolution of non-Abelian $5f_0$  and Abelian $f_0$ oscillations, resulting in spectral peaks at $4f_0$ and $6f_0$. Were the fusion channel to contain a fermion instead, the bulk quasiparticles would be in a different fermion parity state and the overall phase of Abelian oscillations would shift by $\pi$ with no change in the oscillation period. In the second scenario, the fluctuations in the fusion channel -- fluctuations in the fermion parity -- would scramble the $f_0$ component (due to random $\pi$ phase shifts throughout the magnetic field sweep) thus eliminating the beats, resulting in a single spectral peak at $5f_0$. Finally, in the third scenario, the interference of charge-$e/4$ excitations around other $e/4$ excitations would be eliminated entirely: their interference is suppressed for odd numbers of enclosed $e/4$ quasiparticles by their non-Abelian nature and for even numbers by rapid phase fluctuations.

Note that the first and second  scenarios  may actually coexist within a sweep across the entire $\nu=5/2$  plateau: one could envision e.g. a situation whereby the fermion parity is stable near the middle of the plateau while becoming progressively less stable closer to its margins, where the concentration of the bulk quasiparticles becomes larger and hence their typical distance to the edge smaller. The latter scenario would in turn enhance tunneling of neutral fermions between the edge and the localized quasiparticles, scrambling the well-defined fermion parity in the bulk~\cite{Rosenow2008a}. In such a case one would find oscillation peaks at $4f_0$ and $6f_0$ near the middle of the plateau and $5f_0$ closer to its flanks.

We should note, however, that in the Coulomb-dominated regime another mechanism for generating these high-frequency spectral features may arise: specifically, one could imagine a scenario whereby the active area of an interferometer ``breathes'' with the period corresponding to the intoduction of additional Abelian $e/2$ quasiparticles in order to minimize the energy of the QH droplet that defines the active area -- the period exactly matching that of the non-Abelian even-odd effect.  Such a mechanism, which is described in more detail in Section~\ref{sec:S1} of Supplementary Materials, would rely solely on the energetics of the $e/2$ quasiparticles inside the interferometer and thus be oblivious to the nature of charge $e/4$ quasiparticles. Therefore a convincing proof of the non-Abelian statistics of the latter must rule out this scenario; we address this important issue in the following section as well as in Section~\ref{sec:S4b} of Supplementary Materials.

%The aforementioned expectations for the high-frequency spectral features are summarized in Table~\ref{tab:expectations}.

These considerations can also be applied to  $\nu=7/2$, where the charge $e/4$ quasiparticle is similarly expected to obey non-Abelian statistics.  Upon magnetic field sweep the non-Abelian even-odd effect will manifest itself through the resistance oscillations with period corresponding to the magnetic field increment needed to change the number of $e/4$ quasiparticles by two, which in terms of flux corresponds to the period of $2\Phi_0 e^*/(\nu e)=\Phi_0/7$ or frequency $7f_0$ in the Fourier spectrum of the resistance oscillations. These oscillations may be then modulated by the frequency of $1.5f_0$ corresponding to $e/4$ quasiparticles interfering around $e/2$ quasiparticles, according to Eq.~(\ref{eq:interferenece_general}). Thus spectral peaks are expected at either $7f_0$ or $7 f_0 \pm 1.5 f_0$, depending upon the fermion parity stability and the Fourier transform window -- see Table~\ref{tab:expectations} for the summary.  Finally, the spectrum could also display a peak corresponding to $e/2$ interfering around e/2, and from Eq. (2) this frequency would be $3f_0$.  The $7/2$ spectrum could then be comprised of peaks at either $1.5f_0$, $5.5f_0$ and  $8.5f_0$ or  at $1.5f_0$  and $7f_0$ (or, perhaps, at all of those); and in addition there might be a spectral peak at $3f_0$ as well.

%\vspace{0.2cm}
%\noindent
\begin{table}[h!]
\centering
\setbox0\hbox{\tabular{@{}c}
\thead{Non-Abelian\\ even-odd effect}\endtabular}
\begin{tabular}{|c|c|c|c|}
\hline
\multicolumn{2}{|c|}{Mechanism} & $\nu=5/2$ & $\nu=7/2$ \\
\hline\hline
\multirow{3}{*}{\rotatebox{90}{\usebox0}}
& {\makecell{Stable fermionic\\ parity}} & $(5\pm 1)f_0$ &  $(7\pm 1.5)f_0$  \\ \cline{2-4}
& \makecell{Slow parity\\ fluctuations} & $5f_0$ & $7f_0$ \\ \cline{2-4}
& \makecell{Fast parity\\ fluctuations} & --- & --- \\  \hline
%\hline
\multicolumn{2}{|c|}{ $e/4\bigodot e/2$} & $f_0$ & $1.5f_0$ \\ \hline
\multicolumn{2}{|c|}{ $e/2\bigodot e/4,e/2$} & $2f_0$ & $3f_0$ \\ \hline
\multicolumn{2}{|c|}{\makecell{Coulomb-dominated\\ ``breathing''}} & $5f_0$ & $7f_0$ \\ \hline
\end{tabular}
\caption{Possible mechanisms and spectral features of resistance oscillations at $\nu=5/2$ and $7/2$.}
  \label{tab:expectations}
\end{table}

Yet just like in the case of $\nu=5/2$, in order to ascertain the non-Abelian nature of high-frequency oscillations at $\nu=7/2$, one must rule out the possibility of the aforementioned ``breathing'' of the active area of the interferometer, which in the Coulomb-dominated regime could also produce resistance oscillations with frequency $7f_0$. Fortunately, this scenario would not be specific to $\nu=5/2$ and $7/5$; it would result in similar high-frequency oscillations in nearby Abelian fractional quantum Hall states and thus can be eliminated from consideration absent such oscillations. In the next section we provide experimental details that support eliminating this mechanism.

In summary, in the Aharonov--Bohm regime the observed resistance oscillations should be a combination of interference patterns resulting from charge $e/4$ Ising anyon encircling another $e/4$ Ising anyon; a charge $e/2$ Abelian anyon encircling another charge $e/2$ Abelian anyon as well as both types of anyons encircling the other kind. In addition, the phase of the charge $e/4$ quasiparticle interference should depend on the parity of neutral fermions inside the loop. At 5/2 filling the first type of process leads to a resistance that oscillates with magnetic flux with frequency $5f_0$ whereas at $7/2$ the corresponding frequency is $7f_0$. The $e/2-e/2$ interference in these states should result in oscillations with frequency $2f_0$ and $3f_0$ respectively; we will show later that the amplitude of these oscillations can be tuned independently of $e/4$ interference oscillations, effectively allowing for them to be turned on or off.  Finally, the interference of $e/4$ quasiparticles around Laughlin $e/2$ quasiparticles produces oscillations with frequency $f_0$ at $\nu=5/2$ and $1.5f_0$ at $\nu=7/2$; its convolution with the first type of process results in oscillation frequencies $5f_0\pm f_0$ and $7f_0\pm 1.5f_0$ respectively. The stability of the fermion parity should determine whether a measured high-frequency spectral peak is split in this manner or remains centered at $5f_0$ or $7f_0$ (with an additional possibility of both scenarios occurring within the same plateau). Table~\ref{tab:expectations} presents an abbreviated summary of these predictions.

Irrespective of those details, once the Coulomb-dominated regime is eliminated in favor of the Aharonov--Bohm regime, an appearance of such high-frequency spectral features in the magnetic field-driven AB oscillations in the vicinity of $5f_0$ at $\nu=5/2$ and $7f_0$ at $\nu=7/2$ should be an unmistakable signature of non-Abelian nature of those states. Observations of high-frequency spectral peak(s) both in the previous~\cite{Willett2013b} and present studies can then be interpreted not only as a confirmation of the non-Abelian nature of the $\nu=5/2$ state but also as a validation of the results of the earlier side-gate studies as the main conceptual criticism of those was rooted in doubts about the fusion channel stability~\cite{Rosenow2008a}.

\section{Methods}
\label{sec:methods}
Two wafer types have been employed in our study: shielded well and doping well wafers; see Section~\ref{sec:S2} of Supplementary Materials. The shielded well densities are all near $4\times 10^{11}\text{cm}^{-2}$ ;  the single doping well wafer density is $2.7\times 10^{11}\text{cm}^{-2}$. Note that the electron densities of these wafers, and in particular of the shielded well wafers, are substantially larger, by factors of 1.5 to 7, than those used in other edge interferometry measurements~\cite{Mcclure2012a,Zhang2009a,Ofek2010,Nakamura2019,Nakamura2020,Sivan2018}. Wafer mobilities all exceed $25\times 10^{6}\text{cm}^2/Vs$.  Table~\ref{tab:S2-T1} of Supplementary Materials contains details of density and respective wafer mobilities.  All measurements have been performed in dilution refrigerators with base temperatures $20\,\text{mK}$ to $25\,\text{mK}$. Unless otherwise stated the measurement temperature is at or near these base temperatures.

Several essential unique experimental methods contribute to the results in this study and are outlined here, followed by description of interference device operation needed to understand the results.  Specific heterostructure designs and growth features are crucially important: breakthrough improvement in the purity of Al in the GaAs/AlGaAs heterostructure quantum wells, consequently increasing the electron correlation effects, and placement and electron population of conducting layers parallel to the principal quantum well to suppress Coulomb blockade.
Aluminum purity in our heterostructures was improved by first assessing the oxygen impurity levels in the heterostructure AlGaAs layers~\cite{Chung2018}, then developing methods of offline Al-effusion furnace bakes to reduce these charged impurities.  These bakes ultimately resulted in AlGaAs barrier material used in the heterostructures of this study with about eight times fewer impurities than previous material~\cite{Chung2018a}.  In these extreme high purity materials, and in previously grown heterostructures~\cite{Willett2007,Willett2009a,Willett2010a,Willett2013a,Willett2013b}, a unique multiple conduction layering structure was employed for materials used in our interference measurements. Parallel to the principal quantum well both above and below, poorly conducting layers are grown, which suppresses Coulomb blockade or domination of the interferometer’s laterally confined electron layer in the principal well. (A similar charge shielding approach was employed in recent studies~\cite{Nakamura2020} to suppress Coulomb domination in the low density samples.) These parallel layers are populated differentially by the doping layers, and illumination of the samples at low temperatures contributes further to that population: such illumination of the samples is distinctly unique to our method, as is the aluminum purification, versus all others~\cite{Dolev2008a,Radu2008a,Venkatachalam2011,Banerjee2018,Bhattacharyya2019,Mcclure2012a,Camino2005b,Camino2007a, Nakamura2019,Nakamura2020}.
Different illumination, cool-down, and gating histories for a given sample can produce different electron populations in the layers. Each such history is numbered and in the results sections is referred to as the preparation number for each sample.  See Section~\ref{sec:S2} of Supplementary Materials for details on heterostructure construction and illumination and on the Al purification method for this study. Also see Section~\ref{sec:S2} for details of the preparation histories applied to samples used in our measurements.

\begin{figure}[thb]
\centering
    \includegraphics[width=0.6\columnwidth]{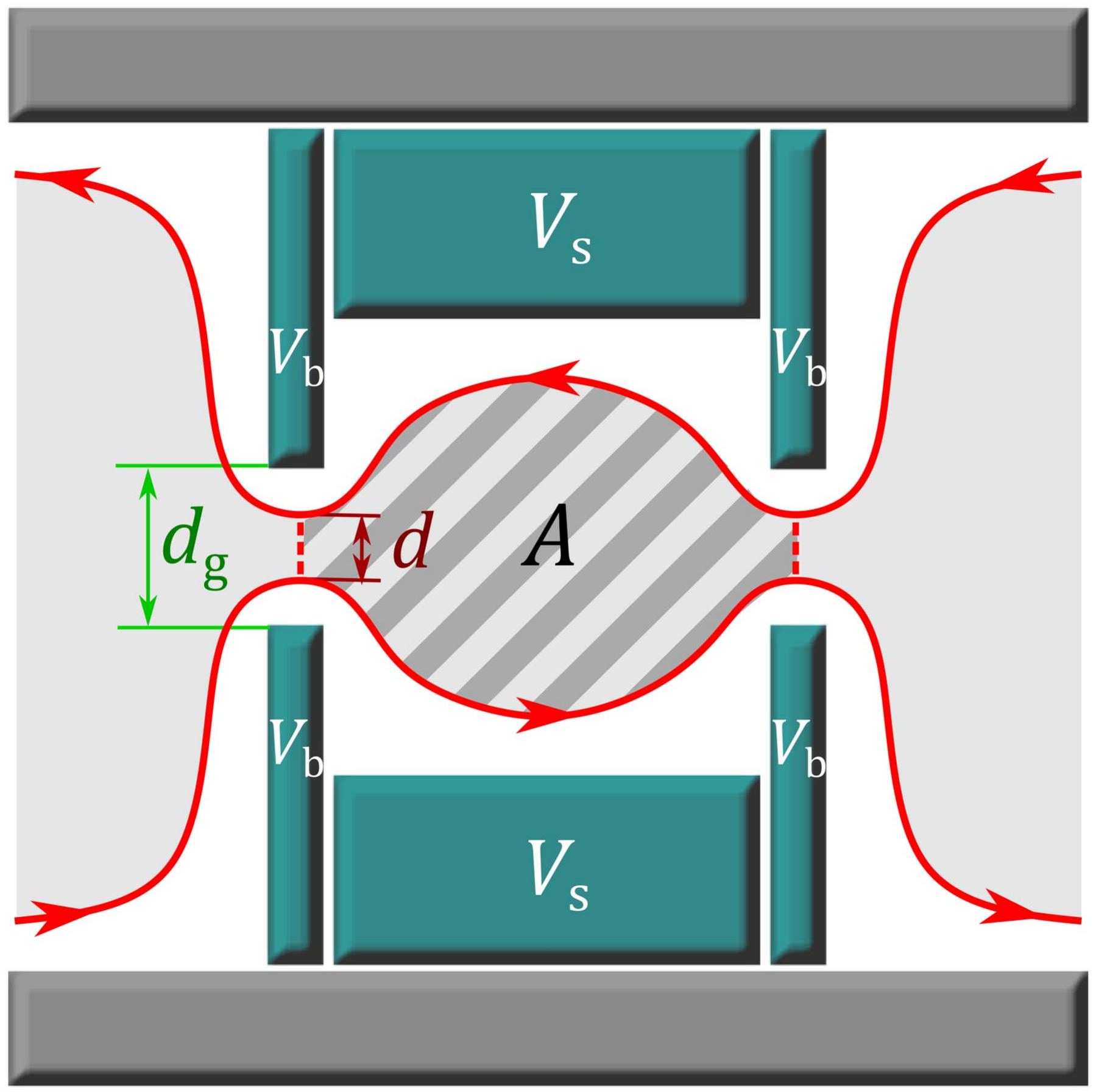}
    \;\;
    \includegraphics[width=0.35\columnwidth]{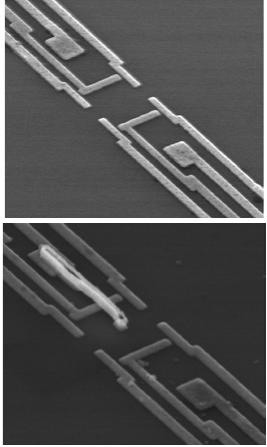}
    \caption{
    % Schematic and images of interferometer devices, transport through the devices in high purity Al heterostrutures,
    % large interference oscillations at 7/2 filling factor.
Schematic and electron-micrograph images of interference devices.  The interferometer is defined by surface gates operated at voltages $V_b$ and $V_s$ that deplete the electron population below them.   Currents propagate along the edges separating the incompressible QH liquid (shaded) and the depleted regions; contacts diffused into the heterostructure away from the device are used to measure resistance across the device longitudinally, $R_L$, and in Hall configuration across the device, $R_D$. Edge current can backscatter at one of the two constrictions resulting in interference of the two paths. The area between the two paths -- the active area $A$ of the interferometer -- is indicated by stripes; the filling factor within the active area is the same as in the shaded bulk regions on either side.  Device images show one device with no structure inside area $A$ (top -- device type a) and another one with a top gate central dot (device type b). In each device the lithographic separation $d_g$ of gates defining the constrictions is $\sim \!\!1\upmu$m. The actual tunneling distance between the edge currents $d$ is controlled by the gate voltage $V_b$.
}
    \label{fig:interferometer}
\end{figure}

An interference device is shown schematically in Figure~\ref{fig:interferometer}. The top gates are charged to a negative voltage sufficient to deplete the underlying electron layer.  At high magnetic fields as prescribed for filling factor $\nu=5/2$ or $7/2$ (filling factor $\nu$ being the ratio of the electron areal density to the magnetic flux density), the currents carrying the excitations of the fractional quantum Hall state will travel along the edge of these depleted areas, surrounding an area of the bulk filling factor $\nu$. The important principal physical property of the interferometer device is two separated locations where these edge currents are brought in proximity of one another.  At these points backscattering from one edge to the other can occur, and with this backscattering two different current paths are established that can interfere, as shown by the dashed lines in the schematic.  The one path encircles the area marked $A$ in the schematic, and changes in the magnetic flux number within area $A$ or changes in the particle number within area $A$ will cause phase accumulation for that path (the Aharonov--Bohm and statistical phase contributions). Interference of that path and the one not entering the area $A$ produce oscillations in the resistance measured across the interference device.  The voltages on the top gates can be adjusted to promote backscattering (gates marked $V_b$) and to change the enclosed area $A$ (gates marked $V_s$).  The separation of the backscattering top gates, distance marked  $d_g$ in Figure~\ref{fig:interferometer}, is sufficiently large that for nominal voltages on $V_b$ the backscattering is weak, an important feature to maintain the 5/2 fractional Hall state contiguously from outside to inside the active area $A$ of the interferometer.  Note also that area $A$ is ultimately the area which is enclosed by the edge states in the quantized Hall systems, and not the lithographic area.  The location of the edge states is determined electrostatically and can be modified by the applied gate voltages.

Tunneling between the inner-most edge currents that surround the region of bulk electron density (depicted as shaded in Figure~\ref{fig:interferometer}) at the two constrictions results in the interference between two possible paths for the backscattered current.
Although the lithographic area is several square microns, the active area $A$ is typically less than one square micron.
The experimental evidence that the desired $\nu=5/2$ and $7/2$ QH states persist inside the active area of the interferometer is shown in Section~\ref{sec:S4} of Supplementary Materials.
In one interferometer device type a small dot is placed centrally in the area $A$ and is accessed by an air-bridge that extends over one of the side gates marked $V_s$ in Figure~\ref{fig:interferometer}. Although such type b devices (i.e. those with a top central gate) are present in several of the samples used in this study, the central gate was kept grounded for all the measurements and preparations for the purposes of obtaining the data presented in this paper.  Also shown are electron micrographs of the interferometers.

Resistance and resistance oscillations are measured using low noise lock-in amplifier techniques.   A constant current (typically 2nA) is driven through the 2D electron system underlying the interferometer top gate structure, and the voltage, and so resistance, is determined with a four-terminal measurement.  The voltage drop along the same edge of the 2D electron system and across the device gives the longitudinal resistance $R_L$; across the device and across the two edges of the 2D system gives diagonal resistance $R_D$.  Similar measurements performed away from the interferometer device yield $R_{xx}$ and $R_{xy}$ respectively.

\begin{figure}[htb]
\centering
    \includegraphics[width=0.85\columnwidth]{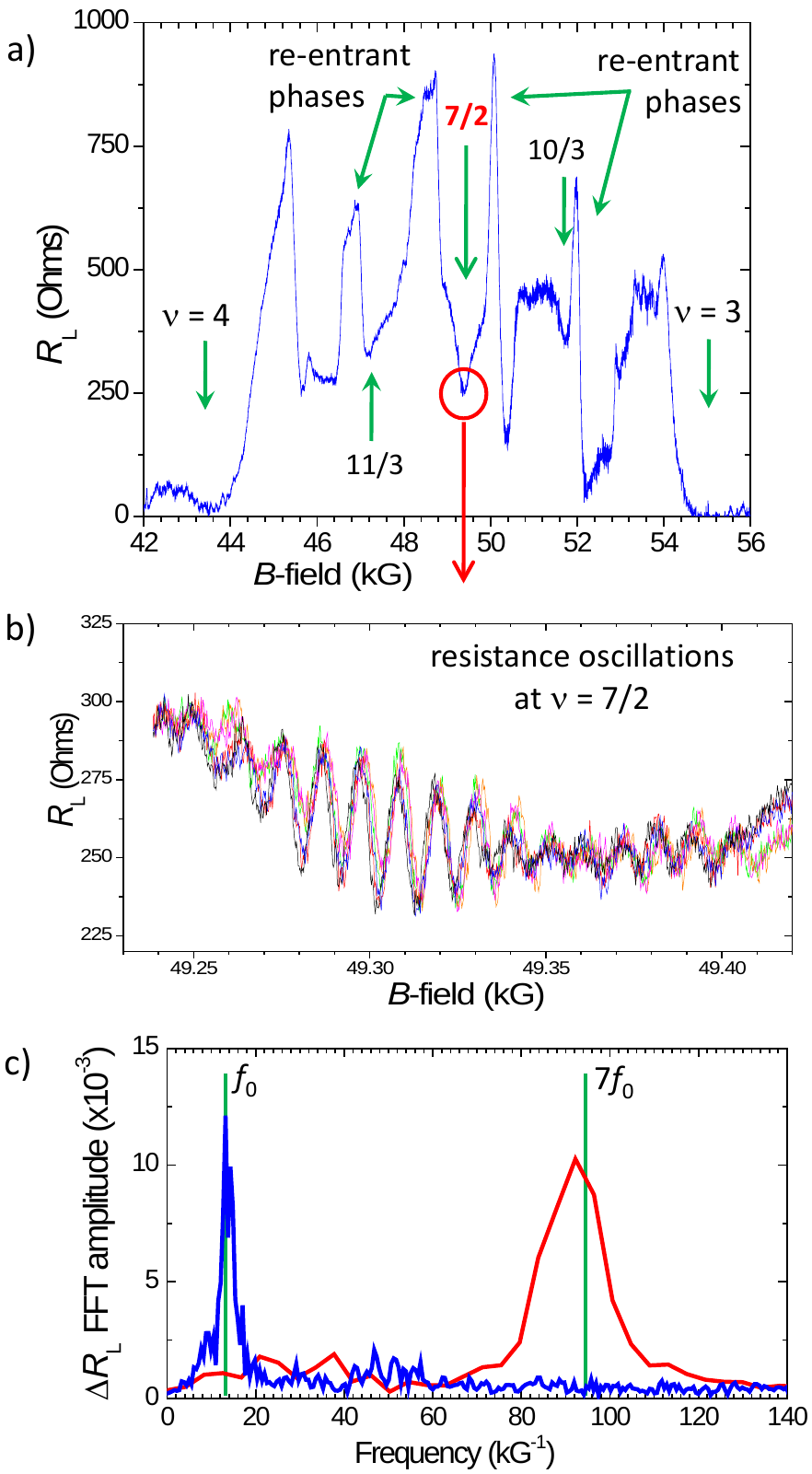}
    \caption{
    (a) Longitudinal resistance $R_L$ measured across the interferometer between filling factors $\nu=3$ and $\nu=4$ in an ultra-high mobility heterostructure.  Integer and fractional quantum Hall states as well as phase-separated (nematic) states  are labeled. Temperature $\sim20$mK, sample 6, preparation 16, device type b.\\
    (b) A blow-up of $R_L(B)$ near the filling factor $7/2$ demonstrating large-amplitude interference oscillations. Trace sets show oscillations near $\nu=7/2$ for  both up- and down- sweeps of magnetic field $B$.  The set of six sweeps shown here covers a total time of 8.75 hours, or 87.5 minutes per each directional sweep. The data demonstrate both a high level of reproducibility and stability for the time and magnetic field ranges over which the data were taken. See also Figures~\ref{fig:S5-2-2}, \ref{fig:add_osc1}, \ref{fig:add_osc2}.\\
    (c) A Fourier transform of these oscillations (red) overlayed with the Fourier transform of oscillations observed at $\nu=3$ (blue). Vertical green lines mark the frequency $f_0$ of the integer spectral peak and its multiple $7f_0$, the expected frequency of the non-Abelian even-odd effect at $\nu=7/2$.
}
    \label{fig:Hall_trace}
\end{figure}

An example of longitudinal resistance $R_L$ across an interferometer in an ultra-high mobility heterostructure characterized by improved Al purity is shown in Figure~\ref{fig:Hall_trace}(a). (Also see Section~\ref{sec:S4} of Supplementary Materials.) In comparison to heterostructures without improved Al purity (see Figure~\ref{fig:power_spectra} and Figure~\ref{fig:S5-2-1}), this resistance trace shows sharper resistance features throughout this filling factor range. A blow-up of the $R_L$ trace near $\nu=7/2$ reveals a set of reproducible oscillations shown in Figure~\ref{fig:Hall_trace}(b). Their Fourier transform reveals a prominent peak at a frequency roughly seven times that of the main spectral peak observed at an integer filling fraction, as shown in Figure~\ref{fig:Hall_trace}(c)~\footnote{In magnetic field sweeps $R_L$ measurements at integer filling factors show AB interference oscillations that correspond to addition of one flux quantum and the period of these oscillations is $\Phi_0$ with the fundamental frequency $f_0=1/\Phi_0$.  For each preparation such a measurement at integer filling factor is performed to establish this fundamental value for comparison.  The fundamental frequency $f_0$ is independent of the integer filling factor (see Figure~\ref{fig:S4-2a} of Supplemental Materials). It is important to note that if a density shift occured in the 2D electron system this would affect the measured $f_0$ making comparison between interference features at different densities untenable.}. A similar set of data for $\nu=5/2$ is shown in Figure~\ref{fig:5_halves_interf} in the Results section below. The analysis of these oscillations, their spectra and their attribution to the non-Abelian even-odd effect is the main focus of this paper.

\begin{figure}[t!]
\centering
    \includegraphics[width=0.9\columnwidth]{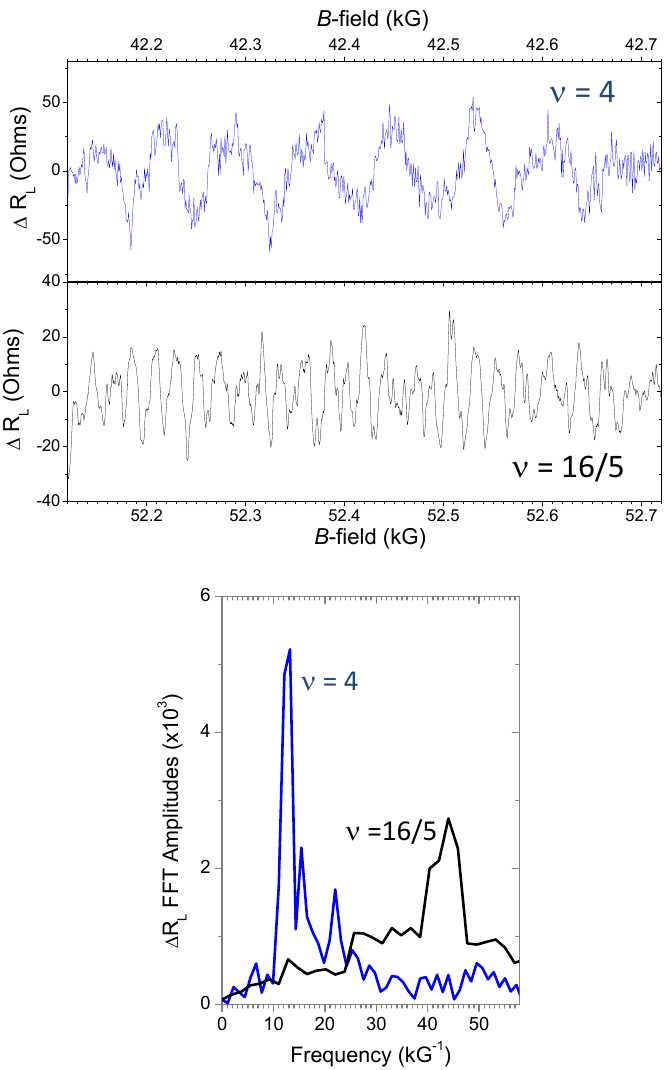}
    \caption{A comparison of magnetic field sweeps and their Fourier transforms at $\nu=4$ and $\nu=16/5$ in the same sample (sample~6, preparation~25, device type~b, $T \sim 20$mK). A dominant spectral feature observed at $\nu=16/5$ is located at approximately three times the frequency of the integer peak, as expected from Eq.~(\ref{eq:AB_phase2}).}
    \label{fig:integer_vs_fractional_peak_positions}
\end{figure}

The interference oscillations in the measured resistance are analyzed by applying fast Fourier transforms (FFT) to the data.  Because the oscillations are observed near the minima of resistance of quantum Hall states, the corresponding background – the shape of the minimum - is subtracted before the FFT is applied.  The subtracted background is determined equivalently by either a polynomial fit or a running large element smoothing of the minimum.
By following this procedure at both integer ($\nu=4$) and fractional ($\nu=16/5$) filling, we can test the validity of our approach and, specifically, confirm the expectation that our interferometers operate in the Aharonov--Bohm and not Coulomb-dominated regime, thus justifying the key assumption used in deriving Eq.~(\ref{eq:AB_phase2}). Specifically, at the $\nu=16/5$ FQH state one expects a Laughlin state with $e^*=e/5$ and $2\theta_{e^*}=2\pi/5$.  Consequently, Eq.~(\ref{eq:AB_phase2}) predicts the phase accumulation of $-6\pi$ per additional flux quantum resulting in the expected AB periodicity of $\Delta\Phi = \Phi_0/3$, which is what we observe experimentally -- see Figure~\ref{fig:integer_vs_fractional_peak_positions}.

\begin{figure}[htb]
\centering
    \includegraphics[width=0.85\columnwidth]{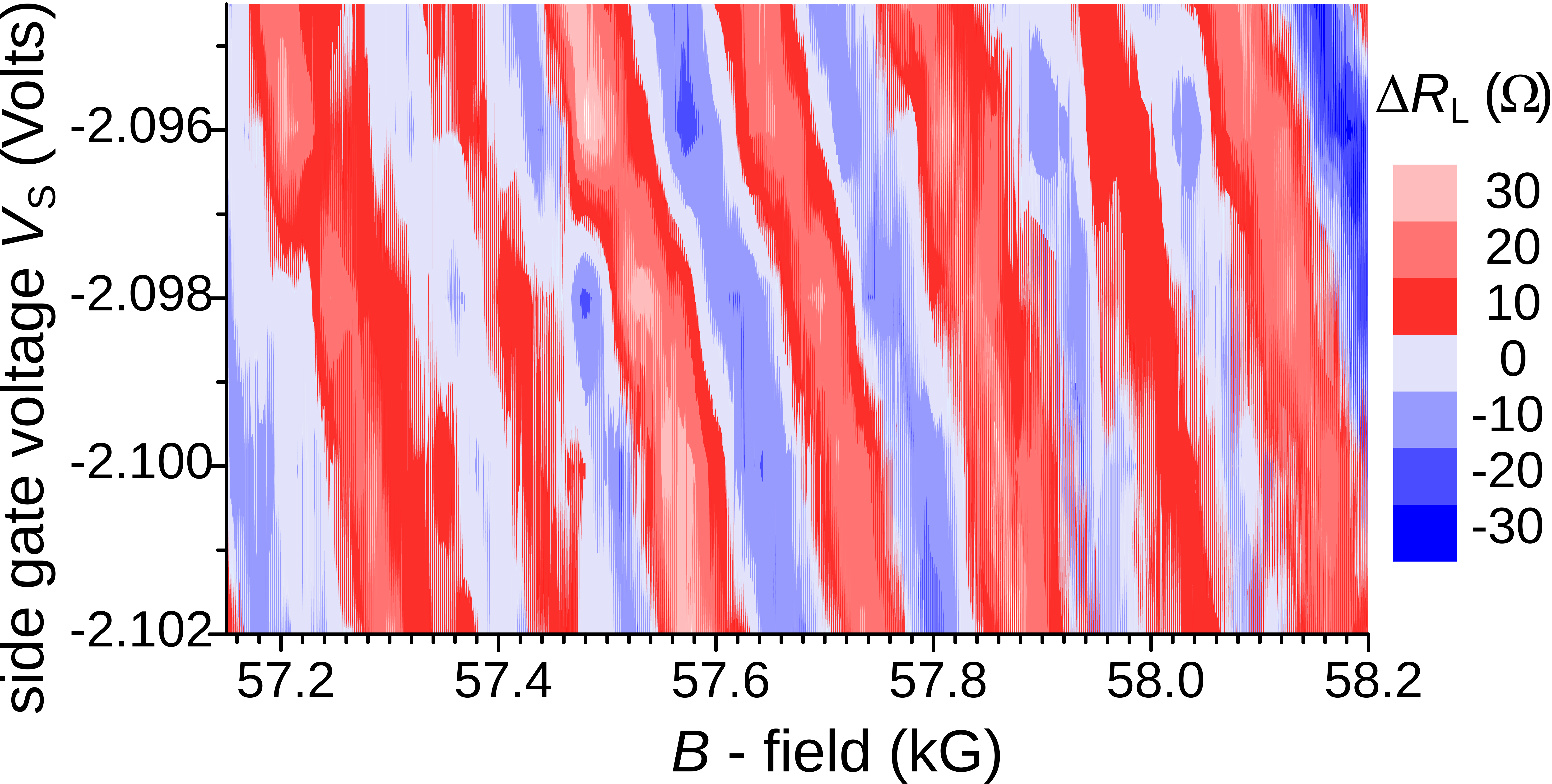}
    \includegraphics[width=0.85\columnwidth]{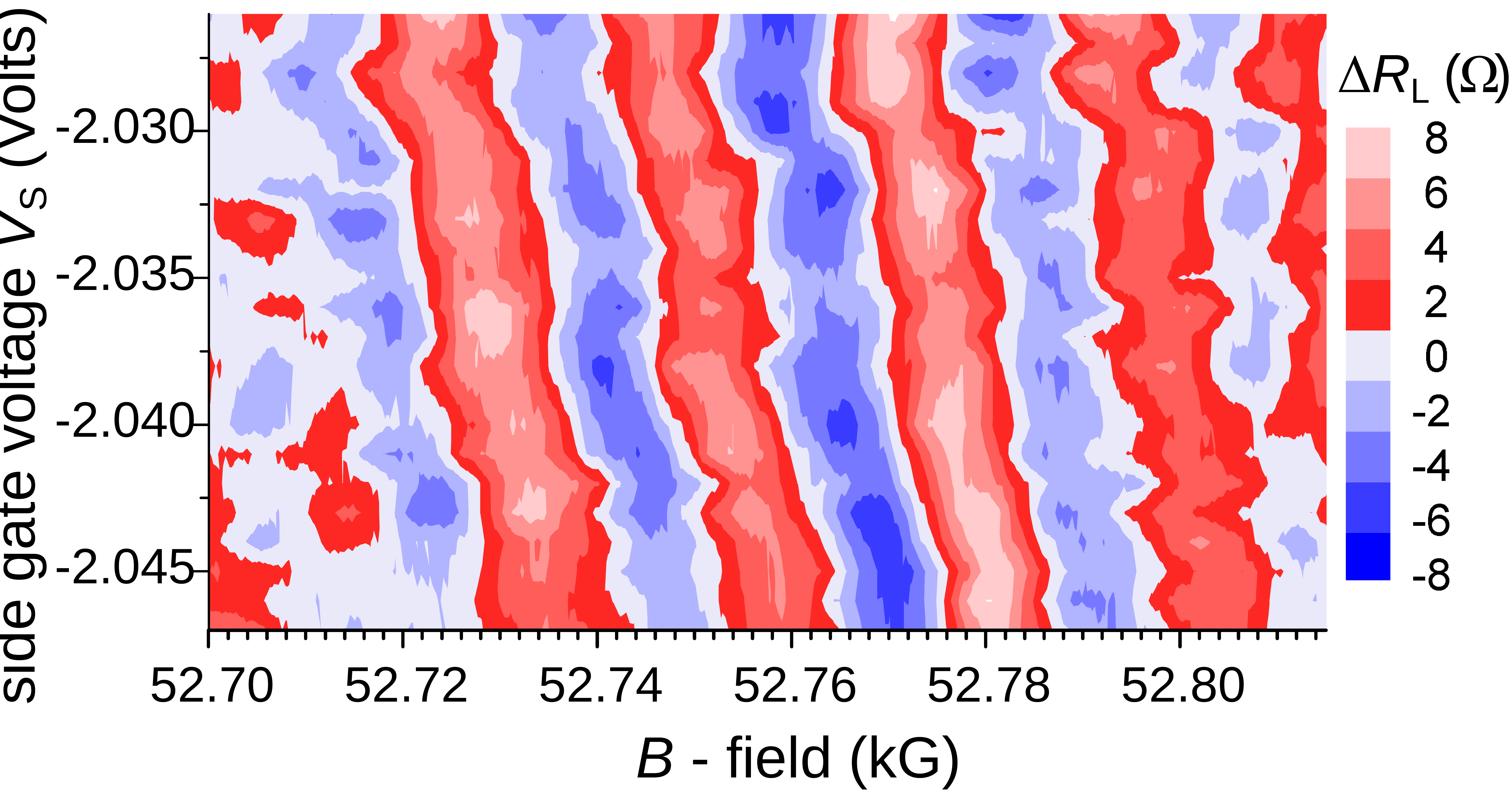}
    \caption{``Pajama plots'' showing the resistance oscillations at $\nu=3$ (top) and $\nu=16/5$ (bottom) as a function of both magnetic field $B$ and side gate voltage $V_s$ consistent with the Aharonov--Bohm as opposed to Coulomb dominated nature of observed oscillations (top panel: sample~6, preparation~3, device~b;  bottom panel: sample~6, preparation~25, device~b, for which the integer oscillations are shown in Figure~\ref{fig:integer_vs_fractional_peak_positions}). Note that the integer period (top panel, preparation~3) indicates an active interferometer area roughly half that of the device used in bottom panel (preparation~25), yet still demonstrating a negative phase slope indicative of the AB regime.}
    \label{fig:pajama_plot}
\end{figure}

Further evidence of the Aharonov--Bohm nature of the observed interference is provided by the ``pajama plots'' allowing one to trace the lines of constant phase in the $B-V_s$ plane. For the Aharonov--Bohm oscillations, one expects these lines to have a negative slope whereas Coulomb domination should result in a positive slope~\cite{Halperin2011a}; the data shown in Figure~\ref{fig:pajama_plot} for both $\nu=3$ and $\nu=16/5$ are clearly consistent with the former.  (Figure~\ref{fig:pajama_plot} shows data for the same sample but different preparations; different preparations can have different active areas $A$ which can be determined from the periodicity of interference oscillations in any of their integer QH states.) Note that while it is not entirely unreasonable to expect different energetics in integer and fractional QH states due to e.g. different width of their edge states, it is much harder to fathom a scenario whereby two nearby fractional QH states would produce drastically different energetics. The $\nu=16/5$ state has been chosen as a prominent FQH state close to $\nu=7/2$ whose expected Abelian nature makes the interpretation of the pajama plot sufficiently straightforward.  More details on the Aharonov--Bohm nature of the interference observed at integer filling factors in our samples are presented in Section~\ref{sec:S4} of Supplementary Materials.

\begin{figure}[thb]
\includegraphics[width=0.95\columnwidth]{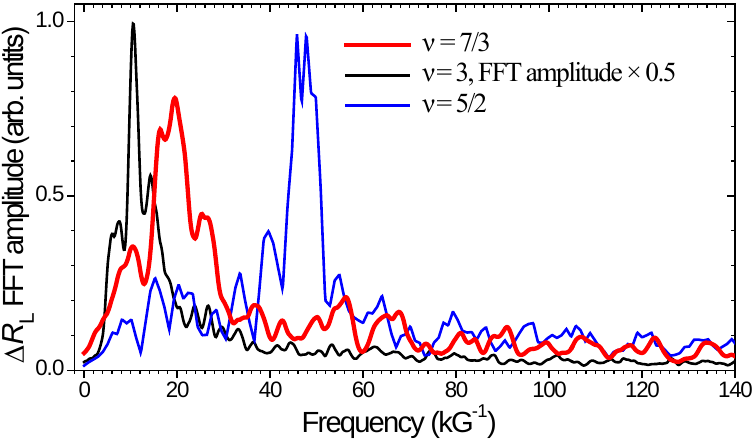}
\caption{Absence of significant high-frequency spectral features in the oscillations at $\nu=7/3$ (red) in the same sample, preparation and magnetic field trace where a marked peak in the vicinity of $5f_0$ is observed at $\nu=5/2$ (blue). The FFT of oscillations at $\nu=3$ in the same sample is shown in black for comparison; it establishes the fundamental frequency $f_0$ (sample~6, preparation~36, $T\sim 25$mK).}
\label{fig:seventhirds}
\end{figure}

Finally, we also present results of a direct test that rules out a mechanism that could produce high-frequency spectral features at $\nu=5/2$ and $7/2$ due to the oscillations of the active area of the interferometer upon changes of the magnetic field. As mentioned in the previous section and further explained in Section~\ref{sec:S1}, such a mechanism relies on the periodic oscillations of the active area of the interferometer upon introduction therein of additional Laughlin quasiparticles. Assuming the energetics behind such a scenario do not change dramatically with the filling fraction, this mechanism would also produce high frequency oscillations at $\nu=7/3$ -- either at $f=7f_0$ or $f=(7\pm2)f_0$ depending on whether the oscillations originate from merely changing the tunneling geometry or from geometric oscillations modulating quantum interference. As can be seen in Figure~\ref{fig:seventhirds}, the FFT spectrum of the oscillations observed at $\nu=7/3$ shows an expected peak near $2f_0$ (as determined from the integer spectrum) due to the interference of charge $e/3$ quasiparticles but no marked features in the vicinity of $7f_0$. Crucially, the same sample shows significant features near $5f_0$ at $\nu=5/2$ in the Fourier transform of \emph{the same magnetic field trace} (shown in Figure~\ref{fig:add_seven_thirds} of Supplementary Materials), which rules out the Coulomb domination as the mechanism behind those oscillations.

To conclude, we would like to reiterate the importance of two fundamental improvements between the new high Al purity shielded wells heterostructures and the previously used shielded well samples: the high Al purity samples can display a ten-fold increase in the amplitude of the interference oscillations at $\nu=7/2$ and $5/2$ compared to those previously observed in shielded-well samples, and, furthermore, the high Al purity materials also demonstrate sharper definition of the fractional states and the reentrant phases.  This later point is shown comparing the $R_L$ data in Figures~\ref{fig:Hall_trace}(a) and \ref{fig:5_halves_interf} versus that of Figure~\ref{fig:power_spectra}(a) (comparison is also made in supplemental Figure~\ref{fig:S5-2-1}). Figure~\ref{fig:power_spectra}(a) shows only continuous evolution in $R_L$ from the re-entrant phases to, for instance, the 5/2 minimum, in stark contrast to the abrupt changes in $R_L$ in sweeping the magnetic field from the re-entrant phases to $\nu=5/2$ shown in Figure~\ref{fig:5_halves_interf}. In Figure~\ref{fig:Hall_trace}(a) the transitions from re-entrant phases to $\nu=7/2$ are also distinct.  It is posited but not proven that this relative sharpening of the re-entrant features, less mixing with the target $\nu=5/2$ and $\nu=7/2$ states, may contribute to the larger amplitude of the oscillations at those FQH states.
% The dramatically larger amplitude of the interference oscillations in the high Al purity materials can be compared to those of the shielded wells used in Figures 10 and 12 from the data in Ref.~\cite{Willett2013b}.

\section{Results}
\label{sec:results}
%The results are presented in the following progression:\\ i) introduction of high stability interference oscillations at $\nu=7/2$ attributed to non-Abelian $e/4-e/4$ braiding -- Figure~\ref{fig:Hall_trace}(b), ii) interference oscillation spectra at $\nu=5/2$ showing previously undocumented low frequency peaks associated with Abelian braiding, in addition to the peaks associated with non-Abelian braiding -- Figure~\ref{fig:power_spectra}, iii) spectral peak identification -- Figure~\ref{fig:peak_identification}; the spectral properties of the $7/2$ interference oscillations -- Figures~\ref{fig:7_halves_interf} and \ref{fig:interf_model}(b),  iv) direct measurement of $\pi$ phase jumps in interference oscillations attributed to fermion parity change -- Figure~\ref{fig:fermionic_parity}, and v) demonstrated control of interference at the frequency attributed to $e/2$ - $e/2$ braiding -- Figure~\ref{fig:2f0_peak}.

\subsection{High-frequency spectral features of interference oscillations at $\nu=7/2$}
\label{sec:results_oscillations}

\begin{figure}[t]
\centering
    \includegraphics[width=0.95\columnwidth]{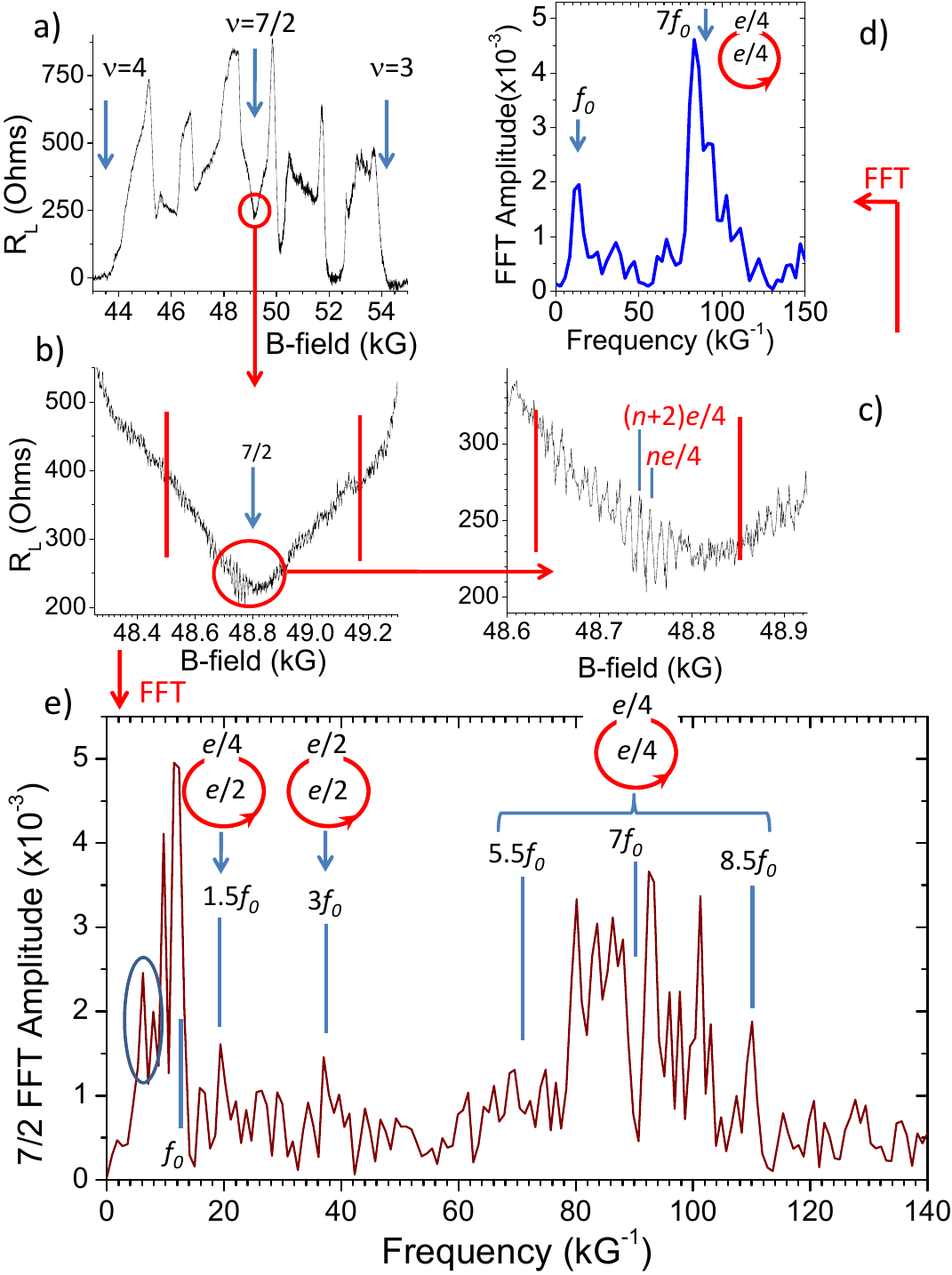}
    \caption{Interference oscillations at $\nu=7/2$ and their power spectra.\\
(a) Transport between filling factors $\nu=3$ and $\nu=4$.\\
(b)\&(c) Successive blow-ups of $R_L(B)$ around the minimum at $\nu=7/2$, demonstrating sets of interference oscillations (sample~6, preparation~8, device type~b, $T \sim 20$mK).\\
(d) The Fourier spectrum of the sweep in panel (c) with the FFT window marked by the red vertical lines. $f_0$ marks the position of the FFT peak at $\nu=3$ for this sample and preparation (see Figure~\ref{fig:S5-2-1}).
\\
(e) An FFT using a larger range of $B$-field centered around $\nu=7/2$ can potentially express the full complement of $e/4$  and $e/2$ braids as shown for $\nu=5/2$ in Figure~\ref{fig:power_spectra}. The FFT window corresponding to $\nu = 3.5\pm 0.03$ is marked by red vertical lines panel (b). The vertical lines marked $1.5f_0$, $3f_0$, $5.5f_0$, $7f_0$, and $8.5f_0$  are the respective multiples of $f_0$ identified at the integer filling, which correspond to expected features at $\nu=7/2$ (see text for details). A broad spectral feature is centered around $7f_0$, with a sharp \textit{minimum} developing at that frequency.
Additional $\nu=7/2$ spectra are shown in Figure~\ref{fig:interf_model}(b) and in Section~\ref{sec:S5-4} of Supplementary Materials, all demonstrating a large spectral feature centered at their respective $7f_0$ frequencies, as expected for non-Abelian $e/4$ at $\nu=7/2$.
}
    \label{fig:7_halves_interf}
\end{figure}
  	
We now turn to our main findings, beginning with the analysis of the interference oscillations for the first time observed at $\nu=7/2$. Figure~\ref{fig:Hall_trace}(a) depicts a representative quantum Hall trace between integer fillings $\nu=3$ and $4$ observed in a higher purity sample. It shows a well-defined feature at filling fraction = $7/2$.  Zooming into the corresponding resistance minima reveals the oscillating behavior of the resistance $R_L$ as a function of the magnetic field. As can be seen in Figure~\ref{fig:Hall_trace}(b), the observed oscillations are both prominent, with the typical amplitude of order $20\Upomega$, and remarkably reproducible – the figure presents the results of six different magnetic field sweeps separated from one another by hours. We should also note the generic nature of these oscillations: they are seen in different preparations and within a significant range of gate voltage settings -- see Figures~\ref{fig:add_osc1} and \ref{fig:add_osc2} of Supplementary Materials. The most significant feature of these oscillations is their frequency. Indeed, as we will demonstrate, it is inconsistent with any Abelian AB interference but is consistent with the non-Abelian even-odd effect.

In order to analyze the origin of these oscillations, we must first establish the reference frequency for our interferometer.  Because we do not know the precise active area $A$ of the interferometer, we rely on the oscillation spectra measured at integer filling factors to extract the reference frequency $f_0$ -- see e.g Figures~\ref{fig:Hall_trace}(c) and~\ref{fig:integer_vs_fractional_peak_positions} or Figure~\ref{fig:S5-2-1} of Supplementary Materials.  Since only electrons can interfere in the integer QH regime, the oscillation period observed as the magnetic field is swept across the integer plateau corresponds to the change of one flux quantum through the active area of the interferometer.
This procedure has been repeated for each sample, each cool-down, and each gate voltage setting (see also Figure~\ref{fig:power_spectra}, which exemplifies this procedure).  Equipped with the knowledge of the electron AB oscillation frequency $f_0$ we proceed to examine the oscillation spectra at filling fractions $7/2$ and $5/2$.

Having established the reference frequency $f_0$, we focus our attention to the interference signal at $\nu=7/2$ shown in Figure~\ref{fig:Hall_trace}. A crucial parameter in  the analysis of  the resistance oscillations is the width of the Fourier transform window. Figure~\ref{fig:7_halves_interf} shows a series of successive zooms into the resistance minimum at this filling and the corresponding Fourier transforms. The measurements presented here have been done using a different preparation of the same sample that was used for the measurements in Figure~\ref{fig:Hall_trace}. Similarly to what has been shown in Figure~\ref{fig:Hall_trace}, using the narrower FFT window (marked in the last of these zooms, Figure~\ref{fig:7_halves_interf}(c)) results in a prominent spectral peak seen in Figure~\ref{fig:7_halves_interf}(d) at $\sim 84\, \text{kG}^{-1}$ (or $\sim 87\, \text{kG}^{-1}$ if measured as a mid-point at half-height), the frequency of easily identifiable reproducible oscillations shown in Figure~\ref{fig:7_halves_interf}(c), as well as~\ref{fig:Hall_trace}(b). The reference frequency $f_0$ has been determined to be $\sim 13.0\pm0.5\, \text{kG}^{-1}$ in this sample, so the observed spectral peak corresponds to $7f_0$ within reasonable precision – the very frequency expected for the even-odd effect in this quantum Hall state. A larger Fourier transform window bounded by vertical red bars in Figure~\ref{fig:7_halves_interf}(b) reveals lower-frequency features in the Fourier spectrum as can be seen in Figure~\ref{fig:7_halves_interf}(e). It also points towards more complex structure of the spectral peak centered around $7f_0$, which will be discussed later.

\subsection{Temporal stability of oscillations}
\label{sec:results_stability}
One of the remarkable features of the observed oscillations attributable to the non-Abelian $e/4$ quasiparticles is their temporal stability. Generally, there are two ways of inferring the fermion parity stability: 1) direct observation of the phase stability of the $5f_0$ and $7f_0$ frequency oscillations respectively at $\nu=5/2$ and $7/2$, and 2) indirect inference via the presence or absence of spectral properties, split peaks at  $(5\pm 1)f_0$ at $\nu=5/2$ and  $(7\pm 1.5)f_0$ at $\nu=7/2$; see Section~\ref{sec:interference} and also Section~\ref{sec:S1} of Supplementary Materials). While there is some indication of such splitting in Figure~\ref{fig:7_halves_interf}(e), it is far from conclusive. In general, such indirect inference is complicated; in order to achieve the required resolution, it requires a sufficiently large magnetic field range that should be employed in the FFT, which at the same time makes the data more susceptible to low-frequency noise -- see Figure~\ref{fig:7_halves_interf}(d) versus \ref{fig:7_halves_interf}(e).

\begin{figure}[htb]
\centering
    \includegraphics[width=0.8\columnwidth]{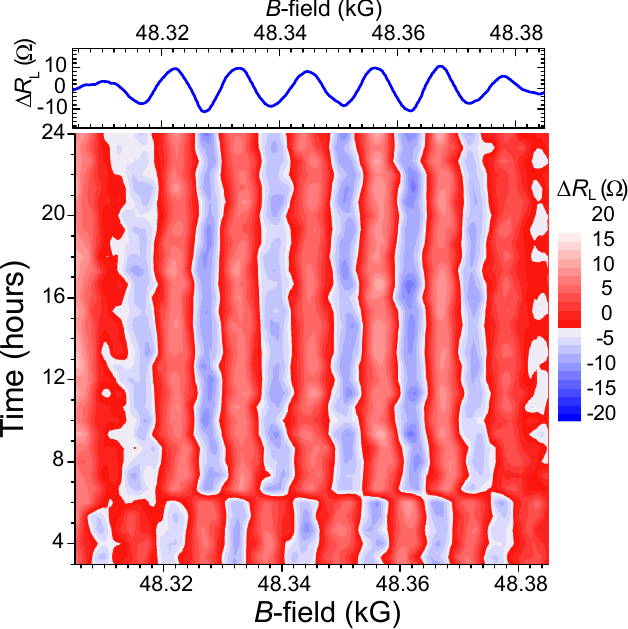}
        \caption{Color plot of $\Delta R_L$ as a function of magnetic field ($x$-axis) and time in hours ($y$-axis) for a representative interference
measurement near $\nu=7/2$. This plot represents the total of 76 up and down magnetic field sweeps similar to those of
Figure~\ref{fig:Hall_trace}(b) (for a different sample preparation) but with the background subtracted, plotted as a function of time to
demonstrate the stability of the oscillations. A $\pi$ phase jump occurs at about 6 hours, and for the
following 18 hours the oscillation set is stable (sample~6, preparation~25, device type~b, $T \sim 25$mK).
}
    \label{fig:fermionic_parity_stability}
\end{figure}

\begin{figure}[htb]
\centering
    \includegraphics[width=0.95\columnwidth]{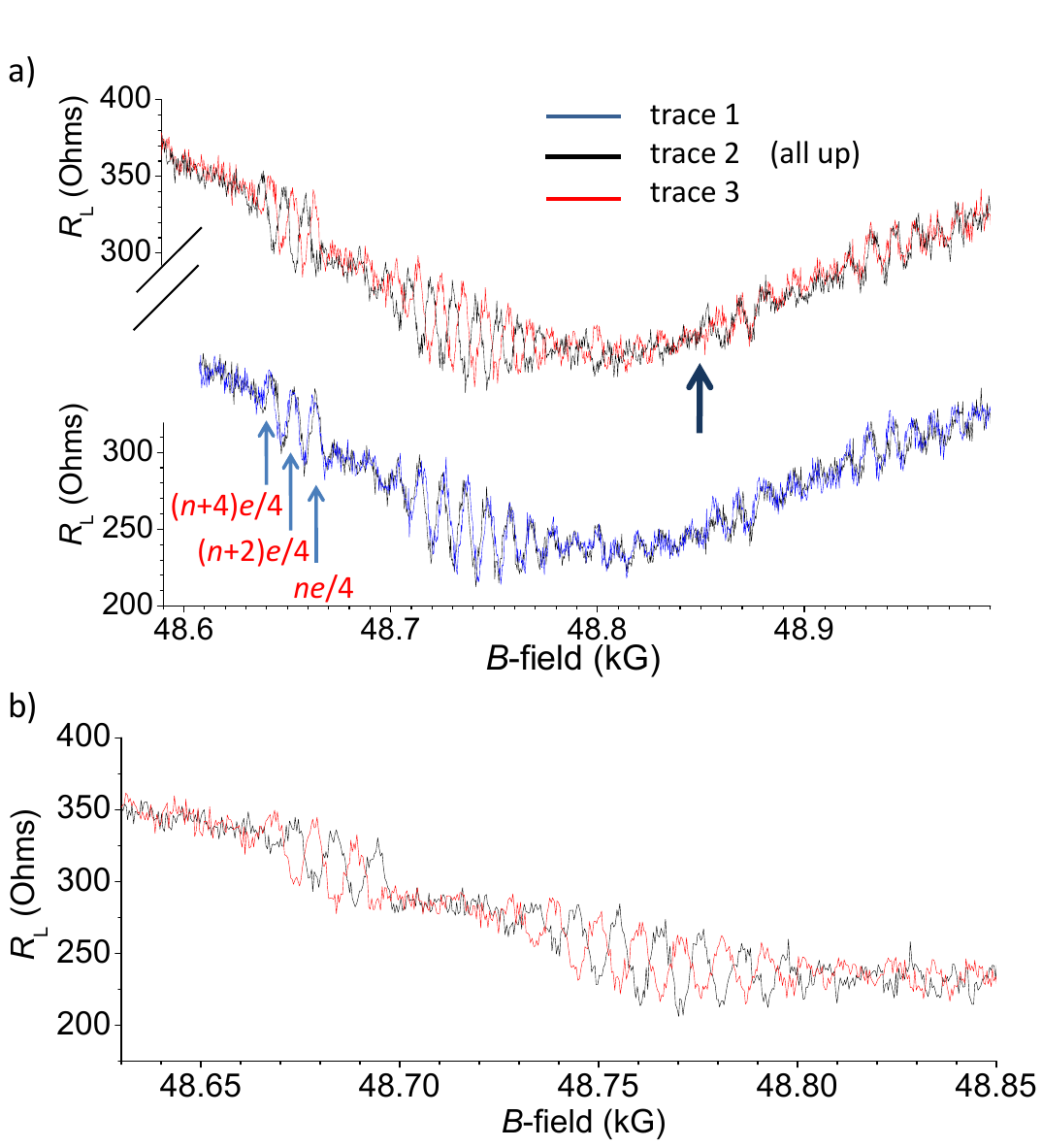}
    \;\;
    \includegraphics[width=0.95\columnwidth]{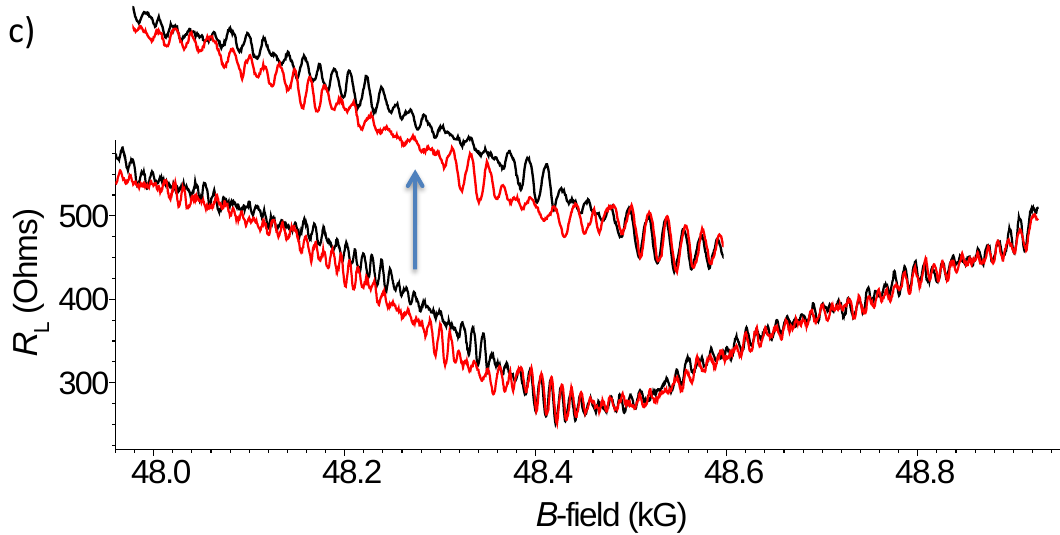}
    \caption{$\pi$ phase shifts in high-frequency oscillations at $\nu=7/2$:\\
    (a) Three consecutive magnetic field up-sweeps taken near $\nu=7/2$ as in Figure~\ref{fig:7_halves_interf} from the same sample but of a different preparation (sample~6, preparation~19, device type~b, $T \sim 20$mK). Traces 1 and 2 (bottom of panel (a)) overlap whereas  in trace~3 (top of panel (a), shown alongside trace~2 repeated for comparison) a $\pi$ phase jump has occurred in the time intervening trace~2 and 3 as this phase difference is apparent from the lowest shown $B$-field up to about 48.85kG (marked by an arrow), at which point the phase reverts to that of trace~2. (See also Supplementary Materials, Section~\ref{sec:S5-6}.)\\
    (b)  Two down-sweeps (same sample and preparation, sweeps separated by days from (a)).  The two traces of $R_L$ oscillations are out of phase by $\pi$. The $\pi$ phase jump seemingly occurred away from the depicted oscillations showing the full extent of the sweep.
    \\
    (c) Consecutive down-sweeps from the same sample and preparation (hours apart between the sweeps but days prior to the sweeps in panels (a) and (b)).
}
    \label{fig:fermionic_parity}
    \end{figure}

Nevertheless, this limitation does not preclude us from studying the fermion parity stability in the temporal domain by repeated sweeps within a limited magnetic field interval, with the results shown in Figure~\ref{fig:fermionic_parity_stability}.
Their reproducibility includes their phase which remains stable over several hours. What is even more remarkable is that when the phase of the oscillations fluctuates, it happens predominantly through phase jumping by $\pi$, as can be seen in Figures~\ref{fig:fermionic_parity_stability} and \ref{fig:fermionic_parity}(a). A $\pi$-shift in the oscillations attributable to the $e/4$ particles can come from two distinct sources. The two scenarios correspond to the fluctuating parity of either neutral fermions or non-Abelian $e/4$ particles inside the interferometer. To better clarify the difference between these mechanisms, we remind the reader that we attribute the high-frequency oscillations not to the AB interference per se but rather to the switching of the interference signal on again, off again when the number of $e/4$ particles inside the interferometer loop changes from even to odd and back to even in response to the changing magnetic field. A stray $e/4$ quasiparticle tunnelling into or out of the active area shifts this pattern by half a period due to the change in the parity of these quasiparticles. Meanwhile, the fermionic parity is the combined property of an even number of $e/4$ quasiparticles; it determines the sign of the interference contribution whenever it is present; its fluctuations would flip this sign. Hence while both mechanisms would result in the oscillations shifted by $\pi$, there is, in principle, a difference in their signatures: fluctuations in the number of $e/4$ particles would result in a purely ``horizonatal'' shift of the oscillations whereas in the case of fluctuating fermionic parity one would expect a ``vertical'' reflection of these oscillations about either their peak or their minimum values.
Such $\pi$ jumps can be seen in Figure~\ref{fig:fermionic_parity}, with the observed jumps shown in Figure~\ref{fig:fermionic_parity}(a,b) seemingly resemble those resulting from the fluctuations in the number of $e/4$ particles. Figure~\ref{fig:fermionic_parity}(a) represents three different magnetic field sweeps taken over eight hours, each individual trace takes about 100 min.   Two of them are perfectly in phase while the third one displays a $\pi$ phase jump whose location is indicated by an arrow. Two more sweeps performed in the same sample clearly show a similar $\pi$ phase jump in Figure~\ref{fig:fermionic_parity}(b). Meantime the $\pi$ shift seen in Figure~\ref{fig:fermionic_parity}(c) appears more consistent with the fluctuating fermionic parity. Definitive discrimination between these mechanisms, however, requires a more systematic study: the out-of-phase regions of oscillations seen in panel (c) also appear to be related to one another by a horizontal shift (rather than vertical reflection); such a study is currently under way.
Similar behavior is also observed at $\nu=5/2$: runs of reproducible oscillations are punctuated by occasional $\pi$ shifts, the predominant instability – see Figure~\ref{fig:5_halves_interf} and Section~\ref{sec:S5-6} of Supplementary Materials, Figures~\ref{fig:S5-6-1} and \ref{fig:S5-6-2}.

This direct observation of the oscillation stability gives us the means of establishing the temporal stability, the magnetic field dependence of that stability by examining different magnetic field intervals, and influence of a host of other parameters, such as different gating configurations.  The fact that we can actually see the rare phase jumps and assign their characteristic timescale is significant: this timescale provides the lower bound for the fermion parity stability which, in turn, not only further validates our theoretical model but may also have profound consequences for the future of such systems in quantum computation. After all, the fermion parity is supposed to encode the state of a topological qubit; its remarkable stability is a key to its potential utility.

Given the potential importance of the $\pi$ phase jumps, it is essential to ask what experimental parameters can control their incidence and prevalence. One parameter conceivably influencing these $\pi$ phase jumps – the density of encircled  $e/4$ quasiparticles – can be estimated from the collected data. By simply counting of the $7f_0$ oscillation periods from the center of the  $7/2$ FQH plateau, and using  $f_0$ to determine area $A$ (see Figure~\ref{fig:power_spectra}), we estimate their density to be $\sim 170 \upmu\text{m}^{-2}$ towards the edge of the plateau. This puts their separation at $\sim 0.07\upmu\text{m} = 70\text{nm}$ for the maximum density, which corresponds to about 5 magnetic lengths at these magnetic fields.  (For more detail on the calculations see Supplementary Materials, Section~\ref{sec:S5-6}.) This suggests that the oscillations occur over a range of reasonable separations between  $e/4$ quasiparticles. Yet it does not provide any indication of a critical density that should trigger phase jumps.
%A more systematic study of these phase jumps is presently underway.

We should comment in passing on the crucial difference between these $\pi$ shifts  and the phase jumps   reported in \cite{Nakamura2020}. The latter jumps are purportedly the consequence of charged Laughlin quasiparticles appearing within the active area of an interferometer as a consequence of changing magnetic field. In our case, due to the different energetics, the number of charge-$e/4$ quasiparticles is changing continuously with magnetic field and is \emph{causing} the observed oscillations. Meanwhile, the $\pi$ jumps occur either due to the fluctuating number of \textit{neutral} fermions or \emph{stray} charge-$e/4$ quasiparticles, which are expected in both $\nu=5/2$ and $\nu=7/2$ states. In the former scenario,  the
number of fermions cannot be systematically changed by varying the magnetic field due to their neutrality. In the latter scenario, this mechanism may be driven by $e/4$ particles trapped at some deep-lying impurity levels. In both scenarios such fluctuations may occur even at a constant field, in which case they will manifest themselves as an extremely low-frequency telegraph noise revealed by a vertical cut through the color plot of  Figure~\ref{fig:fermionic_parity_stability}.

\subsection{Interference oscillations at $\nu=5/2$}
\label{sec:results_five_halves}
\begin{figure}[htb]
\includegraphics[width=\columnwidth]{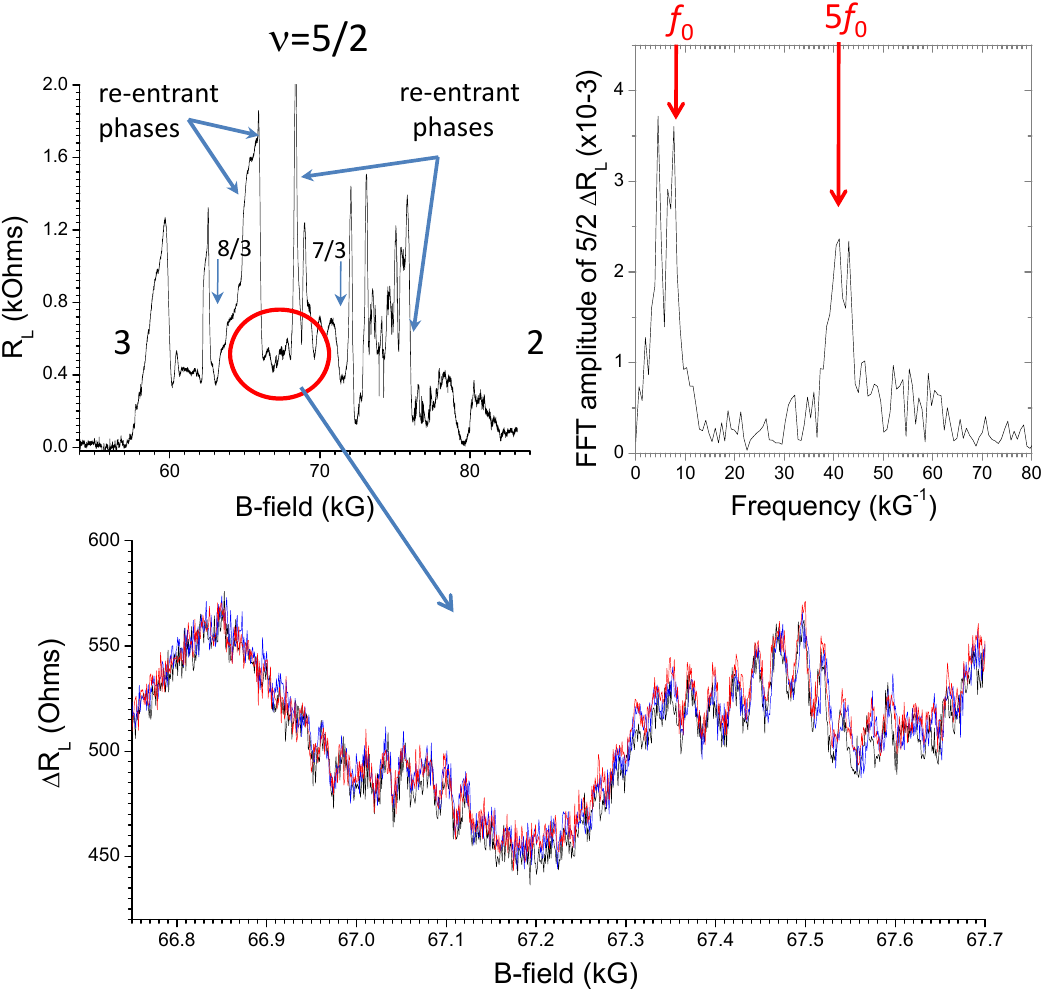}
\caption{Interference oscillations and their spectrum at $\nu=5/2$.\\
The top left panel shows the trace of $R_L$ from filling factor $\nu=3$ to 2;
its blow-up in the vicinity of $\nu=5/2$ (circled area) is shown in the bottom panel, with the corresponding  FFT spectrum  shown above in the right panel. $f_0$ is frequency of oscillations observed at integer filling in the same sample.  \\
The lower panel shows three $B$-field sweeps in a single direction, demonstrating significant overlap of the resistance oscillations.  The frequency of rapid oscillations is $\sim 5f_0$ shown in the power spectrum.  Each trace is 60 minutes, so total data collection time is 6 hours (the opposite sweep direction is not shown).\\
Sample~6, preparation~18, device type~b, $T\sim 20mK$.
}
\label{fig:5_halves_interf}
\end{figure}
We now turn our attention to the interference oscillations observed at $\nu=5/2$.
The interference oscillations observed at this filling fraction in the devices with high Al purification strongly resemble those at $\nu=7/2$.
As shown in Figure~\ref{fig:5_halves_interf}, these oscillations are remarkably stable and occur at five times the reference frequency $f_0$, which is once again indicative of the non-Abelian even-odd effect. Notice that another prominent spectral feature is located near $f_0$ itself.
\begin{figure}[htb]
\centering
    \includegraphics[width=\columnwidth]{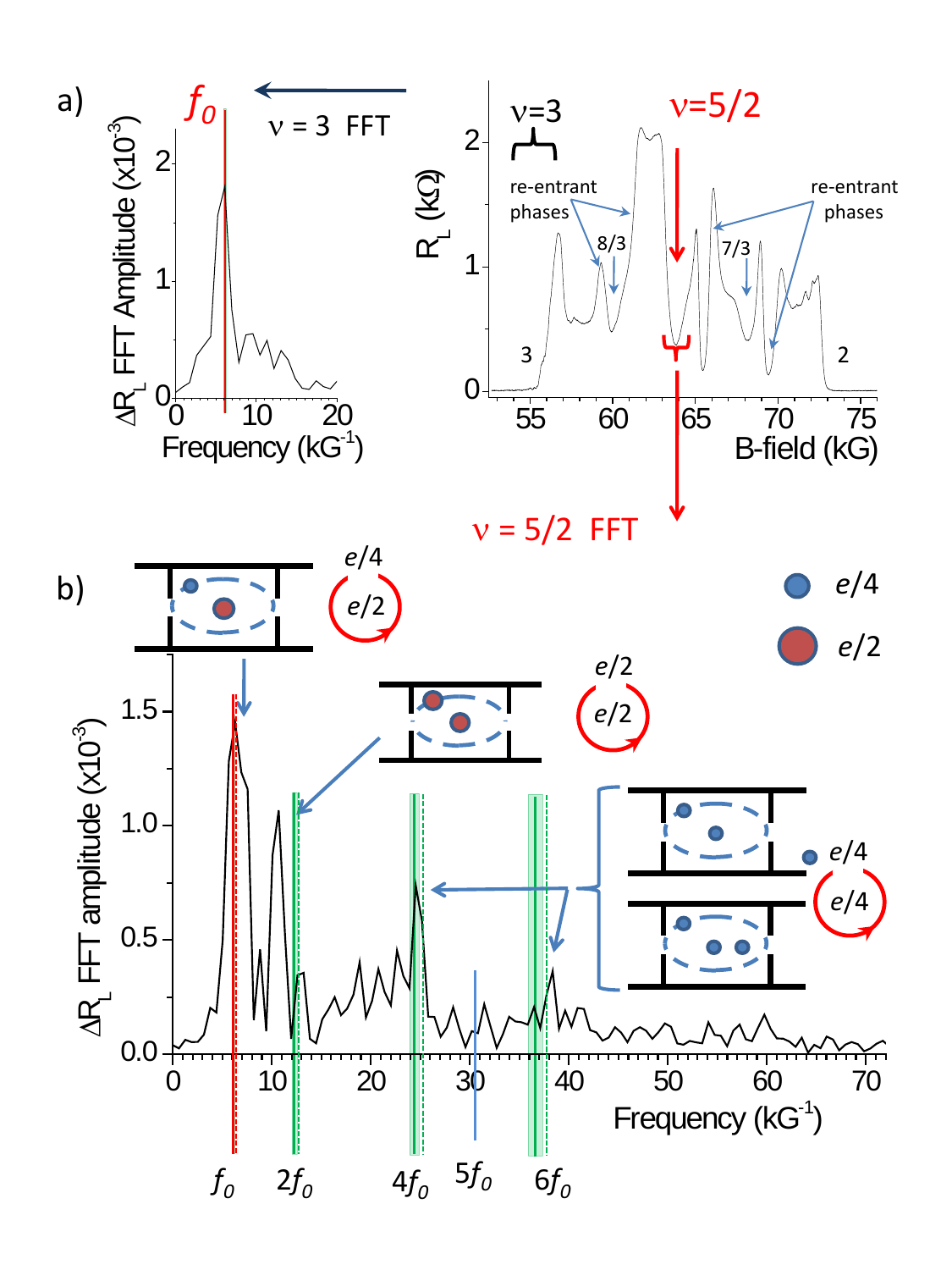}
    \caption{Extracting power spectra in $B$-field sweeps and interference spectra at $\nu=5/2$:\\
    (a) Transport ($R_L$) through an interference device shown in Figure~\ref{fig:interferometer}, bottom micrograph (device type~b), between filling factors $\nu=2$ and $3$ in a high mobility heterostructure without additional Al purification.  The power spectrum is extracted from a $B$-field sweep as follows:  the background resistance near $\nu=3$ is subtracted, after which a fast Fourier transform (FFT) is applied to the residual oscillations, producing the power spectrum that shows a principal peak at roughly $6\,\text{kG}^{-1}$, the frequency identified as $f_0$ whose value is  closely matched at other integral filling factors.  See also Fig.~\ref{fig:S5-2-1}. Temperature $\sim 20$mK, sample~2, preparation~2, device type~b.\\
    (b) Power spectrum of $R_L(B)$ near $\nu=5/2$.  The FFT window is bracketed in panel (a).  After background subtraction and application of the FFT, four dominant peaks located near $f_0$, $2f_0$, $4f_0$ and $6f_0$ can be discerned. Here $f_0$ is the frequency of the Aharonov--Bohm oscillations in the integer quantum Hall state (panel (a)), and marked in panels (a) and (b) by the solid red vertical line. Solid green lines in panel (b) indicate multiples of that frequency with the shaded area around them indicating the potential error margins stemming from the error of $\pm 1/8\;\text{kG}^{-1}$ in determining $f_0$. Meantime the dashed lines indicate multiples of the actual frequency of the first peak in panel (b).}
    \label{fig:power_spectra}
\end{figure}

Before we examine the spectra in more detail, it worth reiterating that the highest predicted oscillation frequency for Abelian AB interference at $\nu=5/2$ is $2f_0$; any significant spectral weight at higher frequencies (specifically, within the range between $4f_0$ and $6f_0$ for $\nu=5/2$) that is not a result of noise is an indicator of a non-Abelian  nature of the state.   Specifically, it should be interpreted as the evidence of the even-odd effect, which is a consequence of the quasiparticles' statistics and not charge; a naive application of Eq.~(\ref{eq:AB_phase2}) to the $e/4$ quasiparticles at $\nu=5/2$ while neglecting their non-Abelian nature would instead result in the oscillations frequency of $2.25 f_0$  in the Moore--Read state and $2.75f_0$ in the anti-Pfaffian state. For the genuinely Abelian candidate states, these frequencies would be $f_0$ in the $K=8$ state and $3.5 f_0$ in the $(3,3,1)$ state~\cite{Bishara2009a}. (At $\nu=7/2$, the respective frequencies would be $3.25f_0$ in the Moore--Read state, $3.75 f_0$ in the anti-Pfaffian state, $1.5 f_0$ in the $K=8$ state and $5f_0$ in the $(3,3,1)$ state, all significantly below $7f_0$.) We can discard the possibility of multiple windings which would, in theory, result in higher frequency of interference oscillations. The amplitude of those contributions would be negligible as they require multiple tunnelling across relatively wide constrictions.

While we empirically attribute the substantially more prominent high-frequency oscillations associated with the even-odd effect to the improved Al purity of the newer heterostructures, we emphasize that samples without this high Al purity can still demonstrate high frequency oscillations at $\nu=5/2$ as reported in the earlier work~\cite{Willett2013b}.  The focus of those studies was only on the oscillations associated with the even-odd effect; meantime recent improvements in those materials and devices (without the Al purity change) have provided higher quality samples with better defined FQH states of interest. This in turn allows for larger Fourier transform windows, letting us focus on the finer details of the oscillation spectra, including their low-frequency features, than was possible in the previous studies. Results of such study of the full spectral frequencies follow here.

\begin{figure}[t]
\centering
    \includegraphics[width=0.9\columnwidth]{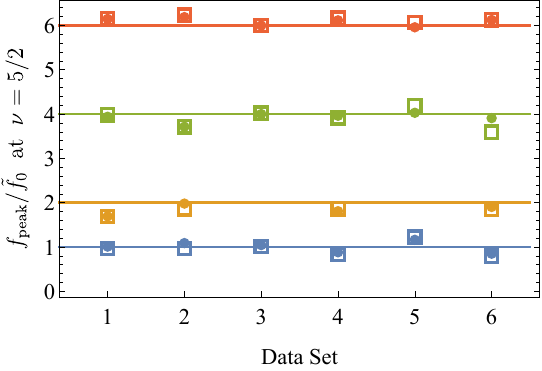}
    \caption{Frequencies corresponding to the most prominent spectral peaks at $\nu=5/2$ in 6 different data sets from different samples and preparations.
    Frequencies are normalized by ${\tilde f}_0$, which is the parameter obtained by applying linear fit to each observed set of peak frequencies (shown in circles) with the assumption that they correspond to integer multiples of this fundamental frequency. Open squares correspond to the frequencies obtained by taking the average value of the two frequencies that mark the width at half-maximum for each peak, normalized by the same coefficients. The data set identification is as follows: 1 - sample~2, preparation~2; 2 - same sample/preparation four days later; 3 - sample~2, preparation~1; 4 - sample~5, preparation~2; 5 - sample~3, preparation~1; 6 - sample~4, preparation~1. Note that two of the samples did not exhibit peaks at or near $2f_0$ -- cf. traces in Fig.~\ref{fig:2f0_peak}(a,b) corresponding to data sets 3 \& 2 shown here. $T \sim 20$mK.}
    \label{fig:sup_peak_fit}
\end{figure}

A representative Fourier spectrum of the oscillations at $\nu=5/2$ is shown in Figure~\ref{fig:power_spectra}(b) along with the possible interpretation of its most prominent features.  The oscillations in $R_L$ around filling fraction $\nu=3$ are used to determine the integer Aharonov--Bohm interference frequency $f_0$; details of the FFT analysis and magnetic field window size used for each FFT in results are described in Section~\ref{sec:S3} of Supplementary Materials. The observed
spectral peaks are located near  $f_0$, $2f_0$, $4f_0$  and $6f_0$, with $f_0$ being the aforementioned reference frequency. (See Section~\ref{sec:S5-7a} of Supplementary Materials for more details on peak identification.) In order to verify this identification of the observed  peaks, in Figure~\ref{fig:sup_peak_fit} we plot their actual locations vs. integer factor $m$ that we assign to these peaks in five different samples, with two different preparations for one of the samples -- 6 data sets altogether. Each set of observed of peak frequencies at $\nu=5/2$ has been rescaled by the coefficient $\tilde{f_0}$ obtained from linear fit for each sample's frequency data.  We can then compare $\tilde{f_0}$ obtained from linear fit to frequency $f_0$ corresponding to spectral peaks measured at $\nu=3$ for each sample/prepapration; the results of this comparison are shown in  Table~\ref{tab:sup_f_0}.

The peak locations shown in Figure~\ref{fig:sup_peak_fit}  correspond to the highest measured values of spectral peaks. In Figures~\ref{fig:peak_identification} and \ref{fig:sup_peak_identification} in Supplementary Materials we present a comparison between two peak identification processes, one identifying peak locations by their highest value and the other using the average of the two points corresponding to their half-values.  The observed features are consistent with the expectations for both Abelian and non-Abelian interference processes due to charge $e/4$ and charge $e/2$ quasiparticles. (We remark that the peak identification procedure used here is the same as was illustrated earlier in Figure~\ref{fig:integer_vs_fractional_peak_positions} showing the comparison of actual traces and their Fourier transforms for $\nu=4$ and $\nu=16/5$, further strengthening our confidence in identification of observed spectral features.)
\begin{table}[bht]
  \begin{tabular}{| c | c | c |}
        \hline
    $\tilde{f_0}\;\;(\text{kG}^{-1})$ & ${f_0}\;\;(\text{kG}^{-1})$ & sample/preparation \\
      \hline\hline
    6.2 & $6$ &  sample 2, prep 2\\ %\hline
    5.8 & $6\frac{1}{2}$ &  sample 2, prep 2, later\\ %\hline
    6.3 & 7 & sample 2, prep 1\\ %\hline
    10.0 & $9\frac{1}{4}$ & sample 5, prep 2\\ %\hline
    4.5 & $4\frac{3}{4}$ &  sample 3, prep 1\\ %\hline
    4.7 & $4\frac{1}{2}$ & sample 4, prep 1\\ %\hline
    \hline
  \end{tabular}
  \caption{Comparison of frequencies $\tilde{f_0}$ obtained from linear fit of the $\nu=5/2$ spectral data to ${f_0}$ observed at $\nu=3$ for each sample/preparation.}
  \label{tab:sup_f_0}
\end{table}

In the power spectrum presented in Figure~\ref{fig:power_spectra}(b) the amplitudes of these four peaks drop progressively with the frequency, a common property of these spectra in our study, but with exceptions (see Supplementary Materials, Figure~\ref{fig:S5-3-1}).  The base frequency, $f_0$, depends on the size of the interferometer, as seen by comparing the power spectra from different devices, but the ratios of the power spectrum positions remain close to  1:2:4:6.  (Note that the base frequency $f_0$ can be measured at the different integer filling factors, a property of AB oscillations, and this is demonstrated in Supplementary Materials, Figure~\ref{fig:S4-1} and \ref{fig:S4-2a}.)  The spectral features shown in Figure~\ref{fig:power_spectra}(b) are sharp, in part due to the low temperature of 20mK, but also due to the details of construction of this particular interferometer.  Larger devices have larger separation in the spectral features (since $f_0\propto A$) and, consequently, better resolution, but the amplitudes of the higher frequency features at $4f_0$ and $6f_0$  are reduced in the largest devices tested. A repetition of this measurement on the same device but days later is shown in Figure~\ref{fig:2f0_peak}(b) (presented there as a part of another study to be discussed later), with the dominant peaks persisting at the same frequencies. Further 5/2 power spectra are displayed in Supplementary Materials, Figure~\ref{fig:S5-3-1} and \ref{fig:S5-3-2}.

\begin{figure}[thb]
\centering
    \includegraphics[width=0.95\columnwidth]{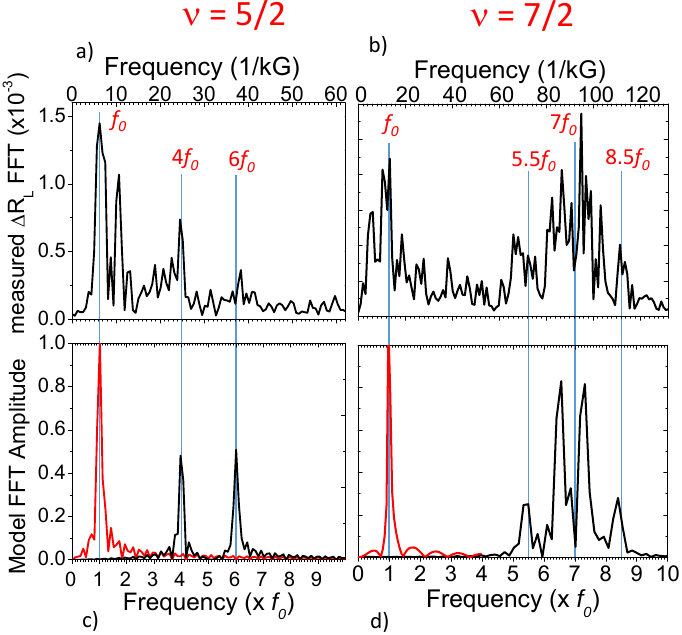}
    \caption{Modelling FFT spectra at $\nu=5/2$ and $7/2$. In a magnetic field sweep at $\nu=5/2$  and $\nu=7/2$  rapid oscillations due to the even-odd effect will be modulated by slower oscillations due to $e/4-e/2$ interference. At $\nu=5/2$ this implies modulating oscillations at $5f_0$ by oscillations at $f_0$, resulting in peaks at $4f_0$ and $6f_0$. At $\nu=7/2$, $1.5 f_0$ modulation of $7f_0$ even-odd oscillations should produce peaks at $5.5f_0$ and $8.5 f_0$. Panels (a) and (b) show measured power spectra at these filling fractions: (a) $\nu=5/2$, sample~2, preparation~2 (same as in Figure~\ref{fig:power_spectra}(b)); (b) $\nu=7/2$, sample~6, preparation~15, device type~b; $T \sim 20$mK in both cases.\\
    The measured power spectra are compared with simple models. The black trace in panel~(c) is the FFT of the expression $\Delta R_L =\cos(2\pi Bf_0)\times\cos(2\pi B(5f_0 ))$ taken using the magnetic field window similar to the one used to obtain power spectra in the experiment.  The red trace is the FFT of $\Delta R_L = \cos(2\pi Bf_0)$ demonstrating the $f_0$ oscillation peak. The spectra are scaled to match the locations and amplitudes of their peaks at frequency $f_0$.\\
    Panel (d) shows a similar modulation of the $7f_0$ even-odd oscillations by $1.5f_0$ oscillations expected for the $e/4-e/2$ interference at $\nu=7/2$.  In addition, the minima of $R_L$ are reproducibly formed away from the re-entrant quantum Hall phases in the high quality material used in this study, resulting in an additional low-frequency modulation at about $0.4 f_0$. To capture these effects we model the resistance as $\Delta R_L = \left[0.25 \cos(2\pi B(0.4f_0)) + 0.75\cos(2\pi B (1.5f_0))\right] \times \cos(2\pi B(7f_0))$. The FFT of this expression results in the spectrum shown in panel~(d), compared directly to measured data at $\nu=7/2$ in panel~(b).  Again, an FFT of $\Delta R_L = \cos(2\pi Bf_0)$ is the red trace panel~(d).  (See also Supplementary Materials, Section~\ref{sec:S5-5} and Figure~\ref{fig:S5-5-2}.)
    }
    \label{fig:interf_model}
\end{figure}

Before attempting a similar analysis of the finer spectral features of interference oscillations at $\nu=7/2$ accessible through a wider Fourier transform window, let us reiterate that
the common observed feature in multiple sample preparations, interference measurements, and their respective spectra such as those shown in Figures~\ref{fig:Hall_trace}(c) and \ref{fig:7_halves_interf} (with another example shown in Figure~\ref{fig:interf_model}(b) and more spectra presented in Section~\ref{sec:S5-4} of Supplementary Materials) is the large spectral weight concentrated around $7f_0$, the frequency of the predicted even-odd effect. Yet upon closer inspection of the spectrum in Figure~\ref{fig:7_halves_interf}(e), another common feature appears to be a sharp minimum right at that frequency $7f_0$. (A similar feature is also observed in samples at $\nu=5/2$:  See e.g. Fig.~\ref{fig:S6-2}.) The most likely origin for this dip is some low-frequency modulation of the even-odd oscillations. Irrespective of its origin, such a modulation would split the high-frequency Fourier peak into two resulting in a spectral \textit{minimum} at the location of the original peak. The visibility of this minimum then becomes a function of the frequency window used in the Fourier transform; this effect can be clearly seen in Figure~\ref{fig:7_halves_interf} where a prominent spectral peak at $7f_0$ seen in panel (d) evolves into a pronounced minimum seen in panel (e) as the size of the FFT window is increased in attempt to tease out finer spectral features from the dominant peak. As has been mentioned in the introduction, the $e/4-e/2$ interference with fixed fermion parity would result in a split peak,  $(7\pm 1.5)f_0$. However, there are also other potential, if less systematic sources of low-frequency modulation of the dominant $7f_0$ frequency such as potential dependence of the oscillation amplitude on the distance from the overall resistance minimum marking the QH state.  Such modulation is discussed in Section~\ref{sec:S5-5} of Supplementary Materials;  the corresponding spectral feature at about  $0.4f_0$ is encircled in Figure~\ref{fig:7_halves_interf}(e). The modulation by  $0.4f_0$,  is then combined with the modulation by $1.5f_0$ due to $e/4-e/2$ interference oscillations. The comparison between the actual observed spectrum and the spectrum modeled by including both of these modulations is shown in Figures~\ref{fig:interf_model}(b) and (d). Note that the model uses a finite Fourier transform window similar to that used in processing the experimental data. For comparison, Figures~\ref{fig:interf_model}(a) and (c) show the observed and modeled spectra at $\nu=5/2$ where only the modulation due to  $e/4-e/2$ interference is taken into account.

\subsection{Abelian interference oscillations at $\nu=5/2$ and $7/2$}
\label{sec:results_abelian}
While the high-frequency oscillations are of the foremost importance for demonstrating the non-Abelian nature of the  $\nu=5/2$ and $7/2$ states, the low-frequency features that we have been able to investigate in this study also reveal some important information.  We should mention that the ability to discern lower-frequency spectral features is another key new element of the present study; in the past they were hard to extract due to the narrow magnetic field range where these oscillations had been seen. Utilizing higher purity heterostructures and somewhat different interferometer design we have been able to better isolate the quantum Hall state at $\nu=5/2$ from the surrounding compressible states, which resulted in more pronounced, broader minima in  $R_L$ supporting longer runs of higher-amplitude interference oscillations than before. This allowed us to observe and analyze those lower-frequency spectral features, thus addressing one of the shortcomings of our earlier study~\cite{Willett2013b}.

It is instructive to compare the low-frequency features between the  $\nu=5/2$  and $7/2$ states. The two states are theoretically expected to correspond to the same topological order, with the same charge and statistics of its excitations. However, due to its different filling factor, the periodicity of AB interference at $\nu=7/2$  is expected to be different from 5/2, as can be seen from Eqs.~(\ref{eq:AB_phase2})--(\ref{eq:interferenece_general}). Specifically, the interference of $e/4$ edge excitations around $e/2$ bulk quasiparticles should now occur with the periodicity of $2\Phi_0/3$ (instead of $\Phi_0$), and thus the expected spectral peak at $1.5f_0$, nicely distinguishing it from potential electron contribution still occurring at $f_0$. Lastly, $e/2-e/2$ interference should manifest itself through a peak at $3f_0$. The oscillation spectrum shown in Figure~\ref{fig:7_halves_interf}(e) is consistent with these expectations.  See also Section~\ref{sec:S5-4} of Supplementary Materials.

\begin{figure}[h!tb]
\centering
    \includegraphics[width=0.75\columnwidth]{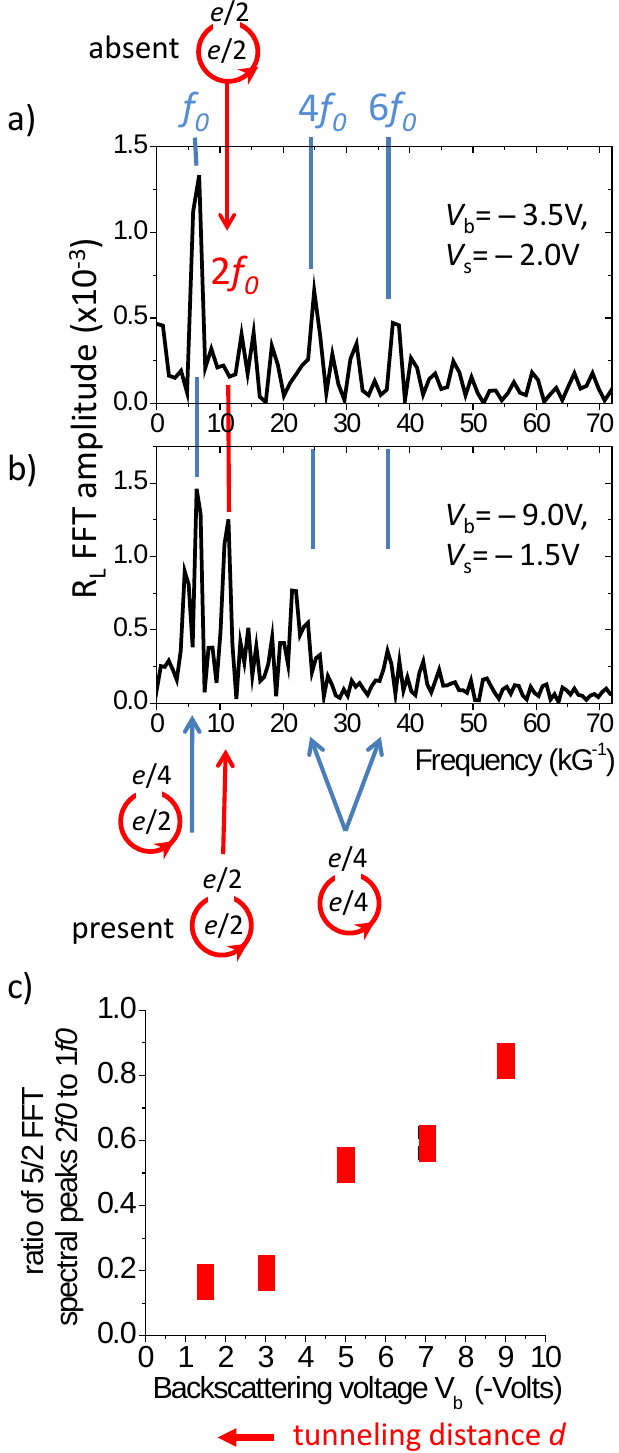}
    \caption{
%    FFT spectral peak identification and back-scattering gate control of $2f_0$ peak ($e/2$ braiding $e/2$) in the 5/2 spectra:\\
(a)-(b) $\nu=5/2$ oscillation spectra at two different gate voltage settings (sample~2, preparations~1 and 2, device type~b, $T \sim 20$mK).  The spectrum in panel~(b) shows four peaks at   $f_0$, $2f_0$, $4f_0$ and $6f_0$, consistent with $e/4$ and $e/2$ interference.  The $2f_0$ peak is absent when the backscattering gate voltage $V_b$ is changed from $-9$V to $-3.5$V, increasing the tunneling distance $d$ for interfereing quasiparticles.  The side gate voltage $V_s$ was adjusted to preserve the area $A$. (Supplementary Materials, Section~\ref{sec:S5-7b}, Figure~\ref{fig:S5-7-2} demonstrate the overall similar transport between $\nu=2$ and $3$ for these gate settings.)\\
(c) Ratio of the peak value at $2f_0$ to that at $f_0$ for a series of backscattering gate voltages $V_b$ in the same device (sample 2, preparations~1 to 5, $T \sim 20$mK).  The increase of the peak amplitude ratio in response to the constriction narrowing (more negative $V_b$) may be reflecting relative tunneling amplitudes for $e/2$ and $e/4$ quasiparticles.
%See also Supplementary Materials, Section S5-7, Fig.~S5-7-1 for a plot of the amplitude of the $2f_0$ peak versus $V_b$, revealing a similar dependence.
}
    \label{fig:2f0_peak}
\end{figure}

The $f_0$ spectral peak at $\nu=5/2$ decays rapidly with increasing temperature up to $T\approx 80$mK, which is roughly the onset temperature of the deep minimum in $R_L$. (See Supplementary Materials, Section~\ref{sec:S5-7c}, Figures~\ref{fig:S5-7-3} and  \ref{fig:S5-7-4} for more details.) This strongly implies that oscillations at this frequency are, indeed, due to the physics of this fractional quantum Hall state. In particular, this hints at the $e/4-e/2$ interference contribution to these oscillations, in addition to the omnipresent electron contribution.

The last process whereby $e/2$ quasiparticles braid around bulk $e/2$ quasiparticles should result in spectral features at $2f_0$ for $\nu=5/2$ and at $3f_0$ for $7/2$. While the observed oscillation spectra at $\nu=7/2$ contain hints of the $3f_0$ feature -- see e.g. Figure~\ref{fig:S5-4-2} of Supplemental Materials for the best example -- Figure~\ref{fig:7_halves_interf} illustrates the problematic nature of reliably discerning this feature from the noisy  background. Meantime its $2f_0$ counterpart in the $\nu=5/2$ spectra obtained in the samples without additional Al purification appears far more prominent, as can been seen in Figure~\ref{fig:power_spectra}. We should note that this feature is systematically seen at frequencies somewhat \textit{below} $2f_0$ -- see e.g. Figure~\ref{fig:sup_peak_fit} showing its observed position for four different samples/preparations. While we are not certain about the reasons for this discrepancy (it is a subject of an ongoing study), the prominence of this spectral feature enables us to implement additional tests.
%Focusing on the $\nu=5/2$ state, we have further investigated this process.
An important property of the observed $\nu=5/2$ oscillation spectra presented here is changing of the magnitude of the $2f_0$ peak as a function of the separation between the inner edge currents in the backscattering constrictions (distance $d$ in schematic of Figure~\ref{fig:interferometer}).  Figures~\ref{fig:2f0_peak}(a) and (b) show two complete power spectra in the same interferometer at $\nu=5/2$ but with different applied voltages $V_b$ and $V_s$: the more negative is the $V_b$ value, the narrower the constrictions acting as ``beam splitters'' are, which should change the backscattering (tunneling) amplitudes. $V_s$ can be adjusted so that for two different $V_b$ values, the interferometer areas $A$ are kept essentially the same.  Standard longitudinal resistance ($R_L$) measurements (see  Supplementary Materials, Section~\ref{sec:S5-7b}, Figure~\ref{fig:S5-7-2}) show little difference for the two different gate configurations.  Note that in the device with voltage $V_b=-3.5$V, the $2f_0$ peak is essentially absent, whereas in the device with $V_b=-9.0$V a large $2f_0$ peak is present.  We therefore conclude that the width $d$ of the constriction (the separation between the QH edges – see Figure~\ref{fig:interferometer}) is an important factor controlling the presence of the peak at $2f_0$.  The $V_b$ voltages were adjusted to a range of values between the two shown in the full spectra of Figures~\ref{fig:2f0_peak}(a) and (b); the resulting plot of the measured ratio between the peak amplitudes at $2f_0$ and $f_0$ as a function of $V_b$ is shown in Figure~\ref{fig:2f0_peak}(c). According to our findings, as $V_b$ becomes more negative (the backscattering distance d becomes smaller) the amplitude ratio increases.  This finding is consistently observed in multiple different heterostructure wafers and multiple different devices. (We expect a similar behavior of the $3f_0$ peak at $\nu=7/2$ but have not performed this study; it is a subject of ongoing research.)

This result can be understood as a consequence of the difference in tunneling probability of $e/4$ and $e/2$ edge excitations as a function of the tunneling distance d between the edges in the constrictions that form the interferometer. Theoretical analysis of the tunneling process of the excitations at 5/2 predicts that the amplitude of the  $e/2$  tunneling process is suppressed by comparison to the $e/4$  tunneling due to the larger momentum transfer required for backscattering at a gated narrowing, with the effect becoming more pronounced as the tunneling distance increases~\cite{Bishara2009a}.  This effect is illustrated in~\cite{Chen2009} where the dependence of the ratio between the $e/4$ and $e/2$ tunneling amplitudes as a function of the tunneling distance has been investigated numerically for the Moore-Read state. Our experimental findings are consistent with this picture, showing the amplitude of the $2f_0$ peak attributed to the $e/2$ interference decreasing much faster with the distance than that of the peak at $f_0$. Note that the physics behind this effect should be insensitive to the precise nature of the $\nu=5/2$ state; it is a consequence of the difference between different quasiparticles’ charges. This should be contrasted with the temperature and bias voltage dependence of these tunneling processes, which is expected to be governed by the quasiparticles’ scaling exponents. Those considerations would lead to different predictions for the anti-Pfaffian vs. both the Moore-Read and the PH-Pfaffian states, potentially allowing for discriminating between these states~\cite{Bishara2009a}. Our study is not sensitive to this distinction.

We should also comment here that contrary to the analysis of~\cite{Keyserlingk2015}, we do not expect to be able to distinguish between the Moore-Read, anti-Pfaffian and PH-Pfaffian states based on the location of the high-frequency spectral peaks. For the reasons mentioned earlier (and further elaborated in the Supplementary Information), the Abelian phase associated with the $e/4-e/4$ braiding (which is different in these states) does not directly contribute to the AB phase observed in the experiment; the interference itself is destroyed if an $e/4$ quasiparticle braids another unpaired one whereas only the overall fusion channel matters for paired ones. Therefore the observed high-frequency peaks are merely indicative of the non-Abelian nature of the state but their positions cannot be used to discriminate between the candidate states. That said, they contain the information about temporal stability of the non-Abelian fusion channels. For the case of stable overall fermion parity at $\nu=5/2$  we expect to see peaks at $4f_0$ and $6f_0$. This scenario appears to be in agreement with the observed enhanced spectral features at these frequencies reported here.  However, in both past~\cite{Willett2013b} and present studies (see Section~\ref{sec:S6} of Supplementary Materials) some sweeps result in only a $5f_0$ peak, consistent with random (but not rapidly fluctuating) fusion channel of additional pairs of $e/4$ quasiparticles introduced into the interferometer. The conditions that affect the temporal stability of the fermion parity are the subject of future investigation.

\section{Summary and Conclusions:}
	Using magnetic field sweeps in quantum Hall Fabry--P\'{e}rot interferometers at $\nu=5/2$ and $7/2$  filling we have observed large amplitude, reproducible interference oscillations at frequencies inconsistent with Abelian Aharonov--Bohm interference but specific for the non-Abelian even-odd effect for the respective quantum Hall states.  With new ultra-high purity heterostructures we have observed such oscillations for the first time in the  $\nu=7/2$ QH state, providing the first evidence for the non-Abelian statistics of its charge–$e/4$ excitations. Furthermore, the oscillations attributed to the non-Abelian $e/4$ quasiparticles have been shown to be remarkably stable, indicating stability of their fusion channel – the fermion parity – which in turn may have profound consequences for their applications for topological quantum computation. This interpretation is further strengthened by the observation of occasional $\pi$ phase jumps, the predominant type of instability observed in our traces. These are consistent with the change of either the fermion parity or the parity of non-Abelian charge-$e/4$ quasiparticles  inside the interferometer, both of which should manifest themselves in a $\pi$ phase shift of the oscillations. Moreover, the fact that we observe them infrequently and can actually associate a timescale (hours!) to the parity flips is perhaps the most promising outcome of our study, demonstrating the potential of such quantum Hall systems for quantum computing.

The Fourier transforms of the magnetic field dependence of the resistance around those filling factors show not only peaks corresponding to the non-Abelian even-odd effect associated with  $e/4$ quasiparticles but also a set of spectral peaks consistent with those anticipated for all possible combinations of interfering quasiparticle types expected to be present in these systems, thus also accounting for the Abelian interference effects.

In our study we also demonstrate means of controlling the interference processes.  The Abelian  $e/2-e/2$ interference is shown to be suppressed by reducing the edge quasiparticle backscattering, thus allowing us to distill different contributions to the interference oscillations.  This gate control is presently the subject of intense investigation, and these are essential steps in an effort to develop devices for topological quantum computation.

\subsection*{Acknowledgments}

We are grateful for some illuminating discussions with D.~E.~Feldman and B.~I.~Halperin as well as with Parsa Bonderson, who has also been our co-author on related earlier projects. We acknowledge the hospitality of KITP supported in part by the NSF under Grant No.~NSF PHY-1748958. We also acknowledge the Aspen Center for Physics, which is supported by the NSF under Grant No.~NSF  PHY-1607611.
The work at Princeton University (L.N.P., K.W.W., K.W.B.)  was funded by the Gordon and Betty Moore Foundation through the EPiQS initiative Grant No.~GBMF4420 and Grant No.~GBMF9615 (L.N.P), and by the National Science Foundation MRSEC Grant No.~DMR 1420541 (Y.J.C.).

\clearpage

\newpage
\newcommand{\beginsupplement}{%
        \setcounter{table}{0}
        \renewcommand{\thetable}{S\arabic{subsection}-T\arabic{table}}%
        \setcounter{figure}{0}
        \renewcommand{\thefigure}{S\arabic{subsection}-\arabic{figure}}%
        \setcounter{equation}{0}
        \renewcommand{\theequation}{S\arabic{equation}}
                \setcounter{subsubsection}{0}
        \renewcommand{\thesubsection}{S\arabic{subsection}}
        \renewcommand{\thesubsubsection}{$\!$\alph{subsubsection}}
     }

%\counterwithin{figure}{subsubsection}
\beginsupplement

\section*{Supplementary materials}

\subsection*{Overview}

%The Supplementary materials focus on six specific areas that provide support for the article’s main fundings:
%1) Theory, focusing on the model for non-Abelian $e/4$ and Abelian $e/2$ and explicit spectral frequency values in interferometry, 2) Methods-materials, heterostructure materials with focus on new high purity Al heterostructures and the shielded well heterostructures used in interferometry, 3) Methods- measurement, describing  data acquisition and data analysis,  4) Methods-interferometry, focusing on the Coulomb dominated vs. Aharonov--Bohm interference regime, presenting data supporting the latter and describing how the interferometers and heterostructures are designed to accomplish this, 5) Data supplementing what has already been presented in the main part, 6) spectra addressing  fermion parity stability; this subsection presents data wherein expression of spectral features indicates the fermion parity stability.

In Section~\ref{sec:S1} we review the theory of quantum Hall Aharonov--Bohm interferometers.

Sections~\ref{sec:S2}--\ref{sec:S4} provide overview of the methods.
Specifically, Section~\ref{sec:S2} addresses material aspects focusing on new high purity Al heterostructures and the shielded-well heterostructures used in the interferometry studies.
Section~\ref{sec:S3} focuses on measurements, describing  data acquisition and data analysis. The focus of  Section~\ref{sec:S4} is interferometry; specifically the experimental signatures on the Coulomb dominated vs. Aharonov--Bohm interference regime; it presents data supporting the latter and describes the relevant design features of the interferometers used in our study.

Finally, Section~\ref{sec:S5} provides additional experimental data supplementing what has already been presented in the main part, whereas Section~\ref{sec:S6} specifically focuses on the evidence of fermion parity stability.

\subsection*{Sections:}
\begin{enumerate}	
\item \ref{sec:S1} Theory
%Methods overview.
\item	\ref{sec:S2} Methods: Heterostructures and Interferometer Designs
\begin{enumerate}
\item	\ref{sec:S2a} Basic heterostructure designs
\item	\ref{sec:S2b} Extreme aluminum purity in heterostructures of this study
\item	\ref{sec:S2c} Sample illumination and gating processes
\item	\ref{sec:S2d} Interferometer designs and heterostructure concerns
\end{enumerate}
\item   \ref{sec:S3} Methods: Measurements
\begin{enumerate}
\item	\ref{sec:S3a} Data acquisition: Signal measurement and signal size
\item	\ref{sec:S3b}	Data analysis: Fast Fourier transform spectra of resistance traces
\end{enumerate}
\item	\ref{sec:S4}	Methods: Interferometry
\begin{enumerate}
\item	\ref{sec:S4a} Oscillations at $f_0$ and the interferometer
\item	\ref{sec:S4b} Coulomb dominated versus Aharonov--Bohm regimes
\end{enumerate}
\item		 \ref{sec:S5} Data: Supplements to individual figures from the main text\\
Sections~\ref{sec:S5-2} to \ref{sec:S5-7c}.
\item		\ref{sec:S6} Spectral dependence on fermion parity stability
\end{enumerate}

\subsection*{List of Supplemental Figures and Tables}

\begin{figure}[tbh]
\centering
    \includegraphics[width=0.8\columnwidth]{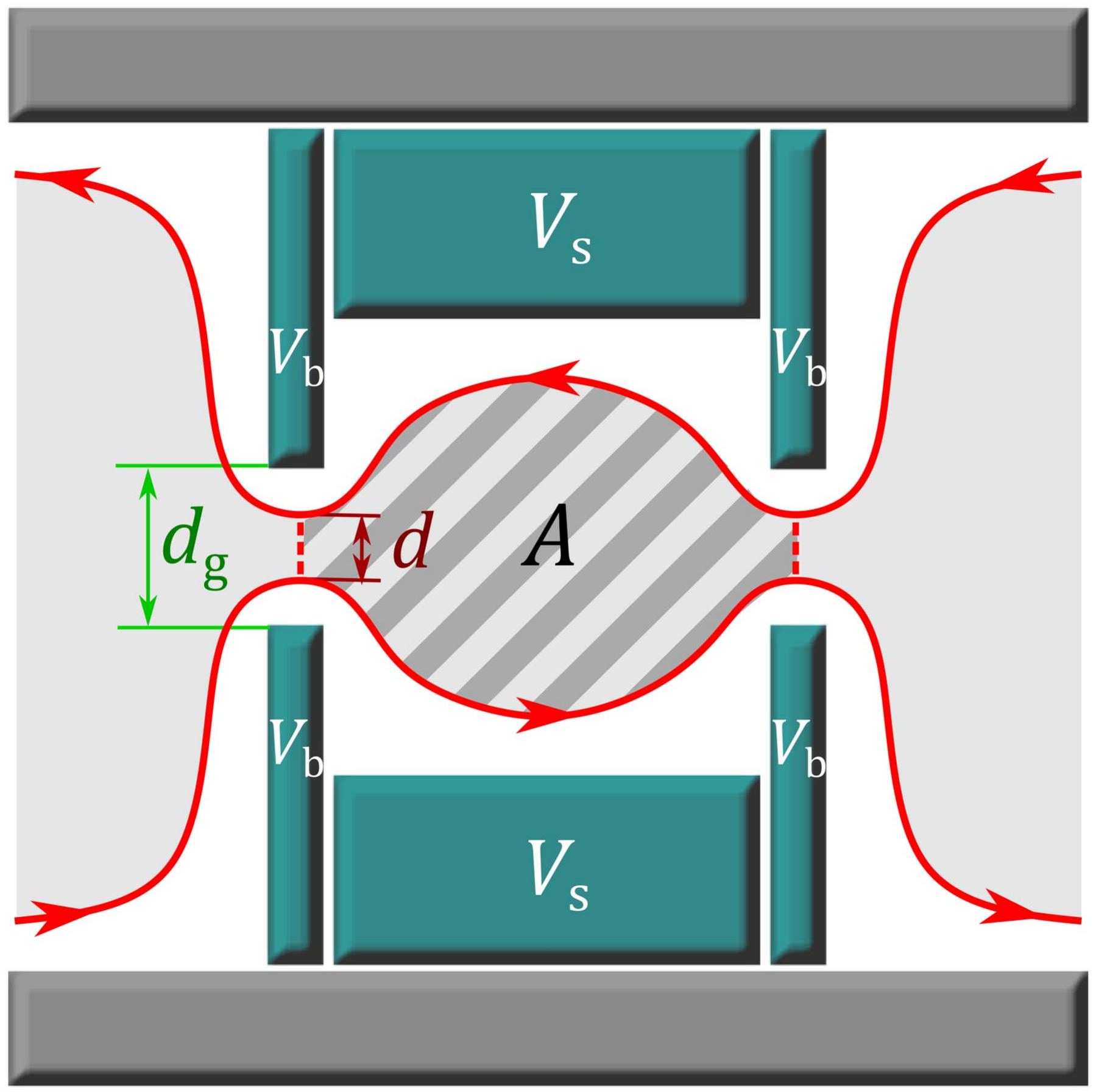}
    \caption{
    % Schematic and images of interferometer devices, transport through the devices in high purity Al heterostrutures,
    % large interference oscillations at 7/2 filling factor.
Schematic of the interference device and current paths displayed in
Figure~\ref{fig:interferometer} of the main text and repeated here for clarity.
}
    \label{fig:S1-1}
\end{figure}

\paragraph*{Figures:}
\begin{itemize}
\item[] \ref{fig:S1-1} Schematic of interference device.
\item[] \ref{fig:S1-2}  Fermion parity and the non-Abelian even-odd effect.
%\item[] \ref{fig:S1-3} Interference at integer and $\nu=3+1/5$ filling as prescribed by Eqs.~(\ref{eq:AB phase}) and (\ref{eq:interferenece_general}) of the main text and 3 and 4 of the Supplementary Materials.
\item[] \ref{fig:S2-1} Heterostructure layering, normal well, shielded well, doping well.
\item[] \ref{fig:S4-1} Interference oscillations near integer filling factors.
\item[] \ref{fig:S4-2a} Different integer filling interference oscillation periods.
\item[] \ref{fig:S4-2b} A waterfall and pajama plots of interference oscillations on magnetic field versus side gate voltage showing negative slope, indicative of Aharonov--Bohm oscillations.
%\item[] \ref{fig:seventhirds} Absence of significant high-frequency spectral features at $\nu=7/3$.
\item[] \ref{fig:S4-5} Transport through interferometers.
\item[] \ref{fig:S4-6} Temperature dependence of coarse transport through an interferometer and energy gaps.
\item[] \ref{fig:S5-2-1} Figure~\ref{fig:Hall_trace} supplement: comparison of transport in standard shielded well and shielded well with high purity aluminum and example FFT determination of  $f_0$, the electron AB frequency.
\item[] \ref{fig:S5-2-2} Figure~\ref{fig:Hall_trace} supplement: oscillation set of  Figure~\ref{fig:Hall_trace}(b) broken down to down- and up- magnetic field sweeps.
\item[] \ref{fig:add_osc1} Figure~\ref{fig:Hall_trace} supplement: oscillations in $\Delta R_L$ at $\nu=7/2$ observed for different preparations and with different voltage settings of both $V_s$ and $V_b$.
\item[] \ref{fig:add_osc2} Figure~\ref{fig:Hall_trace} supplement: additional sets of oscillations in $\Delta R_L$ at $\nu=7/2$ observed for different side gate voltage settings $V_s$ with fixed $V_b$.
%\item[] \ref{fig:S5-2-3} Figure~2 supplement: $\Delta R_L$ at $\nu=5/2$ as a function of side gate voltage $V_s$, showing $e/4$, $e/2$ AB oscillations and non-Abelian $e/4$ even-odd effect.
%\item[] \ref{fig:S5-2.4} Figure~2 supplement: $B$-sweep measurements in lower quality sample
\item[] \ref{fig:add_QH_trace} Figure~\ref{fig:seventhirds} supplement: quantum Hall trace showing minima at $\nu = 7/3$, $5/2$ and $3$.
\item[] \ref{fig:add_seven_thirds} Figure~\ref{fig:seventhirds} supplement: blow-up of resistance oscillations in the vicinity of $\nu = 7/3$.
\item[] \ref{fig:add_five_halves} Figure~\ref{fig:seventhirds} supplement: blow-up of resistance oscillations in the vicinity of $\nu = 5/2$.
\item[] \ref{fig:add_nu_three} Figure~\ref{fig:seventhirds} supplement: blow-up of resistance oscillations in the vicinity of $\nu = 3$.
\item[] \ref{fig:S5-4-1} Figure~\ref{fig:7_halves_interf} supplement: three power spectra at $\nu=7/2$ and their common features.
\item[] \ref{fig:S5-4-2} Figure~\ref{fig:7_halves_interf} supplement: $\nu=7/2$ power spectra averages, normalized to have the same  $f_0$ or  $7f_0$.
\item[] \ref{fig:S5-4-4} Figure~\ref{fig:7_halves_interf} supplement: $\nu=7/2$ power spectra, same device and sample, different magnetic field windows.
\item[] \ref{fig:S5-4-3} Figure~\ref{fig:7_halves_interf} supplement: $\nu=7/2$ power spectra, comparing single FFT vs. average of 4 FFTs.
\item[] \ref{fig:S5-4-5} Figure~\ref{fig:7_halves_interf} supplement: $\nu=7/2$ power spectrum averaged over two preparations of the same sample.
\item[] \ref{fig:S5-6-1} Figures~\ref{fig:fermionic_parity} \&~\ref{fig:5_halves_interf} supplement: $\nu=5/2$ power spectrum and stable large amplitude even-odd oscillations.
\item[] \ref{fig:S5-6-2} Figures~\ref{fig:fermionic_parity} \&~\ref{fig:5_halves_interf} supplement: $\nu=5/2$ transport and focus on stable even-odd oscillations with a single $\pi$ phase jump.
\item[] \ref{fig:S5-3-1} Figure~\ref{fig:power_spectra} supplement: Power spectra at $\nu=5/2$ in interferometers from different heterostructure types, doping well and shielded well.
\item[] \ref{fig:S5-3-2} Figure~\ref{fig:power_spectra} supplement: direct comparison of $\nu=5/2$ oscillation spectra of Figures~\ref{fig:power_spectra}(b) and \ref{fig:2f0_peak}(b).
\item[] \ref{fig:S5-3-3} Figure~\ref{fig:power_spectra} supplement: direct comparison of oscillation spectra at an integer filling and at $\nu=5/2$.
\item[] \ref{fig:peak_positions} Figure~\ref{fig:sup_peak_fit} supplement: FFT spectra corresponding to the six data sets used in Figure~\ref{fig:sup_peak_fit}
\item[] \ref{fig:peak_identification} Figure~\ref{fig:sup_peak_fit} supplement: Identification of FFT peaks at $\nu=5/2$ using frequencies corresponding to their highest values.
\item[] \ref{fig:sup_peak_identification}  Figure~\ref{fig:sup_peak_fit} supplement: Identification of FFT peaks at $\nu=5/2$ using midpoints of half-value frequencies.
\item[] \ref{fig:S5-5-1} Figure~\ref{fig:interf_model} supplement: $\nu=7/2$ power spectrum details of Figure~\ref{fig:interf_model}(b).
\item[] \ref{fig:S5-5-2} Figure~\ref{fig:interf_model} supplement: $\nu=7/2$ spectrum, $R_L$ and $\Delta R_L$ window, properties \& evidence for the $0.4f_0$ peak.
\item[] \ref{fig:S5-7-1} Figure~\ref{fig:2f0_peak} supplement: a) transport in the vicinity of $\nu=5/2$ for different voltages $V_b$; b)  $2f_0$, FFT peak amplitude as a function of $V_b$.
\item[] \ref{fig:S5-7-2} Figure~\ref{fig:2f0_peak} supplement: $R_L$ transport filling $\nu=2$ to $\nu=3$ for two different gate configurations.
\item[] \ref{fig:S5-7-3} Temperature dependence of the $f_0$ peak at 5/2 and integer filling.
\item[] \ref{fig:S5-7-4} Temperature dependence of $R_L$ at $\nu=5/2$.
\item[] \ref{fig:S6-1} Fermion parity stability:  5/2 power spectrum ($f_0$, single $5f_0$ peak).
\item[] \ref{fig:S6-2} Fermion parity stability:  5/2 power spectrum ($f_0$, $2f_0$, $5f_0$ complex with $5f_0$ minimum).
\item[] \ref{fig:S6-3} Fermion parity stability:  $7/2$ power spectrum ($f_0$, single $7f_0$ peak, no $1.5f_0$, no $3f_0$).
\item[] \ref{fig:S6-4} Fermion parity stability:  $7/2$ power spectrum ($f_0$, $1.5f_0$, $3f_0$, $7f_0$ complex with $7f_0$ minimum).
\end{itemize}

\paragraph*{Tables:}
\begin{itemize}
\item[]\ref{tab:S2-T1} Wafer list and properties: electron densities, mobilities, and wafer types.
\item[]\ref{tab:S2-T2} Sample device properties: substrate wafer, areas, backscattering gap sizes, and device types.
%\item[]\ref{tab:sup_f_0} Comparison of frequencies $\tilde{f_0}$ from linear fit of the $\nu=5/2$ data and ${f_0}$ from measurements at integer filling.
\item[]\ref{tab:S6-T1} Fermion parity stability times and expected power spectrum peaks.
\end{itemize}

\clearpage
\subsection{Theory: Quantum Hall Aharonov--Bohm Interferometry}
\label{sec:S1}

In this section we provide the theoretical background for the Aharonov--Bohm interferometry in the QH regime. The geometry we are focusing on is that of the Fabry-P\'{e}rot interferometer whose two arms correspond to two constrictions created in a QH bar by e.g. electrostatic gating, which bring the opposite edge states of the QH liquid into close (but not too close) proximity of one another thus allowing for the tunneling between these edge states. In contrast with the Coulomb blockade regime, the area between these constrictions is not pinched off into an isolated droplet, hence the electric charge inside the interferometer loop is not quantized. (This, in particular, implies relatively weak tunneling across the constrictions.)
The relative phase difference acquired by quasiparticles of charge $e^*$ traversing the two ``arms'' of the interferometer upon the insertion of additional flux $\Delta\Phi$ has two contributions, one being the actual Aharonov--Bohm phase due to the change in the enclosed magnetic flux and the other being the statistical phase acquired by the quasiparticle on the edge encircling additional quasiparticles created in the bulk:
\begin{equation}
\Delta \gamma_{e^*}  =2 \pi\left(\frac{\Delta\Phi}{\Phi_0} \right)\left(\frac{e^*}{e}\right)+2\theta_{e^*}\Delta N_{e^*}
\label{eq:suppl_interf_phase1}
\end{equation}
where $\Delta N_{e^*}$ is the change in the number of enclosed quasiparticles and $\theta_{e^*}$ is the statistical angle, e.g. the phase acquired by the wavefunction upon counter-clockwise exchange of two identical quasiparticles. Note that $e$ in the above expression is the absolute value of the electron charge whereas $e^*$ should be negative for a quasiparticle in a QH liquid made of electrons carrying charge $-e$. However, the values of the statistical angles quoted in the literature invariably refer to those applicable to holomorphic wavefunctions, which would correspond to positively charged electrons in a positive magnetic field or negatively charged electrons in a negative field. In other words, increasing the strength of the field would result in $\Delta\Phi<0$ for negatively charged electrons/quasiparticles, meaning that the first term in Eq.~(\ref{eq:suppl_interf_phase1}) would remain positive. It is therefore conceptually and notationally simpler to think about positively charged electrons and quasiparticles in a positive magnetic field and this is the convention we use in this manuscript; the final results for the interference phases are unaffected by this choice.

An increase in magnetic field reduces the magnetic length, ``shrinking'' the wavefunction. The system can accommodate this in one of three ways: (i) by shrinking the size of the quantum Hall puddle inside the interferometer, (ii) by increasing the electron density inside the interferometer while keeping both its active area and the filling fraction intact or, (iii), by creating quasiholes, which can be visualized as ``punctures'' in the otherwise incompressible liquid. Which of these scenarios is actually realized (and they are not mutually exclusive) is determined by the underlying energetics; see~\cite{Halperin2011a,Keyserlingk2015,Rosenow2020}. The first scenario is favored when the energy profile near the edges is relatively flat, which would in turn entail low velocities for edge excitations and hence is generally detrimental for quantum coherence. Let us therefore assume that the edges of the quantum Hall fluid inside the interferometer are energetically pinned to certain locations, as would be the case in the limit of steep confining potential. Under this assumption, the edges do not move in response to changes in the magnetic field, meaning that the area of the interferometer remains fixed, restricting us to the second and third scenarios.  We shall note that the second case, the ``pure'' AB regime, entails accumulation of the net charge inside the interferometer, with the excess charge drawn through the constrictions. While this regime appears to have been recently observed~\cite{Nakamura2019}, we note that this scenario is unlikely to be realized in large interferometers as both the amount of excess charge and the capacitance of the active part of the interferometer scale linearly with its area and hence so does the charging energy. Should this energy exceed the barrier for nucleating a quasihole (or bringing it from outside), a sudden change in the quasihole number would result in a phase jump. Since the energetics can, in principle, depend on the magnetic field actross the plateau~\cite{Rosenow2020}, a transition from  the second to the
the third scenario can occur via several such jumps -- an effect apparently recently observed at $\nu=1/3$~\cite{Nakamura2020}.

Given the relatively large size of the interferometers in our study, we do not anticipate the accumulation of charge and instead expect the third scenario to be realized.
%(In principle, some combination of these regimes may occur -- this is the so-called Coulomb dominated regime~\cite{Halperin2011a,Keyserlingk2015}; we will comment on this later.)
In such a scenario the change in the number of bulk quasiparticles in response to the change in flux $\Delta\Phi$ is given by $\Delta N_{e^*}=-(\Delta\Phi/\Phi_0 )(\nu e/e^* )$, resulting in
\begin{equation}
\Delta \gamma_{e^*}=\left(\frac{\Delta\Phi}{\Phi_0}\right)\left[2 \pi\left(\frac{e^*}{e}\right)-2\theta_{e^*}  \left(\frac{\nu e}{e^*}\right)\right].
\label{eq:suppl_interf_phase2}
\end{equation}
The reason for the negative sign in the second term is that an increase in magnetic field creates \textit{quasiholes}  resulting in the negative change of the quasiparticle number (since $N_{e^*}$ counts quasiparticles rather than quasiholes).

A word of caution is in order. Let us consider an application of this expression to a Laughlin state at $\nu=1/m$~\cite{Chamon1997}. Its quasiparticles carry the charge of $e^*=e/m$ and their statistical angle is $\theta_{e/m}= \pi/m$ resulting in a cancellation of the two terms in Eq.~(\ref{eq:suppl_interf_phase2}) and a seeming prediction of absence of any oscillations. This is, of course, wrong since the first term in the original Eq.~(\ref{eq:suppl_interf_phase1}) changes continuously with the increasing field whereas the number of quasiparticle inside the interferometer need not. Hence it is only meaningful to think of the cancellation in Eq.~(\ref{eq:suppl_interf_phase2}) whenever the magnetic flux is increased by a flux quantum resulting in  $\Delta N_{e^*}=-1$. This would result in the periodicity of one flux quantum
in accordance to both the Byers-Yang theorem~\cite{Byers1961} and the arguments given in Ref.~\onlinecite{Chamon1997}.

These equations can be easily generalized to the case of (Abelian) edge quasiparticles of type $a$ encircling bulk quasiparticles of type $b$:
\begin{multline}
\Delta \gamma_{ab}=2 \pi\left(\frac{\Delta\Phi}{\Phi_0} \right)\left(\frac{e_a}{e}\right)+2\theta_{ab}\Delta N_b
\\
=\left(\frac{\Delta\Phi}{\Phi_0} \right)\left[2 \pi\left(\frac{e_a}{e}\right)-2\theta_{ab} \left(\frac{\nu e}{e_b}\right)\right],
\label{eq:suppl_interf_phase3}
\end{multline}
where $2\theta_{ab}$ is the argument of the monodromy matrix element, i.e. the statistical phase resulting from a full braid of anyon $a$ around anyon $b$.

Discounting for the moment a possibility of Coulomb domination, which we will address later, we apply these considerations to the $\nu=5/2$ state. The possible charged excitations in this state are $e/4$ quasiparticles, $e/2$ Laughlin quasiparticles and electrons. As usual, electrons have trivial braiding with anything else, resulting in the expected Aharonov--Bohm oscillations with period $\Delta\Phi=\Phi_0$. Meantime $e/2$ quasiparticles have $2\theta_{e/2}= \pi$ (irrespective of the candidate state) and hence whenever they interfere around other $e/2$ quasiparticles, the period of Aharonov--Bohm oscillations is $\Delta\Phi=\Phi_0/2$. However, should they interfere around $e/4$ quasiparticles instead, the monodromy becomes $2\theta_{e/2,e/4}= \pi/2$ (also irrespective of the candidate state,~\cite{Bishara2008a,Bishara2009a}), and therefore
\begin{multline}
	\Delta \gamma_{e/2,e/4}=\frac{\Delta\Phi}{\Phi_0}\,\left[2 \pi\left(\frac{e/2}{e}\right)- \frac{\pi}{2} \left(\frac{5e/2}{e/4}\right)\right]\\
=-4 \pi\left(\frac{\Delta\Phi}{\Phi_0}\right)
\label{eq:suppl_phase_half_quarter}
\end{multline}
yielding the periodicity of $\Delta\Phi=\Phi_0/2$ once again.
Finally, there are $e/4$ quasiparticles, which are the most interesting kind since they are expected to be non-Abelian. Since they are Abelian with respect to the $e/2$ quasiparticles, we can use the aforementioned considerations to find
\begin{multline}
	\Delta \gamma_{e/4,e/2}=\frac{\Delta\Phi}{\Phi_0}\,\left[2 \pi\left(\frac{e/4}{e}\right)- \frac{\pi}{2} \left(\frac{5e/2}{e/2}\right)\right]\\
=-2 \pi\left(\frac{\Delta\Phi}{\Phi_0}\right)
\label{eq:suppl_phase_quarter_half}
\end{multline}
which in turn yields the periodicity of $\Delta\Phi=\Phi_0$ -- the same as that for interfering electrons.
Finally, we turn to the $e/4$ quasiparticles interfering around other $e/4$ quasiparticles. For this process, the statistical angle $2\theta_{e/4}$ acquired by the quasiparticles actually depends on the candidate state; the possible values for various such states can be found in Table II in~\cite{Bishara2009a}. (Note the notational discrepancy: what is denoted as twist factor $\theta$ in~\cite{Bishara2009a} corresponds to $\exp(\theta)$ in the notations of this manuscript, as well as those of~\cite{Willett2013b}.) For the purpose of this paper, we will focus on the two most likely non-Abelian states, specifically the Moore-Read (Pfaffian) and its particle-hole conjugate anti-Pfaffian states. (We shall also briefly comment on the recently proposed particle-hole-symmetric Pfaffian (‘PH-Pfaffian’) state~\cite{Bonderson2013b,Chen2014,Son2015,Zucker2016}). The statistical angle $2\theta_{e/4}=\pm \pi/2$   for the Pfaffian/anti-Pfaffian states while it can take on either value for the PH-Pfaffian state, depending on the fusion channel of the quasiparticles forming the braid. According to Eq.~(\ref{eq:suppl_interf_phase2}), the periodicity of the Aharonov--Bohm oscillations should be $4/9\,\Phi_0$ for the Moore--Read Pfaffian state and $4/11\, \Phi_0$ for the anti-Pfaffian, meantime both periods should be present for the PH-Pfaffian state. This argument, however, is misleading as it completely ignores the non-Abelian nature of the $e/4$ quasiparticles in these states. The predicted even-odd effect~\cite{Stern2006a,Bonderson2006a} entails the absence of interference of the $e/4$ quasiparticles whenever an odd number of such quasiparticles/holes is contained within the interferometer. This would in turn mean that the interference signal would disappear and reappear with the period corresponding to bringing an additional pair of quasiparticles in, or $\Delta\Phi=2\Phi_0 e^*/(\nu e)=\Phi_0/5$ at $\nu=5/2$ (and with $\Delta\Phi=\Phi_0/7$ at $\nu=7/2$). When the interference signal reappears, its phase is not governed by Eq.~(\ref{eq:suppl_interf_phase2}) as in this case the edge  $e/4$ quasiparticles invariably braid around an even number of bulk $e/4$ quasiparticles and one should focus on their net fusion channel instead. Let us assume that the even number of bulk quasiparticles always fuse in the trivial (bosonic) fusion channel. That means that the edge  $e/4$ quasiparticles effectively braid around a collection of $e/2$ quasiparticles in the bulk. Therefore the phase difference is governed by Eq.~(\ref{eq:suppl_phase_quarter_half}) which yields the periodicity of $\Delta\Phi=\Phi_0$ irrespective of whether the state is Moore-Read, anti-Pfaffian or PH-Pfaffian, contrary to what was claimed in~\cite{Keyserlingk2015}. Consequently one would expect to see two peaks at frequencies $f_\pm = (5\pm 1)f_0$ corresponding to the convolution of slow $f_0$ oscillations with the rapid on-off pulses at $5f_0$ – a manifestation of the even-odd effect. Here $f_0$ is the ``fundamental'' frequency corresponding to the periodicity of one flux quantum, $\Delta\Phi=\Phi_0$.

\begin{figure}[htb]
\centering
    \includegraphics[width=0.95\columnwidth]{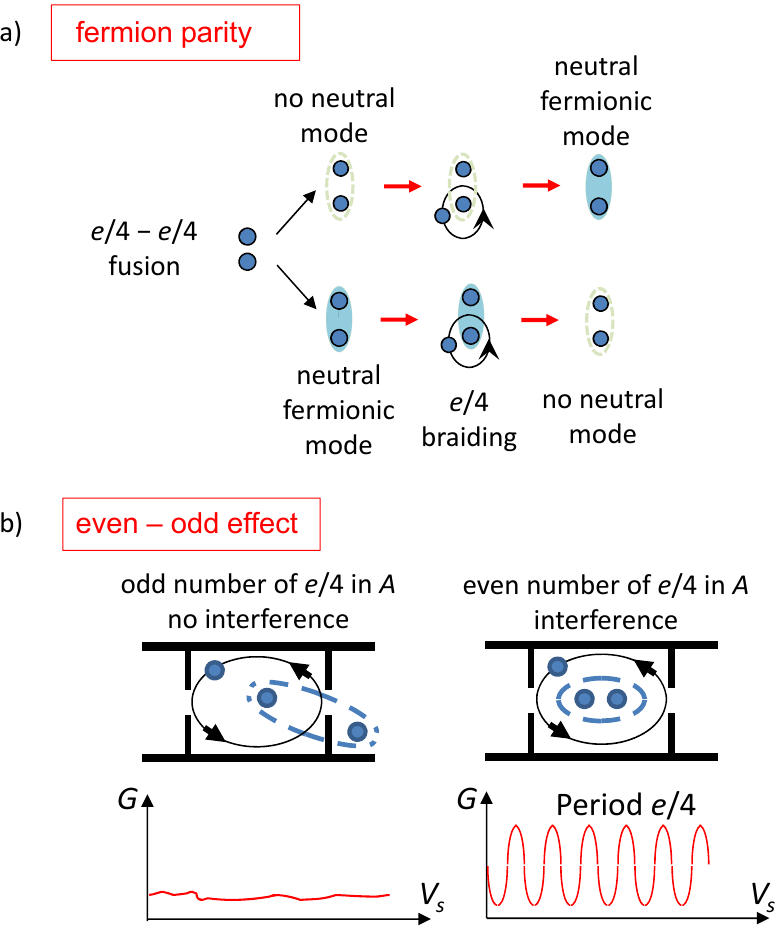}
    \caption{
    % Schematic and images of interferometer devices, transport through the devices in high purity Al heterostrutures,
    % large interference oscillations at 7/2 filling factor.
Fermion parity and the non-Abelian even-odd effect:\\
(a) Fermion parity describes the fusion of two non-Abelian $e/4$ quasiparticles, one with and one without a neutral fermionic mode. Braiding another $e/4$ quasiparticle with one of the fused $e/4$'s changes that fermion parity, which is the key to the non-Abelian even-odd effect.\\
(b) The even-odd effect manifests itself as the absence of inteference of non-Abelian $e/4$ quasiparticles whenever there is an odd number of such quasiparticles inside the interferometer. A quasiparticle propagating through one, but not the other arm of the interferometer invariably changes the fermion parity inside, thus leaving the system with a ``record'' of the path taken and preventing the interference. No such effect occurs when the number of $e/4$ quasiparticles inside the interferometer is even. The bottom plots show schematically the longitudinal resistance traces corresponding to these scenarios. In reality, some background signal should be observed in both cases due to the potential interference of other, Abelian quasiparticles.
}
    \label{fig:S1-2}
\end{figure}

So far we have ignored the other crucial feature of the non-Abelian states, namely the existence of neutral fermionic modes~\cite{Milovanovic1996} – one of the two possible fusion channels for a pair of  $e/4$ quasiparticles (see Figure.~\ref{fig:S1-2}). Again, two fusion channels available for a pair of $e/4$ quasiparticles can be either 1 (the identity channel; even fermion parity), which physically corresponds to an $e/2$  quasiparticle, or $\psi$ (odd fermion parity), in which case the  $e/2$  quasiparticle is accompanied by a neutral fermion.  Introducing such a fermionic quasiparticle within the interference loop shifts the overall phase of the $e/4$ interference by $\pi$~\cite{Fradkin1998,Stern2006a,Bonderson2006a,Bonderson2006b}. (It does not affect the interference of other types of quasiparticles). Hence, if the fusion channel of an even number of bulk $e/4$ quasiparticles is $\psi$ (instead of being trivial), i.e. if the fermion parity inside the interferometer is always odd whenever the interference is observed, one should again see the peaks at frequencies $f_\pm= (5\pm 1)f_0$. In other words, the same two peaks should be seen regardless of the fermion parity as long as it remains fixed during a magnetic field sweep. If, however, the fermion parity fluctuates on the timescale of a sweep, we need to distinguish two regimes. The first regime corresponds to intermediate time-scale fluctuations. Specifically, if the fermion parity fluctuates at the rate comparable to the rate at which the flux inside the interferometer is changed by one flux quantum $\Phi_0$ but remains stable on the scale of $\Phi_0/10$ (i.e. the flux change needed to change the parity of bulk $e/4$ excitations), one would expect the oscillations with the frequency of $f_0$ to become suppressed due to the random $\pi$ phase shifts, exposing the frequency of $5f_0$, the hallmark of the non-Abelian even-odd effect. Physically, such a scenario would be realized if e.g. the fusion channel were selected randomly every time a new pair of $e/4$ excitations is introduced into the bulk but then remained stable until another $e/4$ is nucleated inside. Another regime corresponds to rapid fluctuations whereby the fermion parity fluctuates within the time required to change the flux inside the interferometer by $\Phi_0/10$. In this case no high-frequency spectral peaks should be observed at all: for an odd number of bulk $e/4$ excitations the interference is suppressed by the non-Abelian even-odd effect whereas for their even number the interference is suppressed by rapid $\pi$ phase shifts.
Whether or not the fermion parity is fixed or fluctuates is a non-universal feature which is expected to depend on a particular sample.

Finally, turning our attention to the possibility of Coulomb domination~\cite{Halperin2011a,Keyserlingk2015}, we note that its chief effect is that the active area of the interferometer does not remain constant during the magnetic field sweeps. As the magnetic field is ramped up, the active area decreases until quasiholes are introduced into the active area whereupon the area is restored to its initial size and the process repeats. In fact, the quasiholes need not be introduced into the interference loop one-by-one, depending on the energetics one may envision a scenario whereby a clump of quasiholes is introduced before the active area is restored. It is not hard to see that the latter scenario alone (i.e. clumping of quasiholes together) can only result in suppression of certain periodicities but cannot possibly introduce new ones. However, combined with the effect of the fluctuating active area it may result in the appearance of doppelg\"anger spectral peaks of the interference oscillations. Specifically, let us consider a scenario whereby charge-$e/4$ quasiholes enter the interference loop in clumps of two or, equivalently, only charge-$e/2$ quasiholes can enter the loop. The interference signal is given by
\begin{multline}
\delta R_L \propto \cos \gamma_{e/4,e/2}
= \cos \left[\frac{\pi\Phi}{2\Phi_0}+2\theta_{e/4,e/2}N_{e/2}\right]\\
=  \cos \left[\frac{\pi AB}{2\Phi_0}+2\theta_{e/4,e/2}N_{e/2}\right]
\label{eq:suppl_coulomb_domination}
\end{multline}
in agreement with Eq.~(\ref{eq:suppl_interf_phase3}). While we have already analysed the periodicity of this signal -- see Eq.~(\ref{eq:suppl_phase_quarter_half}) and the discussion thereafter -- we note that in the case of Coulomb domination neither term in the argument of the cosine is linear with magnetic field $B$; both $A(B)$ and $N_{e/2}(B)$ contain a Fourier component with the period corresponding to adding one $e/2$ quasihole into the active area. It is easy to see that one additional flux quantum produces five quasiholes at $\nu=5/2$ and seven quasiholes at $\nu=7/2$ and therefore the argument of the cosine contains harmonics with periods of $\Phi_0/5$ and $\Phi_0/7$ (and their unit fractions) at $\nu=5/2$ and $\nu=7/2$ respectively. This would lead to the appearance of frequencies of $5f_0\pm f_0$ (as well as $10f_0\pm f_0$ etc.) at $\nu=5/2$ and $7f_0\pm 1.5 f_0$ at $\nu=7/2$, seemingly mimicking the spectral features due to the non-Abelian even-odd effect. (We are grateful to B.~I.~Halperin and D.~E.~Feldman for bringing this scenario to our attention.)

While we cannot definitely discount such a possibility, we believe the preponderance of experimental evidence does not support Coulpmb domination as the source of the high-frequency oscillations observed in the experiments presented here. Firstly, the sample preparation technique reported in this paper had reducing the likelihood of Coulomb domination in our devices as one of its chief goals; its success appears to be validated e.g. by the nature of the ``pajama plots'' shown in Figs.~\ref{fig:pajama_plot} and \ref{fig:S4-2b} -- see also an extended discussion in Section~\ref{sec:S4b} below. Furthermore, a number of observed features associated with the high-frequency oscillations require progressively convoluted fine-tuning of the Abelian Coulomb domination scenario for their explanation. E.g., the observed $\pi$ phase jumps (see Figs.~\ref{fig:fermionic_parity} and \ref{fig:S5-6-2}) can still be explained in this scenario by the tunnelling of the neutral fermionic mode in and out of the interference loop yet it is unclear why it would only flip the sign of the fast component of the oscillation and not of the whole interference signal given by Eq~\ref{eq:suppl_coulomb_domination}. Furthermore, while one can imagine a scenario whereby the high frequency resistance oscillations due to the oscillations of the active area are of purely energetic rather than quantum interference origin (e.g. due to the variations of the tunnelling distance with the active area), attempts to reconcile such a scenario with the observed $\pi$ phase jumps become even more strained.

Last, but not least, there is a direct experimental test of the Coulomb-domination scenario. Specifically, in any quantum Hall state with the filling fraction $\nu=p/q$, the amount of flux required to introduce a quasihole of charge $e/q$ is equal to $\Phi_0/p$. Assuming that the energetics does not change dramatically within a reasonable range of filling fractions, one could test a nearby, better understood FQH state for the presence of the small period $\Phi_0/p$ resistance oscillations. The $\nu=7/3$ state presents such a convenient test case. The AB oscillation period for the $e^\ast=e/3$ quasiparticles is $\Phi_0/2$ according to Eq.~(\ref{eq:suppl_interf_phase2}) whereas the periodicity of introducing these quasiparticles/quasiholes into the active area is $\Phi_0/7$. Observing a high-frequency spectral feature at $f=7f_0$ would therefore support a Coulomb-domination scenario. We have tested this experimentally and have not found any discernable features at either that frequency or at $f=(7\pm 2)f_0$ -- see Figure~\ref{fig:seventhirds}. (The latter frequency would correspond to the Coulomb modulation of the AB signal.)
%For a more detailed analysis of this phenomenon and its consequences the reader is referred to;%
In light of the presented arguments, we feel confident that Coulomb domination is an exceedingly unlikely explanation for  the spectral features actually seen in the experiment and presented in this manuscript although further tests aimed at quantifying the importance of Coulomb effects in these devices are called for as a part of device characterization.

Having analyzed the set of possible spectral peaks associated with the quasiparticle interference, we should also mention their expected relative prominence and temperature dependence. Those features depend primarily on the type of interfering edge quasiparticles since that determines their tunneling across the constrictions. Theoretically, at low voltages and low temperatures the tunneling amplitude is governed by their scaling exponents~\cite{Wen1995}. These scaling exponents have been tabulated for $\nu=5/2$ candidate states (as well as for some other QH states) in~\cite{Bishara2009a}, where the temperature dependence of possible interference signals have also been discussed. For the purposes of this manuscript, it suffices to mention that for all possible non-Abelian $\nu=5/2$ and $7/2$ states with the exception of anti-Pfaffian (but including PH-Pfaffian), the dominant tunneling process at low temperatures is that of $e/4$ edge excitations, with $e/2$ and electron contributions being progressively suppressed. In the anti-Pfaffian state, the $e/4$ and $e/2$ edge excitations have the same scaling exponent and hence both processes should have the same temperature scaling. This, in principle, allows us to use the temperature dependence to both analyze the origin of each spectral peak and try to glean the nature of the state. However, there is one caveat: even for the peaks associated with the interference of $e/4$ edge excitations, the aforementioned issue of fermion parity may complicate the analysis of the high-frequency peaks since we do not have a good model for how the stability of that parity is affected by the temperature.

\subsection*{Sections~\ref{sec:S2}--\ref{sec:S4}: Methods overview}

Here we first review more generally the methods pertinent to the measurements described in the main text, and in the Sections~\ref{sec:S2}--\ref{sec:S4} delve into even more details and background specific to the different methods.

The two-dimensional electron gases (2DEG) used in this study include GaAs/AlGaAs heterostructure quantum wells with different doping properties. All are modulation-doped and of the highest quality:  all have greater than $20\times 10^6 \text{cm}^2/\text{V}\text{s}$ mobility. Doping is from delta doped silicon layers on both sides of the quantum well, with an intermediate narrow ($<2\,$nm) shielding well of GaAs between the dopant layers and the principle quantum well (nominal width 20nm), or the silicon doping is contained within a narrow GaAs well (1nm) removed from and on either side of the principal quantum well, referred to as a doping well sample (see Figure S2-1 for structures).  The samples range in density from $3$ to $4.5 \times 10^{11} \text{electrons/cm}^2$.  More than 10 samples from 5 different wafers have been used for this study.  A key aspect of this study is use of a new higher-purity class of GaAs/AlGaAs heterostructure quantum wells that have improved  resistance oscillation signals in the interferometers, enabling  observation of the stable oscillations at $\nu=7/2$ attributed to the non-Abelian even-odd effect due to charge $e/4$ quasiparticles.  These structures make use of highly purified Al. The details of the heterstructure devices are described below in Section~\ref{sec:S2} -- Methods: Heterostructures and Interferometer Designs.

In both classes of heterostructures, contacts formed from Ni\textbackslash{Au}\textbackslash{Ge}\textbackslash{Ni} layering are diffused into the perimeter of a mesa containing the 2DEG, and the areas between the contacts are left for deposition of the top gate structures that form the interferometers.  After contact formation and diffusion, a 30nm layer of amorphous SiN is deposited on the mesa to further insulate the 2DEG from the top gate structures.   The top gates are formed from Al and Au layers, Ti and Au and Al layers, or from Ti and Au layers.  The top gate layers do not exceed 120nm in thickness.
	The samples are illuminated after mounting and cooling in a dilution refrigerator but prior to charging the top gates.  The temperature at which illumination is applied ranges from room temperature to the base temperature of $\sim\!20\,$mK, and varies dependent upon the doping structure of the heterostructure to optimize transport quality.  Illumination is crucial to achieve maximum potential mobilities, and in particular to achieve the maximum 5/2 state energy gap. The illumination and gating sequences are referred to as the preparation. In this study the sample number and preparation number are given for each data set; the sample number is a specific device, and the preparation number is the specific illumination and gating preparation for that data collection.
	An interference device is shown schematically in Figure~\ref{fig:S1-1}.   The top gates are charged to a negative voltage sufficient to deplete the underlying electron layer.  At high magnetic fields as prescribed for filling factor $\nu=5/2$  (filling factor $\nu=$electron areal density/magnetic flux density), the currents carrying the excitations of the fractional quantum Hall state will travel along the edge of these depleted areas, surrounding an area of the bulk filling factor $\nu$. The important principal physical property of the interferometer device is two separated locations where these edge currents are brought in proximity (marked 1 and 2 in the Figure).  At these points backscattering from one edge to the other can occur, and with this backscattering two different current paths are established that can interfere, as shown by the dashed lines in the schematic.  The one path encircles the area marked $A$ in the schematic, and changes in the magnetic flux number within area $A$ or changes in the particle number within area $A$ will cause phase accumulation for that path (the Aharonov--Bohm and statistical phase contributions). Interference of that path and the one not entering the area $A$ produce oscillations in the resistance measured across the interference device.  The voltages on the top gates can be adjusted to promote backscattering (gates marked $V_b$) and to change the enclosed area $A$ (gates marked $V_s$).  The separation of the backscattering top gates, distance marked  $d_g$ in Figure~\ref{fig:interferometer}, is sufficiently large that for nominal voltages on $V_b$ the backscattering is weak, an important feature to maintain the 5/2 fractional Hall state contiguously from outside to inside the active area $A$ of the interferometer.  Note also that area $A$ is ultimately the area which is enclosed by the edge currents in the quantized Hall systems, and so will not be the lithographic area but rather the electrostatically determined edge consequent to the applied gate voltages. The lithographic area in all samples used is several square microns, but the active area $A$ is typically less than one square micron. Note that the lithographic area of the devices in the electron-micrographs shown in Figure~\ref{fig:interferometer} is roughly $9\, \upmu\text{m}^2$. Area $A$ is the area enclosed by the innermost edge current, the backscattered current, and has the same filling factor as the bulk outside the interferometer given the interferometer design and appropriate gate voltages. The remaining area between this innermost edge current and the lithographically defined top gates is populated by a cascade of compressible and incompressible states between the bulk filling (the filling factor of area $A$) and zero filling.  The standard interferometer device contains no overlay gate structure. In another interferometer device type a small dot is placed centrally in the area $A$ and is accessed by an air-bridge that extends over one of the side gates marked $V_s$ in Figure~\ref{fig:interferometer} of the main text and Figure~\ref{fig:S1-1} of Supplementary Materials. Electron micrographs of both of these two interferometers are shown in Figure~\ref{fig:interferometer}, right-hand panels.  See Section~\ref{sec:S2} \textit{Methods: Heterostructures and Interferometer Designs} for more interferometer details.

 	Two important points in the construction of the samples work to avoid Coulomb dominated effects and promote Aharonov--Bohm interference in the devices; as described above large separation between the backscattering gates (distance $d_g$ in Figure~\ref{fig:interferometer} of the main text and Figure~\ref{fig:S1-1}), and specific layering of the heterostructure itself.  The large gate separation allows tuning of the backscattering around small reflection amplitudes, and so a nearly open geometry. The second important point in heterostructure design to avoid Coulomb domination effects is use of the shielding wells above and below the principal 2D quantum well: these crucially serve to inhibit charge accumulation. These wells display nominal conduction over large dimensions at high testing temperatures ($\sim\!300\,$mK) which can appear as a background to the resistance. Measurement of the integer filling interference in these samples (see Section~\ref{sec:S4} of Supplementary Materials) demonstrates constant negative phase slope consistent with Aharonov--Bohm interference and inconsistent with Coulomb domination. See Section~\ref{sec:S4} \textit{Methods: Interferometry} for detailed data on Aharonov--Bohm versus Coulomb dominated effects.

	Resistance and resistance oscillations are measured using low noise lock-in amplifier techniques.   A constant current (typically 2nA) is driven through the 2D electron system as laterally defined by the surface interferometer gate structure, and the voltage, and so resistance, is determined with a four-terminal measurement.  The voltage drop along the same edge of the 2D electron system and across the device gives the longitudinal resistance $R_L$; across the device and across the two edges of the 2D system gives diagonal resistance $R_D$.  Similar measurements not across the interferometer device yield $R_{xx}$ and $R_{xy}$ respectively. An example of longitudinal resistance $R_L$ across an interferometer is shown in Figure~\ref{fig:Hall_trace}(a).

	The interference oscillations in the measured resistance are examined for their frequencies by applying fast Fourier transforms (FFT) to the data.  Because the data are from a fractional quantum Hall resistance minimum, the minimum background is subtracted before the FFT is applied.  The background that is subtracted is determined equivalently by either a polynomial fit or a running large element smoothing of the minimum.   Figure~\ref{fig:power_spectra}(a) shows an example of this process; there the interference oscillations in $R_L$ around filling factor 3 are used to determine the integer Aharonov--Bohm interference frequency.
	The magnetic field range over which the FFT is performed is particularly important here in exposing all the predominant interference frequencies that can be associated with the different charge braids.  In order to extract the entire complement of frequencies the background subtraction and transform should be applied to the full extent of the 5/2 or 7/2 minimum. In some samples and preparations distinct changes in the resistance mark that extent, as with the ultra-pure, Al cleaned heterostructures; in others this is not the case.  Using as large an extent of the 5/2 or 7/2 minimum as possible is essential since modulations of the important frequencies (the even-odd effect frequency) can occur, and this modulation is appreciated only when the full pattern of interference is transformed. See Figure~\ref{fig:7_halves_interf} and Section~\ref{sec:S3} \textit{Methods: Measurements} for more information on measurement and data analysis.

\subsection{Methods: Heterostructures and Interferometer Designs}
\label{sec:S2}
\setcounter{figure}{0}

\subsubsection{Basic heterostructure designs}
\label{sec:S2a}
The fundamental layering for heterostructures used in this study is displayed schematically in Figure S2-1.  The 2D electron system resides in the GaAs layer that is typically $\sim 24$nm wide.  On either side is an $\text{Al}_x\text{Ga}_{1-x}\text{As}$ layer: the $x$ in this layer is predominantly $x=24\%$.  These sides are flanked by thick layering of $\text{Al}_x\text{Ga}_{1-x}\text{As}$ of higher Al concentration, $x=32\%$.  Within this highest Al section on both sides of the GaAs well are the Si dopant sheet layers: the dopant layer on the substrate side donates electrons to the GaAs well, and the top Si dopant layer donates to both the GaAs well and the states at and near the surface of the heterostructure.

\begin{figure}[htb]
\includegraphics[width=0.95\columnwidth]{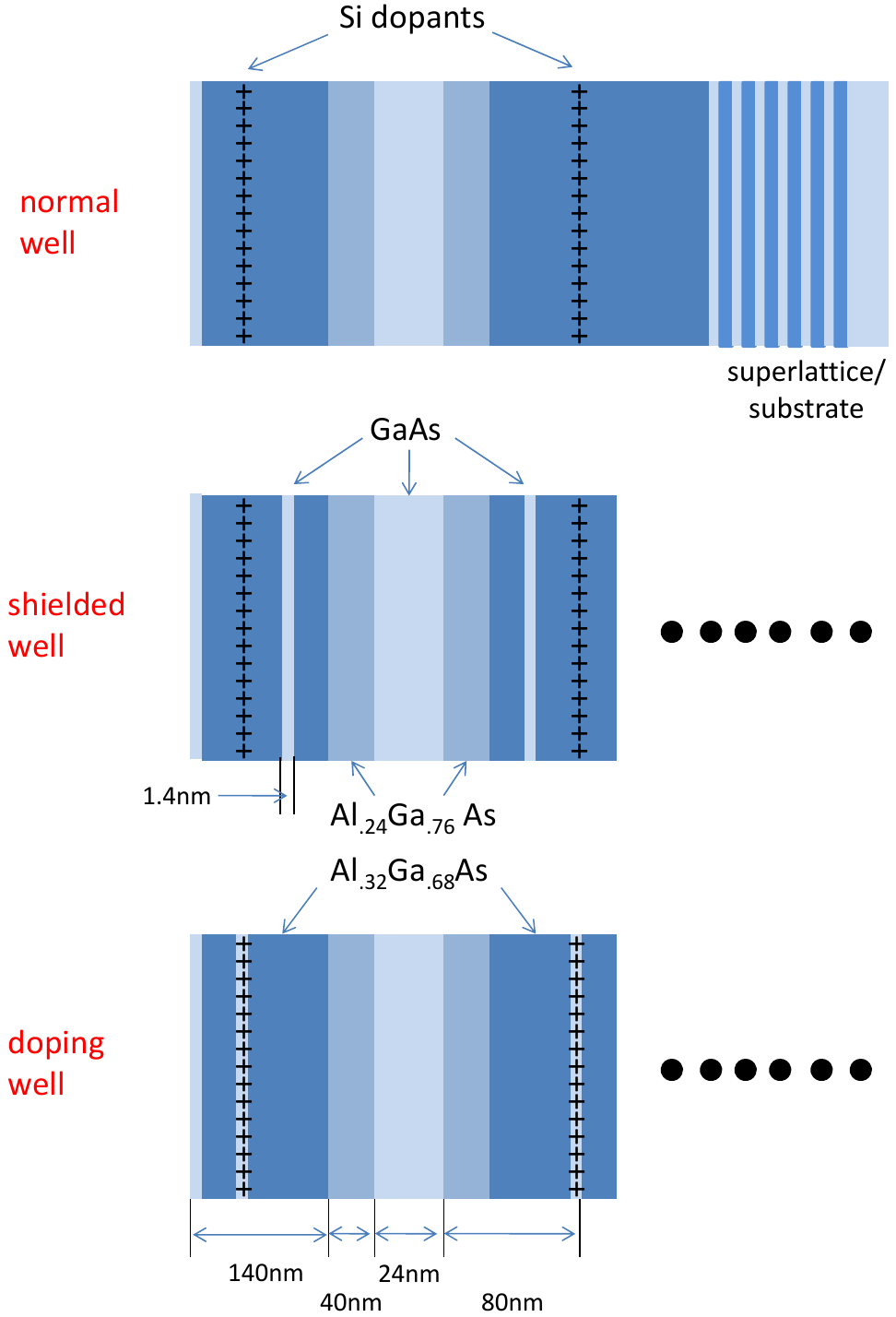}
\caption{Schematics of three different heterostructure designs displaying the layering structure in GaAs and $\text{Al}_x\text{Ga}_{1-x}\text{As}$ materials. The substrate level is on the right, top of the wafer is on the left.  These schematics provide only a general picture of the wafer constructions, as the growth conditions (temperatures, stops, etc.) are not included. }
\label{fig:S2-1}
\end{figure}

Given this basic design, there are two variants used in this study that provide high mobility, strong sets of fractional quantum Hall states, and 2D systems that can be laterally controlled with top gates. These variants are also shown in Figure~\ref{fig:S2-1}.  The first variant, called a shielded well, has a small but important addition to the fundamental normal well design.   Between each doping layer and the quantum well but within the $32\%$ Al concentration layer are placed thin wells of GaAs ($\sim 1$nm).  These layers, on either side of the 2D system in the central 24nm quantum well, accept a low density of electrons and demonstrate some parallel conduction in magneto-transport measurements (confined to low magnetic fields).   These layers provide two important functions.  First, the shield layers seem to provide a smoothing effect on the potentials presented by the ionized doping layers, resulting in stronger electron correlations in the central well.  This manifests as high mobility and larger correlated state gaps.   Second, the shield layers effectively shield the central well itself, minimizing the Coulomb domination.  This is a particularly important effect since the interferometer devices impose lateral confinement to the 2D system, which in small interferometer areas could induce Coulomb blockade or domination of the device by Coulomb effects.  The shielding wells soften this central well charge accumulation; in a small device with larger edge to central area ratio perturbation of the total charge is less energetically possible.

The second variant to the basic heterostructure design used in this study is referred to as a doping well, and it is also shown schematically in Figure~\ref{fig:S2-1}. In this design, the Si dopants are placed in narrow GaAs wells within the high Al concentration layers.  These heterostructures also demonstrate high mobilities and large fractional quantum Hall state gaps, but have one difficulty in use for interferometry: surface gates do not modulate the underlying 2D electron system density in a usable way at lowest, or operational temperatures.  The only method by which these structures can be used is to set the gates at working voltages at high temperatures (4K) and operate at base temperatures performing only magnetic field sweeps.

The description of the heterostructures above is highly simplified.   In order to produce truly functional heterostructures the precise parameters of the structures must be empirically determined, and the parameter set to be established is large.  This set includes doping levels (top and bottom), shielded well and doping well widths, intermediate Al concentration (shrouds) widths, central well widths, temperatures for deposition at each of these stages, and cap layer thicknesses, to name but some.  This all assumes that a proper substrate and substrate platform has been grown.
%\onecolumngrid
%\twocolumngrid

\subsubsection{Extreme aluminum purity in heterostructures of this study}
\label{sec:S2b}

This study relies in part on the development of a new higher-purity class of GaAs/AlGaAs heterostructure quantum wells, that have improved the resistance oscillation signals from our quantum interferometers, and enabled the observation of the stable oscillations at $\nu = 7/2$ attributed to the non-Abelian even-odd effect.   With materials used in the previous studies, the amplitude of the measured interference oscillations had never exceeded $\sim 3\,\Upomega$.  With the new higher purity material, we find interference oscillations in resistance as large as $50\,\Upomega$, as can be seen in Figure~\ref{fig:Hall_trace}(b) of the main text.

Central to the overall improvements in our heterostructures are modifications that have been made to our MBE chamber. One example of this is the installation of a $\sim 17$K cold plate very close to the sample growth space to establish a cleaner growth environment. In addition, there are large-area $\sim 19$K cryogenic cold plates located in the vacuum sump of the MBE growth chamber.  Residual gas analysis shows that these reduce the nitrogen impurities in the MBE vacuum by a factor of twenty, and reduce the water impurities by a factor of three.

However, the improved quality of the samples is most likely linked to our recent efforts to purify the Al source material itself~\cite{Chung2018}.  A theoretical investigation on the influence of residual impurities on quantum scattering rates supports this view, showing that a high-purity AlxGa1-xAs barrier layer is essential to achieve large quantum lifetimes~\cite{Sammon2018}.

A novel MBE surface segregation technique allows one to infer the number of oxygen and other impurities in the AlGaAs barrier layers~\cite{Chung2018}.  With this impurity assessment, we have shown a method by which the charged oxygen impurities in the AlGaAs can be reduced using offline bakes of the Al-effusion furnaces~\cite{Chung2018}.  These techniques were used to grow wide AlAs quantum wells having transport mobilities that are eight times higher than previous work~\cite{Chung2018a}. This suggests that the AlAs material in these wells has on the order of eight times fewer impurities than previous material, and by extension that AlGaAs barrier material used in the heterostructures of this study would have similarly fewer charged impurities.

According to the calculation in~\cite{Sammon2018}, the transport mobility in the highest mobility GaAs quantum wells ($\mu\sim 30\times 10^{6}\text{cm}^2 /\text{V}\text{s}$) is largely limited by an estimated $2\times 10^{13}$ impurities/cm$^3$ located in the GaAs channel. The density of charged impurities in the adjacent AlGaAs barriers of these samples is estimated at $2\times 10^{14}$ impurities/cm$^3$. Because there are two barriers, each typically much thicker than the channel, there could be as many as 100 times more charged impurities in the barriers than in the conducting channel. Transport lifetimes are dominated by the impurities located within the conducting channel that typically cause carrier scattering at large angles.  For this interferometer study, the important lifetimes are quantum lifetimes, which are sensitive to all scattering events at any angle, including those from the large numbers of relatively remote charged impurities in the barrier.

\subsubsection{Sample illumination and gating processes}
\label{sec:S2c}

Both shielded well devices and doping well devices in this study are illuminated in the process of sample preparation.  Illumination is accomplished by exciting a red LED that is positioned over the sample in the dilution refrigerator.  Both the duration of illumination and the temperature(s) at which the light is applied are important and specific to the devices used for achieving optimal operation.  For example, the doping well samples are typically continuously illuminated during cool-down from room temperature to 4K, at which point the illumination is stopped.  A similar procedure can be used in shielded well samples.  However, in certain versions of the shielded well samples, illumination can be applied over a large range of possibilities. This range extends from LED on from room temperature to 4K, or to 1K, or to near base temperature (less than 100mK: base temperature will not be achieved with illumination on).  Or, in the other extreme of exposure duration, the LED is turned on only at base temperature for a brief period (about one minute), where after the system is allowed to re-cool to base (20mK).   Again, within this range of possibilities, the optimum illumination process will be specific to the heterostructure, with its own structure specifics.  In general, the shielded well samples are more illumination process tolerant than doping wells.  However, it is absolutely crucial to illuminate both types of heterostructures in their respective modes to achieve optimum $\nu=5/2$ and 7/2 states.

Given that the heterostructures are necessarily illuminated in order to function, the next process is energizing the gates.  As described above, the doping well samples’ gates must be charged at 4K, and this is performed always with the LED off: no illumination is ever applied during application of voltages to any gates in any type of heterostructure.  The shielded well sample gates can be energized in a range of procedures with respect to illumination.  Always considering that the illumination is applied before the gate voltages are applied, illumination can be applied at any temperature followed by gate charging, either immediately after illumination is off or after any level of time delay. It has been observed for some samples that if illumination is applied at low or base temperatures, transport subsequent to gating can be improved if the gates are charged while the sample is still warmed from the illumination.  It has also been observed that in some materials gating not temporally close to illumination but at base temperature later with a waiting period after gating before application of magnetic field may also improve sample operation.  All these illumination/gating time and temperature processes are best explored with each heterostructure type, following the general guidelines above for doping or shielded wells.

An important property of illumination in shielded well samples with respect to gating is the ``annealing'' or ``reset'' effect of illumination on these heterostructures.  In general, a top gate device can be energized, and any residual effect from such gating can be erased by re-illumination.  This provides a convenient means for testing devices in that following a gating procedure to some set of voltages, the device can be ``reset'' rapidly by returning to zero $B$-field, turning off all gate voltages, and re-illuminating the sample.  Gates can then be recharged according any chosen prescription without any residual effects from the prior gating.  Note again that illumination is always at zero gate bias, but also at zero magnetic field.

We refer to the thermal cycling, illumination, gating and large $B$-field excursion history of each sample as a preparation, and for each presentation of data list both the sample and preparation numbers.  Details of each wafer are displayed in Table~\ref{tab:S2-T1}. The samples are listed in Table~\ref{tab:S2-T2} below, referring to each wafer on which the device is fabricated, interferometer device type, area, and backscattering constriction width ($d_g$ in Figure~\ref{fig:S1-1}).  Below we also list the preparations (with some commentary) for each sample preparation relevant to the results presented in the main text. Note that proper delineation of the preparation includes the history back to cooling from room temperature.
\begin{table}
  \begin{tabular}{| c | c | c | c |}
    \hline
    \multicolumn{4}{|c|}{Wafer parameters} \\
        \hline
    wafer & density & mobility & wafer type \\
     { } & $\left(\times 10^{11}\text{cm}^{-2}\right)$ & $\left(\times 10^{6}\text{cm}^2/\text{V}\text{s}\right)$ &{ } \\
      \hline\hline
    a & 4.0 & 28 & shielded well\\ %\hline
    b & 4.1 & 27 & shielded well\\ %\hline
    c & 3.8 & 29 & shielded well\\ %\hline
    d & 2.7 & 34 & doping well\\ %\hline
    e & 3.9 & 26 & shielded well Al-plus\\ %\hline
    f & 4.4 & 23 & shielded well Al-plus\\ %\hline
    g & 3.5 & 21 & shielded well\\ %\hline
    \hline
  \end{tabular}
  \caption{Wafer parameters.  The wafer number corresponds to specific different wafers used in this study, with the electron density, mobility, and wafer type listed for each. Note that wafer type includes doping wells and shielded wells -- see Figure~\ref{fig:S2-1}. Notation ``Al plus'' is used for the extreme high-purity aluminum heterostructure wafers. These wafer labels (a--g) are used in Table~\ref{tab:S2-T2}.}
  \label{tab:S2-T1}
\end{table}

\begin{table}
  \begin{tabular}{| c | c | c | c | c |}
    \hline
    \multicolumn{5}{|c|}{Sample/device parameters} \\
        \hline
    sample & wafer & area ($\upmu\text{m}\times \upmu\text{m}$) & $d_g$ ($\upmu\text{m}$) & device type \\
   %  { } & $\left(\times 10^{11}\text{cm}^{-2}\right)$ & $\left(\times 10^{6}\text{cm}^2/\text{V}\text{s}\right)$ &{ } \\
      \hline\hline
   1  & b & $2.5\times 2.5$ & 1.1 & standard\\ %\hline
   2  & c & $3.6\times 3.6$ & 1.1 & central dot\\ %\hline
   3  & c & $3.6\times 3.6$ & 1.1 & standard\\ %\hline
   4  & d & $3.6\times 3.7$ & 1.1 & standard\\ %\hline
   5  & b & $5.7\times 5.7$ & 1.1 & standard\\ %\hline
   6  & e & $3.6\times 3.6$ & 1.2 & central dot\\ %\hline
   7  & g & $2.5\times 2.5$ & 1.2 & standard\\ %\hline
   8  & c & $3.6\times 3.6$ & 1.0 & top gate\\ %\hline
   9  & f & $4.6\times 4.6$ & 1.0 & standard\\ %\hline
   10 & d & $5.7\times 5.7$ & 0.9 & standard\\ %\hline
   11 & e & $4.6\times 4.6$ & 1.2 & central dot\\ %\hline
   12 & e & $3.6\times 3.6$ & 1.2 & standard\\ %\hline
    \hline
  \end{tabular}
  \caption{Sample and device parameters.  The sample number corresponds to the labeling used throughout the manuscript.  Area is the lithographic area of the devices as defined by the inner dimension of the gate set bounded by the $V_b$ and $V_s$ gates.  The constriction width marked $d_g$ in Figure~\ref{fig:S1-1} is the lithographic dimension presented in this table.  The device type refers to one of two electron micrographs of Figure~\ref{fig:interferometer}, standard (top), central dot (bottom). Also listed is a top gate structure where all but the perimeter of the enclosed area defined by the top  $V_b$ and $V_s$ gates is covered. }
  \label{tab:S2-T2}
\end{table}

Figure~\ref{fig:Hall_trace}: sample 6, preparation 16.  (1) light (a) on room temperature (RT) to 4K,  (2) at base temp. ($\sim\!20\,$mK) $V_s$ to $-2.0$V, $V_b$ to $-3.0$V,  (3) warmed to 4K, IVC poisoned, gates off,  (4) light (a) on for 10 minutes, wait 1 day,  (5) $V_s$ to $-2.0$V, $V_b$ to $-3.0$V at base temp, (6) gates to 0, light (a) on for 1 minute, wait overnight, (7)  $V_s$ to $-2.0$V, $V_b$ to $-3.0$V,  (8) $V_s$ to $-2.07$V.

Figures~\ref{fig:integer_vs_fractional_peak_positions} and \ref{fig:fermionic_parity_stability}: sample 6, preparation 25.  After 4th cycle to RT, (1) light (a) on, RT to 4K, (2) later cycles to 4K ($\times 2$), last with light on for 30 minutes, low gate voltages ($V_s=-1.5$V, $V_b=-2.0$V,) progress to $-2.0$V \& $-4.0$V respectively.

Figure~\ref{fig:pajama_plot}: sample 6, preparation 3. Second cycle to RT, (1) light (a) on, RT to 4K, gates on at base temp, (2) cycle to 4K, 10 min. light (a) on, gates on ($-2.0$V \&$-3.0\text{V} = V_b$), (3) at base temp., light (a) on for 1 minute, gates ($-2/-3$) on immediately.

Figure~\ref{fig:seventhirds}: sample 6, preparation 36. RT to 4K with light. At base temp. ($\sim\!20\,$mK) $V_s$ to $-2.0$V, $V_b$ to $-2.0$V; re-illuminate; gates $V_s$ to $-2.0$V, $V_b$ to $-4.0$V. Sweeps: gates $V_s$ to $-2.0$V, $V_b$ to $-2.0$V.

Figure~\ref{fig:7_halves_interf}: sample 6, preparation 8.  (1) light (a) on room temperature (RT) to 4K,  (2) at base temp. ($\sim\!20\,$mK) $V_s$ to $-2.0$V, $V_b$ to $-3.0$V, (3) large $B$-range sweeps (filling factor 2 to 7) for 1 week, see stabilization in $R_L$.

%Fig. 4c and 4d. Sample 6, preparation 16.  See above Fig. 2b and c.

Figure~\ref{fig:fermionic_parity}: sample 6, preparation 19. (1) light (a) on room temperature (RT) to 4K,  (2) -- (4) 3 cycles to 4K with light on for 10 to 15 minutes each time over 3 month period, (5) at 4K light (a) on for 10 minutes and (b) for 40 minutes, cool to base and then $V_s$ to $-2.0$V, $V_b$ to $-3.$0V. (6) light (a) on for 1 minute base, immediate $V_s$ to $-2.0$V, $V_b$ to $-2.0$V, (7) hours later $V_s$ to $-2.$0V, $V_b$ to $-6.0$V, (8) two days later $V_s$ to $-2.0$V, $V_b$ to $-4.0$V. (8) $B$-sweep filling 4 to 2 for 1 week, (9) within 7/2 minimum 2 days.

Figure~\ref{fig:5_halves_interf}: sample 6, preparation 18. (1) light (a) on, RT to 4K, (2)-(3) 2 cycles to 4K, 10 min. light (a) on each time, (4) at base temp., light (a) on for 1~minute, gates ($-2/-3$) on.

Figure~\ref{fig:power_spectra}: sample 2, preparation 2.  (1) RT to 4K, light (a) on for 30 minutes, cool to base,  (2) $V_s$ to $-3.0$V, $V_b$ to $-3.0$V,  (3) light (a) on at base for 30sec., back at base $V_s$ to $-2.0$V, $V_b$ to $-3.5$V,  (3) T mix. chamber to 1.5K, light (a) on for 20  sec. , to base and $V_s$ to $-1.5$V, $V_b$ to $-3.0$V, (4)  no more light but $V_b$  to $-4.0$V, day, $-5.0$V, day, $-6.0$V, day, $-7.0$V, day, ground then $-3.0$V. (5) light (a) on for 30 sec., immediate  $V_s$ to -1.5V, $V_b$ to $-3.0$V,  (6) light (a) on for 30 sec., immediate  $V_s$ to $-1.5$V, $V_b$ to $-5.0$V, (7) light (a) on for 30 sec., immediate  $V_s$ to $-1.5$V, $V_b$ to $-7.0$V, (8) light (a) on for 30 sec., immediate  $V_s$ to -1.5V, $V_b$ to $-9.$0V.

Figure~\ref{fig:interf_model}(a) --  see the information for Figure~\ref{fig:power_spectra}  above;  Figure~\ref{fig:interf_model}(b):  sample 6 preparation 15. (1) light (a) on room temperature (RT) to 4K,  (2) at base temp. ($\sim\!20\,$mK)  ramped $B= 0$ to filling factor 2, back to $B=0$, gates grounded, (3) applied $V_s$ to $-2.0$V, $V_b$ to $-3.0$V at base,  (3) small $B$-range sweeps (filling factor 7/2 minimum) for 1 week.

Figure~\ref{fig:2f0_peak}(a) --  see the information for Figure~\ref{fig:power_spectra}  above;  Figure~\ref{fig:2f0_peak}(b): same but with gate voltages changed at $B=0$ to $V_s=-2.0$V and $V_b=-3.5$V.

Figures~\ref{fig:S4-5} and \ref{fig:S4-6}: sample 12, preparations 34--38. RT to 4K with illumination on,  gates on sequentially as listed in Figures at base temperature.

Figure~\ref{fig:add_osc1}, top panel: sample 6 preparation 18. Warm to 4K after preparation 17, illumination at 4K, set gates to $V_s=-1.5$V, $V_b=-6$V, then without more illumination to $V_b=-4$V, $V_s=-2$V.\\
Middle panel: sample 6, preparation 17. (1) after preparation 15, illumination, gates immediately set to $V_b=-3$V, $V_s=-2$V, (2) repeat illumination and gates set to $V_b=-3$V, $V_s=-2$V.\\
Bottom panel: sample 6, preparation 27. Room temperature to base, back to 4K with 30 minutes light, wait to base temperature and set gate voltages $V_b=-2.5$V, $V_s=-1.5$V.

Figure~\ref{fig:add_osc2}.  Sample 6, preparation 20.  (1) after preparation 19, re-illumination at base temperature and immediate application of gate voltages $V_b=-4$V, $V_s=-2$V, followed by steps in side-gate voltage $V_s$.

\subsubsection{Interferometer designs and heterostructure concerns}
\label{sec:S2d}

The two critical aspects of interference device construction are the total active area of the device and the structure of the back scattering top gates.  In addition, the orientation of the interferometer on the heterostructure with respect to the crystallographic axes influences device operation.  These points are reviewed below.

The total active area of the interferometer is smaller than the lithographic area shown in Figure~\ref{fig:S1-1}.  With charging of the gates that define the area, gates $V_b$ and $V_s$, the lateral potential from the edges of those gates will deplete the 2D electron system away from the edges and toward the center of the area marked $A$.  The depletion from these edges extends inward dependent upon the depth of the 2D system and the magnitude of the depletion voltage applied to the gates. This depletion away from the gate edges and into area $A$ results in an active area much smaller than the lithographic area.  The active area is the area with the edge current of the charge corresponding to the bulk filling factor, which is both the bulk filling factor outside the interferometer and what should be the same filling factor through the center point of the constrictions and the center of the area $A$. Note that a series of edge currents will populate the space from the inner most edge, the bulk edge current, to the fully depleted edge, and these currents will correspond to the states between the bulk filling factor and zero filling factor.  Proper operation of the interferometer produces reflection only of the inner most edge currents.

The interference devices employed in this study have lithographically defined borders ranging from $2\upmu\text{m}\times 2\upmu\text{m}$ to $6\upmu\text{m}\times 6\upmu\text{m}$. The depth of the 2D electron system in all devices is approximately 200nm from the surface. The applied gate voltages are between $-2$ and $-10$ Volts; $-2$ Volts accomplishes full depletion in a sample of density $4\times 10^{11}\text{cm}^{-2}$.  These voltages and dimensions determine that the active areas are on the order of or less than one square micron.

The backscattering top gate dimensions are important in interferometer operation, both the lithographic separation marked $d_g$ in Figure~\ref{fig:S1-1}, and the width of the gates. The width of the gate defines the path length over which the edge currents are brought into proximity, and in conjunction with the separation $d_g$ define the potential tunneling paths for the edge currents.   In QH interferometry at $\nu=5/2$, it is crucial to ensure that the state extends contiguously through the constrictions serving as backscattering ``mirrors'' and through the active area. The backscattering gate width and separation $d_g$ will determine this transparency. The voltage applied to these gates nominally must completely deplete the underlying 2D electrons, and the separation dimensions must accommodate this condition but still facilitate the $\nu=5/2$ state transparency.  It is important to reiterate that the backscattering gates are not operated as quantum point contacts: the region within the constriction sustains the gapped QH state and is not depleted to the point that transport is dominated by charging energy inside the active area. In practice, a large separation $d_g$ can be used and compensated by a large range in applied gate voltage: typical $d_g$ is $1\upmu\text{m}$, with $-4$ Volts applied.

Finally, the crystallographic orientation of the heterostructure wafer and the interferometer direction from backscattering to backscattering gates must be considered.  Reentrant quantum Hall states surround the $\nu=5/2$ states and suppressing these improves interferometry for that state.  The reentrant phase can be affected by aligning the $(1\overline{1}0)$ orientation with the backscatter to backscatter gate direction.

\subsection{Methods: Measurements}
\label{sec:S3}

\subsubsection{Data acquisition: Signal measurement and signal size}
\label{sec:S3a}

The measurements are of resistance either across interferometer devices or bulk 2D systems (longitudinal $R_L$ or $R_{xx}$, or Hall $R_D$ or $R_{xy}$ respectfully) using four terminals on the sample. They are constant current measurements where a fixed AC voltage drives current through a load resistor (typically $100\,\text{M}\Upomega$): the resultant currents used are 2nA or less. The relatively large separation ($d_g$) of the interferometer constrictions employed here allows larger currents to be applied as heating is not as big a threat as would be the case for a true quantum point contact. The AC voltage driving the current provides a narrow frequency band excitation to which lock-ins are used to sense the voltage drops across two terminal sets on the sample. Low frequency excitations are used:  typically from 7Hz to less than 100Hz.  Gate voltages are applied using low noise voltage sources that are further filtered, and any leakage current is assessed during measurement.  Gate failure through leakage into the 2D electron system is not a subtle finding.

Signal size of the interference resistances at $\nu=5/2$ has increased with the quality of heterostructure materials used and has posed challenges in early measurements.  The overall resistance value of a longitudinal measurement around the $\nu=5/2$ complex is around $2\,\text{k}\Upomega$, but the interference oscillations are much smaller:  in less optimal material, the $f_0$ oscillations are around $10\,\Upomega$, and the non-Abelian oscillations ($4f_0$ \& $6f_0$ or $5f_0$ for $\nu=5/2$) are roughly $2\,\Upomega$.  In optimal to date material, the amplitude of oscillations at $f_0$ is $10$ to $30\,\Upomega$, but more importantly the amplitude of oscillations attributable to the non-Abelian even-odd effect can be $20$ to $50\,\Upomega$ (see Figures~\ref{fig:Hall_trace}(c), \ref{fig:fermionic_parity} and \ref{fig:S5-2-2}), a factor of ten improvement.

Note the noise levels in the majority of the measurements shown in this study are conservatively at about $0.5\,\Upomega$.   In the low amplitude version, non-Abelian amplitude of $\sim 2\,\Upomega$, signal filtering would be typically as follows:  a lock-in sensitivity of 5 to $10\,\upmu\text{V}$ would use a lock-in time constant of 1, 3, or 10 seconds, with a magnetic field sweep rate of less than 20 Gauss/min.  This would provide some latitude for digital data averaging, but this averaging is not typically exercised – the data presented in this study are with no digital averaging or no averaging using multiple different $B$-field sweeps unless explicitly stated. The magnetic field sweep data is presented as single sweeps that are representative of sets of repeated sweeps: averaging different sweeps offers the risk that unappreciated data could be lost in the presentation. The better heterostructure material interferometers displaying non-Abelian oscillation signals $20$ to $30\,\Upomega$ still have noise levels of $\sim 0.5\,\Upomega$, and because of this the measurement time constants have been dramatically reduced; time constants of 0.3 seconds are used with magnetic field sweep rates at maximum 100~Gauss/min but typically much less. Again, single sweep data is shown in these cases as well.

\subsubsection{Data analysis: Fast Fourier transform spectra of resistance traces}
\label{sec:S3b}

Data analysis for the examination of power spectra follows the simple steps of selecting the proper range of this data around $\nu=5/2$ or 7/2 to examine, subtracting the background from $R_L$ or $R_D$ data, and finally performing the fast Fourier transform.

Determining the proper range of data around the target filling factor, $\nu=5/2$, 7/2 or other fractional or integer filling involves examining the raw $R_L$ or $R_D$ data and establishing the high and low magnetic field limits to the correlated states.  In $R_L$ data this is usually a simple case of establishing the magnetic field range of the overall minimum, although it can be difficult for $\nu=5/2$ and 7/2 due to the adjacent re-entrant phase-separated Hall states.  Examples are Figure~\ref{fig:S5-2-1}(a): the $R_L$ data there show only minor distinct transitions from the lateral re-entrant phases to the $\nu=5/2$ and 7/2 states.  In such cases multiple ranges are selected centered around the minimum, extending from small to larger magnetic field extents, and the transition points away from the fractional states can usually be detected by inspection; if not the most typical spectrum is used. In most circumstances however, a distinct change in slope of the resistance can be found, and these points typically delineate the boundaries of the $\nu=5/2$ or 7/2 state range used for the FFT spectrum.
The background is determined by either a polynomial fit to the $R_L$ or $R_D$ data or by using multiple point smoothing. Both are iterative processes.  For instance, in the smoothing method a guess is made about the lowest frequency oscillation present (often apparent in the $R_L$ or $R_D$ data), the number of points over which the smoothing is applied is a number that is larger than the number of points in that lowest frequency, and the smoothing and subsequent subtraction of the smoothed data from the original data are applied. Then another smoothing of a larger number of points is then applied to the original $R_L$ or $R_D$ data to see if lower frequency components are exposed beyond those present in the first subtraction. Several of these checks are applied to rule out missed spectral peaks.
The fast Fourier transform is applied to the  $R_L$ or $R_D$ data selected range that has its background subtracted.  The standard FFT applied uses a rectangular window method, and these resultant power spectra are usually checked against other window methods for consistency: Welch, Hanning, Hamming, and Blackman. Power spectra from single magnetic field traces are presented throughout this study, and single power spectra are also shown; no power spectra averaging is used unless explicitly stated.
The range of the magnetic field over which the FFT is performed can influence the details of the resulting FFT for small differences in window size, as illustrated below in Figure~\ref{fig:S5-4-4}.

\subsection{Methods: Interferometry}
\label{sec:S4}
\setcounter{figure}{0}

\subsubsection{Oscillations at $f_0$ and the interferometer}
\label{sec:S4a}
\begin{figure*}[hbt]
\centering
\includegraphics[width=0.9\textwidth]{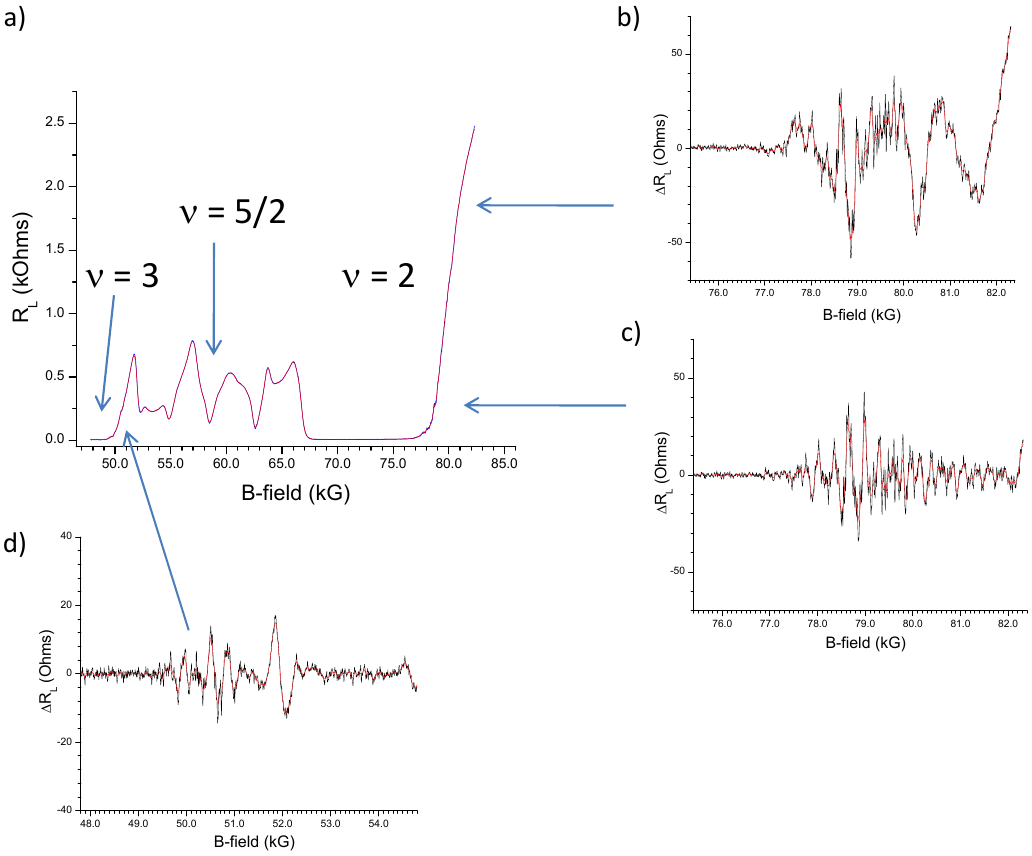}
\caption{(a) Resistance through an interferometer $R_L$ from filling factor $\nu=3$ to less than $2$.\\ Panel b) to d) $\Delta R_L$ over the respective $B$-fields of each plot.  Panel b) subtracts a moving average of $\sim 1.4$kG from  $R_L$, while panels c) and d) subtract a moving average of $\sim 0.3$kG.   Note the large period oscillations associated with the highest resistance displayed, and also note the similar small period oscillations on the high $B$-field sides of both filling factors $\nu=2$ and 3.  Temperature is 25mK, sample 7, preparation 2.}
\label{fig:S4-1}
\end{figure*}

Closer examination of the interference process at the borders of the integer filling factor can help to understand how charge traverses the interferometer. Figure~\ref{fig:S4-1}(a), shows $R_L$ transport through an interferometer from filling factor $\nu=3$ to below 2, and importantly along the high magnetic field side of $\nu=2$. Note that as $B$-field is increased to filling factor below the integer state at $\nu=2$, resistance increases well above that in the $\nu=3$ to $\nu=2$ complex around $\nu=5/2$.  This resistance increase indicates more and more backscattering from one side of the 2D electron system to the other counter current flow side: this is true even in the absence of the interferometer device, albeit at $B$-field values further from $\nu=2$ than shown here. The interference oscillations of the device can be used to demonstrate progression of the backscattering process as this high field side of $\nu=2$ $R_L$ increase occurs. Panels b) and c) show the $B$-field range that includes the resistance minimum at $\nu=2$ and at higher $B$: in both panels a background has been subtracted to show the interference oscillations.  In panel b) an averaging over about 1.4kG is subtracted; this is a large range subtraction that reveals just the longest wavelength oscillations of the  $R_L$ trace, and due to the large range has some systematic error over the $R_L$ section at which the resistance turns up abruptly (near 78 to 79 kG). However, this subtraction reveals large period and magnitude oscillations above 79 to 80kG.  Panel c) uses an averaging range of about 0.3kG which is subtracted, and this reveals a distinct single small period oscillation set with maximum amplitude at less than 79kG, and diminished oscillations at higher magnetic fields.  These small period oscillations are due to reflection – backscattering – at the constrictions ($V_b$) of the interferometer. This period oscillation is pervasive throughout the integer Hall states, with an example of that period oscillation shown in panel d) but from near filling factor $\nu=3$.  This periodicity is much smaller than that which is shown above 80kG in panel b; both are Aharonov--Bohm oscillations, but the small period is consistent with backscattering at the constrictions, and so consistent with an active area $A$ of the interferometer. The large period AB oscillations of panel b) show backscattering encircling a smaller area (by about a factor of 4 to 5). This indicates backscattering within the area $A$.  This is the same backscattering or edge to edge charge transport process observed in the bulk away from integer filling.

	One can consider that the backscattering gates $V_b$ facilitate charge transport from one side to the other of the device, and so the Hall system; different processes can mediate this transport.  This process can occur through tunneling through a gapped state system between the current edge states, or from percolative scattering through a filling factor background that is not gapped.   Both can produce AB oscillations.  The $V_b$ gates in both cases provide the highest probability position for the backscattering to occur.  Away from the quantized Hall states the backscattering can be promoted at particular paths not between the $V_b$ gates.   These paths can be due to anisotropies in the heterostructure materials, such as those that produce the asymmetric surface morphologies observed in these systems.  These ``rogue'' paths are a reason that the larger an interferometer can be made, the better to inhibit errant backscattering, limited clearly by the finite coherence length of the edge charges.
\onecolumngrid

\twocolumngrid

\clearpage

\subsubsection{Coulomb dominated versus Aharonov--Bohm effects}
\label{sec:S4b}

An important issue in quantum Hall interferometry is whether the observed oscillations are due to the Aharonov--Bohm (AB) effect or due to Coulomb domination: is the electron system keeping the same area as $B$-field is swept, so that quasiparticle charge injection is into a fixed area state, or is the area changing upon sweeps to accommodate the system energy altered by the changed magnetic field.

\begin{figure}[hbt]
\includegraphics[width=0.95\columnwidth]{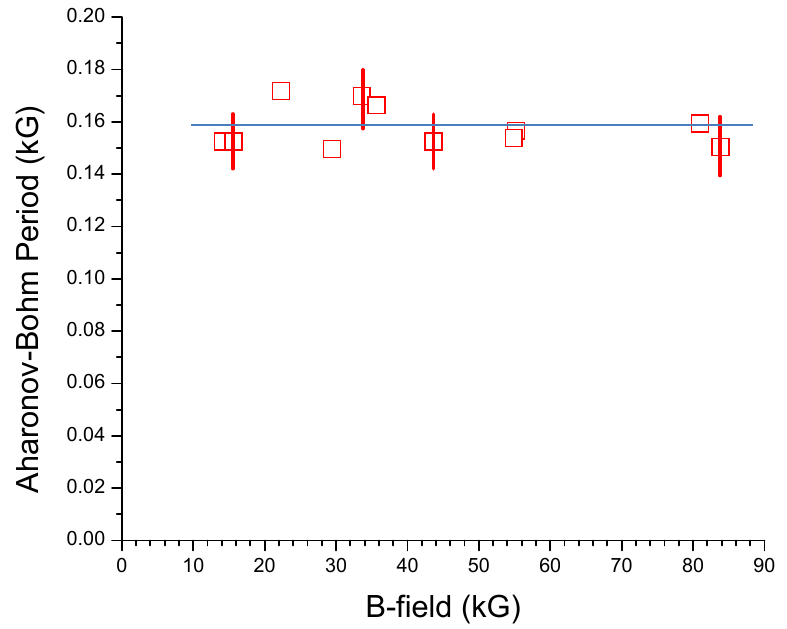}
\caption{Interference periods at integer filling factors from $\nu=2$ to $\nu=12$ (sample~6, preparation~3, $T\sim 25$mK).}
\label{fig:S4-2a}
\end{figure}

\begin{figure}[hbt]
\includegraphics[width=\columnwidth]{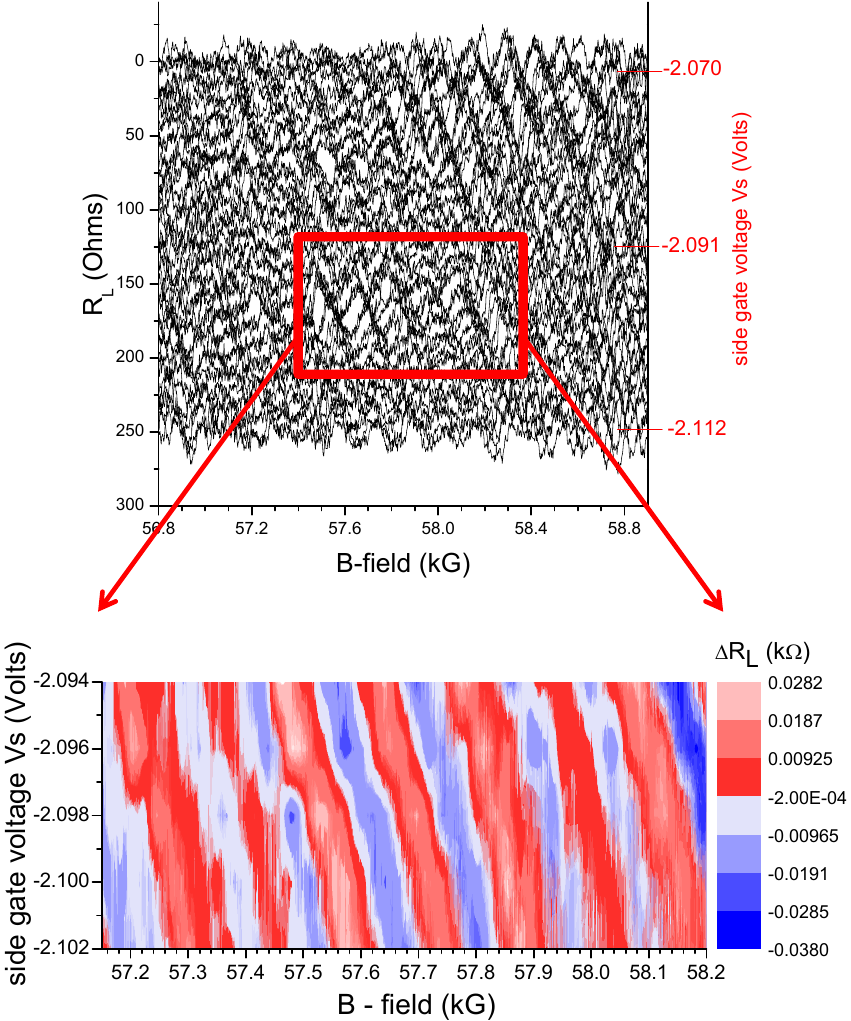}
\caption{Resistance oscillations at/around $\nu=3$ for $B$-field sweeps ($x$-axis) and for different side gate ($V_s$) voltages ($y$-axis): for each different voltage $V_s$, the $B$-sweep data is offset in the y-direction.  The magnetic field sweeps include up and down excursions, the $V_s$ range is shown on the right vertical axis, and the resistance scale is on the left. The top gate voltages ($V_b$ and $V_s$) are comparable to those used in the measurements shown in the data of the main text and Supplementary data.  $V_b = -3.0$V, $V_s$ as shown. The region corresponding to the pajama plot shown in the main text and reproduced here is framed in red. (Sample~6, preparation~3, $T\sim 25$mK.)}
\label{fig:S4-2b}
\end{figure}
	The interferometer devices used in this study are lithographically at least $2\upmu\text{m}\times 2\upmu\text{m}$ in area, and typically larger so that Coulomb domination is not likely.  However, the edge potentials from the gates produce substantial lateral depletion, resulting in small active areas, typically less than $1\upmu\text{m}^2$. In order to determine that the AB effect is at play, two measurements can be made at integer filling: a) in the AB effect, the period of oscillations for the different integer fillings (2, 3, 4, 5, \ldots) should be the same; in Coulomb dominated systems, the period changes linearly with magnetic field, and b) in the AB effect, maps of constant phase when examining plots of resistance $\Delta R_L$ versus magnetic field versus side gate voltage $V_s$ should show a negative slope; addition of area and of flux are the same.  Both of these conditions have been displayed for our sample sets.  We have measured resistance oscillation periods at integer filling factors from $\nu=2$ to 12; Figure~\ref{fig:S4-2a} shows the period plotted as a function of magnetic field.  The period does not change significantly over this large range of filling factors, The periods do not change significantly over this range, consistent with Aharonov--Bohm oscillations and in contrast to Coulomb dominated oscillations. Figure~\ref{fig:S4-2b} shows a plot of resistance oscillations at a single integer filling factor ($\nu=3$) as $B$-field and side gate voltage are changed; the $y$-axis is a cascade of the resistance over steps in side gate voltage and sweeps of the $B$-field.  This waterfall plot (or the pajama stripe plot shown in the same figure)  reveals a negative slope of the constant phase lines in the  $B - V_s$ plane; a negative slope of the constant phase lines is expected for Aharonov--Bohm oscillations, a positive slope for Coulomb effects. In different samples/devices used in this study, further consistency with AB rather than Coulomb effects is demonstrated: in Figure~\ref{fig:S4-1}, the periods for filling factors 2 and 3 are the same in the panels c) and d); a better representation of this fixed period is in reference~\cite{Willett2013a}.  The negative slope of constant phase plot for a shielded well sample as used in this study is also demonstrated in reference~\cite{Willett2013a}.

	The nature of the heterostructures used here and their preparation are crucial to understanding how they demonstrate AB oscillations and not Coulomb domination.  The structure of the shielded well sample, that which is used throughout this study, is such that it intrinsically inhibits charge accumulation needed for Coulomb domination.   As shown in Figure~\ref{fig:S2-1}, the shielding wells are grown on either side of the principal quantum well, and these wells collect some charge from the dopant layers, facilitated by illumination of the sample.  In the selection of the proper growth parameters for these heterostructures, dopant levels are adjusted to just eliminate parallel conduction observed in transport measurement over large areas.  This titration of the doping will certainly leave a nominal charge movement in the shielding well layers that will act to counter any charge accumulation in the principal well.  It is this screening or shielding mechanism that produces the ultra-high mobility of the heterostructure, and also the elimination of charge accumulation during $B$-field sweeps.  Both the shielded well structure and the process of illumination to populate the shielded wells are necessary to screen charge.

\subsubsection{Transport through the interferometer and maintenance of density and 2DES quality}
\label{sec:S4c}

A key element in the interference studies reported here is maintaining continuity of the target filling factor from the bulk on one side of the interferometer, across the active area of the interferometer, connecting without interruption to the bulk on the opposite side of the device.   A completely open geometry where inner-most counter-flowing edge currents on the opposing sides of the device are not in proximity is clearly possible, but proximity of these edges is necessary in an interferometer to achieve the backscattered interfering paths.   Low-probability backscattering at both constrictions can provide that necessary interference current and this backscattering should occur across an electron density contiguous with the bulk regions external to the interferometer and with the active area $A$ of the interferometer, and where the edge current of that bulk density is continuous across the device.   (An example of non-continuous edge currents can be found e.g. in the experiments of ref.~\cite{Sivan2018}.)

\begin{center}
\begin{figure}[htb]
%\centering
\includegraphics[width=0.85\textwidth]{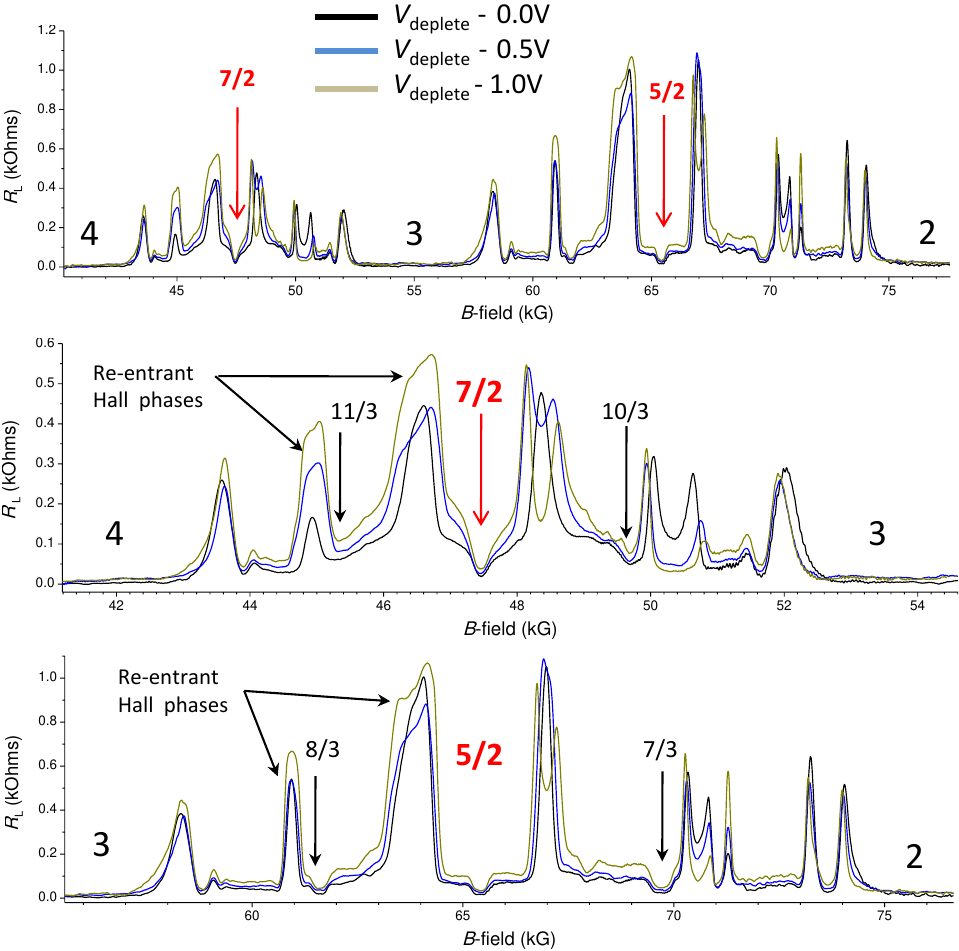}\\
\vspace{0.2cm}
\includegraphics[width=0.9\textwidth]{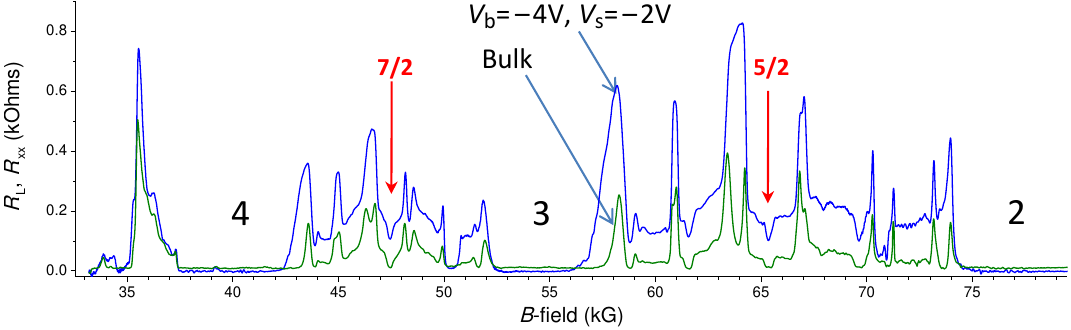}
\caption{Transport through interferometers using nominal $V_b$.\\
Top panel: $R_L$ through the interferometer with no extra voltage beyond nominal depletion,  $-0.5$V in addition to nominal depletion, and $-1.0$V in addition to nominal depletion, providing a picture of device transport in its most ``open'' $V_b$ configuration and near that.  The second panel is a blow-up of the filling factor range between $\nu=3$ and $\nu=4$. Note that the re-entrant phases change substantially but the minimum at $\nu=7/2$ does not.  Third panel is a blow-up of the filling factor range between $\nu=2$ and $\nu=3$. (Sample 12, preparations 35-37, $T=25$mK.)
Bottom panel compares $R_L$ at $V_b=-4$V, $_s=-2$V to $R_{xx}$ measured in the bulk, that is not across the device but using  an adjacent contact set not encompassing the device. Note as with the top two panels the center of each integer filling is unchanged, the $B$-field positions of the fractions 7/2, 5/2 are not changed, but increased backscattering is present over the non-integer quantum Hall states.   In all panels the sets of fractional quantum Hall states are preserved and not shifted in magnetic field as would have been expected if a density shift had occured.  (Sample 12, preparation  38, $T=25$mK.)
}
\label{fig:S4-5}
\end{figure}
\end{center}

We test the edge continuity by examining the longitudinal transport $R_L$ across the device, in particular over a large range of magnetic field that includes multiple filling factors. Such data for our principal wafer material and device type is shown in Figure~\ref{fig:S4-5}. In the top panels the interference gate voltages are applied that just shut down errant current paths under the gates: note again that these heterostructures use illumination before gating so a reduced density results under the gates due to shadowing.  Displayed is a trace with that gate voltage needed to stop the errant paths, and two other traces in which $-0.5$V and $-1.0$V are also applied.  Note that the $\nu=7/2$ and $5/2$ minima are preserved as the backscattering gate voltage is increased, while the overall background between the integer filling factors increases.  This includes changes in the re-entrant phase peaks. Notably the magnetic field positions of the fractional Hall states do not change.

This is further displayed in the bottom trace which directly compares standard device gate voltages used to generate measurable interference oscillations with a bulk transport measurement.   As in the small gate value data of the top traces, no substantial shift in the magnetic field position of the fractional quantum Hall states is apparent, differences in the re-entrant phase peaks are observed, and the overall background now changes even more but the fractional states are preserved.

\begin{figure}[t]
\centering
\includegraphics[width=\columnwidth]{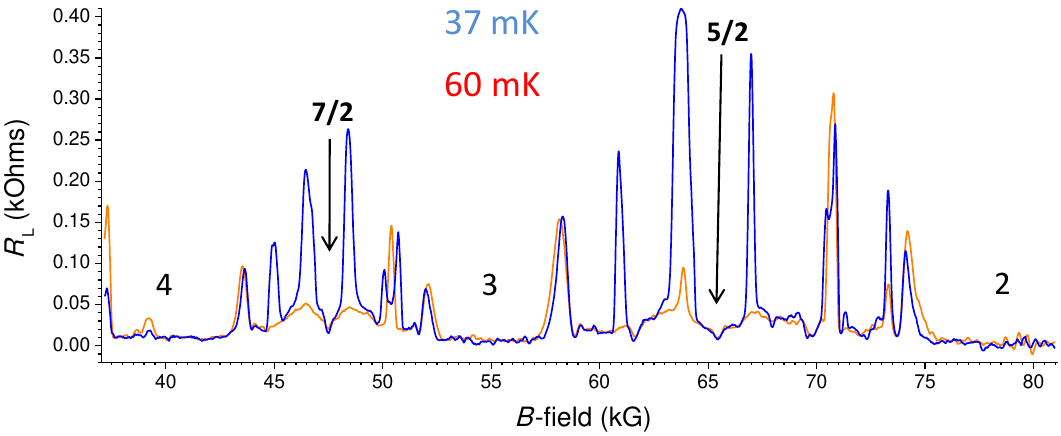}\\
\vspace{0.2cm}
\includegraphics[width=0.7\columnwidth]{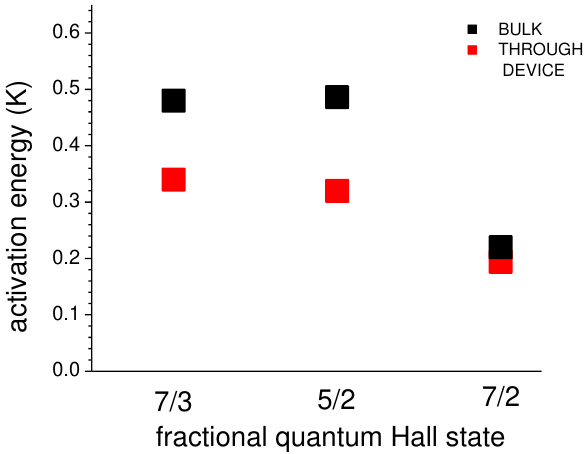}
\caption{Temperature dependence of coarse transport through an interferometer and energy gaps.\\
Top panel shows $R_L$ at two temperatures, $37$ and $60$mK extending over filling factor range from $\nu=2$ to $\nu=4+$.  Note the diminished re-entrant phase peaks at $60$mK and the nominal change in $\nu=5/2$ and $\nu=7/2$ minima.\\
Bottom panel is measured energy gaps for $\nu=7/3$, $7/2$ and $5/2$ as measured with transport in only bulk (not through interferometer device) or through the interferometer.  These data points are derived from Arrhenius plots using the top panel data plus traces at multiple higher temperatures.  (Sample 12, preparation 34.)
}
\label{fig:S4-6}
\end{figure}

The temperature dependence of transport through the device is shown in Figure~\ref{fig:S4-6}. Here two traces at nominal top gate voltages are taken at two different low temperatures, $37$mK and $60$mK.  Note that in this range the re-entrant peaks are diminished for the $60$mK trace as is commonly seen but little difference in the 7/2 and 5/2 minimum is apparent.  Higher temperatures were applied to this preparation and activation energies $\Delta$ were derived from the Arrhenius plots ($R\propto e^{-\Delta/2kT}$) with results displayed in the lower panel of Figure~\ref{fig:S4-6}.   The energy gaps are large and attest to the preserved quality of the material even with the device confinement and processing.

\onecolumngrid
\twocolumngrid

\clearpage
%\onecolumngrid
%\twocolumngrid

%\clearpage

%Section 5  D
\subsection{Data: Supplements to individual figures in the main text}
\label{sec:S5}
\setcounter{figure}{0}
\setcounter{table}{0}

\subsubsection{Supplements to Figure~\ref{fig:Hall_trace}}
\label{sec:S5-2}

Figure~\ref{fig:Hall_trace} of the main text presents the longitudinal resistance trace and its blow-up near $\nu=7/2$. The latter demonstrates remarkably stable, large amplitude interference oscillations at a frequency seven times that of AB interference frequency observed at integer filling fractions, consistent with the non-Abelian even-odd effect.  This new result is a consequence of greatly improved heterostructure quality.  The Supplementary data presented in this section serve to (i) show the transport property improvement with these new heterostructures and how an important parameter, the electron AB interference, is measured,  (ii) provide more detail on the large, stable interference oscillations observed at $\nu=7/2$.

\begin{center}
\begin{figure}[htb]
%\centering
\includegraphics[width=0.8\textwidth]{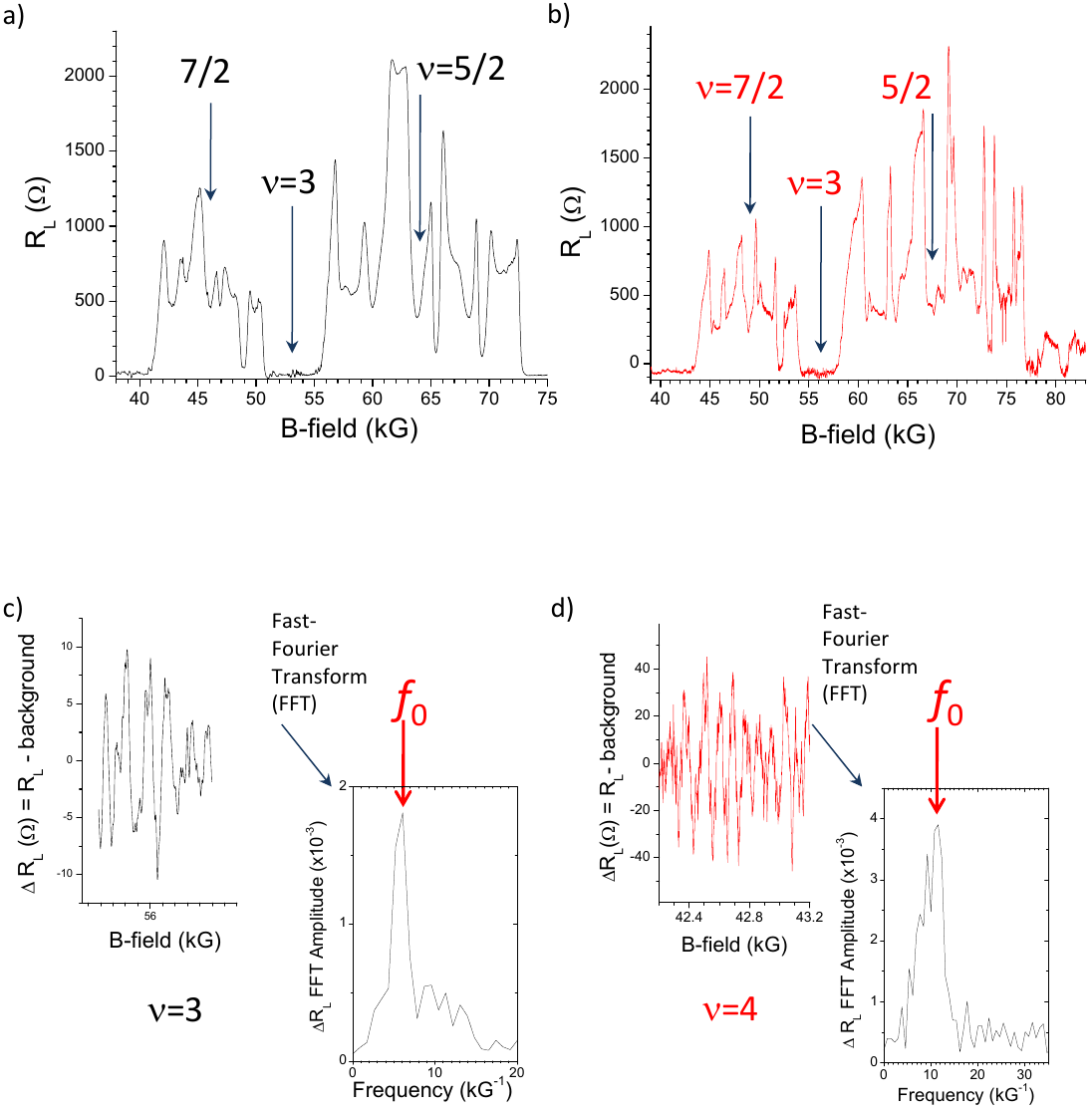}
\caption{Transport through two different devices on the two different wafer types used in this study, with and without aluminum purification, and  determination of $f_0$.\\
(a) Transport $R_L$ measured across device/heterostructure with narrower backscattering aperture and no additional Al purification; these are the population of devices/heterostructures used in previous studies (references~\protect\cite{Willett2007,Willett2009a,Willett2010a,Willett2013a,Willett2013b});\\
(b) Transport in the newer device with larger aperture using a heterostructure with highly purified Al.  Note the sharper large resistance features, consistent with narrower definition of the re-entrant quantum Hall phases. This smaller $B$-field extent of those mixed integer Hall re-entrant states allows larger range of expression for both the $\nu=5/2$ and 7/2 states when compared to previously used heterostructures/devices shown in panel (a).\\
(c) \& (d). Extraction of the electron interference periodicity at integer filling for both device/heterostructure types.  Background resistance determined by a moving average of data points is subtracted from the original $R_L$ trace revealing oscillations at integer filling, and FFTs of these oscillations sets produce power spectra with peaks marked $f_0$ corresponding to the frequency of the Aharonov--Bohm oscillations of the electrons in each. Note that these are two different samples from two different wafers, which results in two different reference frequencies. The measurement shown in panel (d) provides the  $f_0$ corresponding to the oscillation set in Figure~\ref{fig:7_halves_interf} of the main text.\\
%, which has as the principal oscillation frequency $7f_0$.\\
Panels (a) and (c): sample 2, preparation 2. Panels (b) and (d): sample 6, preparation 8. $T=20$mK.}
\label{fig:S5-2-1}
\end{figure}
\end{center}

Figure~\ref{fig:S5-2-1} shows the transport in the two different major heterostructure/device types of this study.
Presented here is the comparison of transport through interferometer devices with large backscattering apertures using newer heterostructures with highly purified Al (figure~\ref{fig:S5-2-1}(b)) versus transport through devices with smaller backscattering apertures using heterostructures without the additional Al purification (figure~\ref{fig:S5-2-1}(a)); see references~\cite{Chung2018,Chung2018a} for Al purification. Newer devices are characterized by
sharper large resistance features associated with the re-entrant compressible quantum Hall states. Their smaller $B$-field extent provides more room for both the $\nu=5/2$ and 7/2 states when compared to previously used heterostructures/devices shown in panel (a).
However, these newer heterostructures can also demonstrate small density shifts at large magnetic fields – note the $B$-field range near filling factor $\nu=2$. These shifts may be due to the parallel shielded wells also employed in these devices, but unfortunately hinder direct comparison of interference oscillations at $\nu=5/2$ to those at $\nu=7/2$. This effect is currently under study.

Figures~\ref{fig:S5-2-1}(c,d) present FFT spectra of integer-filling oscillations for both devices, exemplifying the process of FFT determination of $f_0$, the electron AB frequency.

\begin{figure}[htb]
\includegraphics[width=0.6\textwidth]{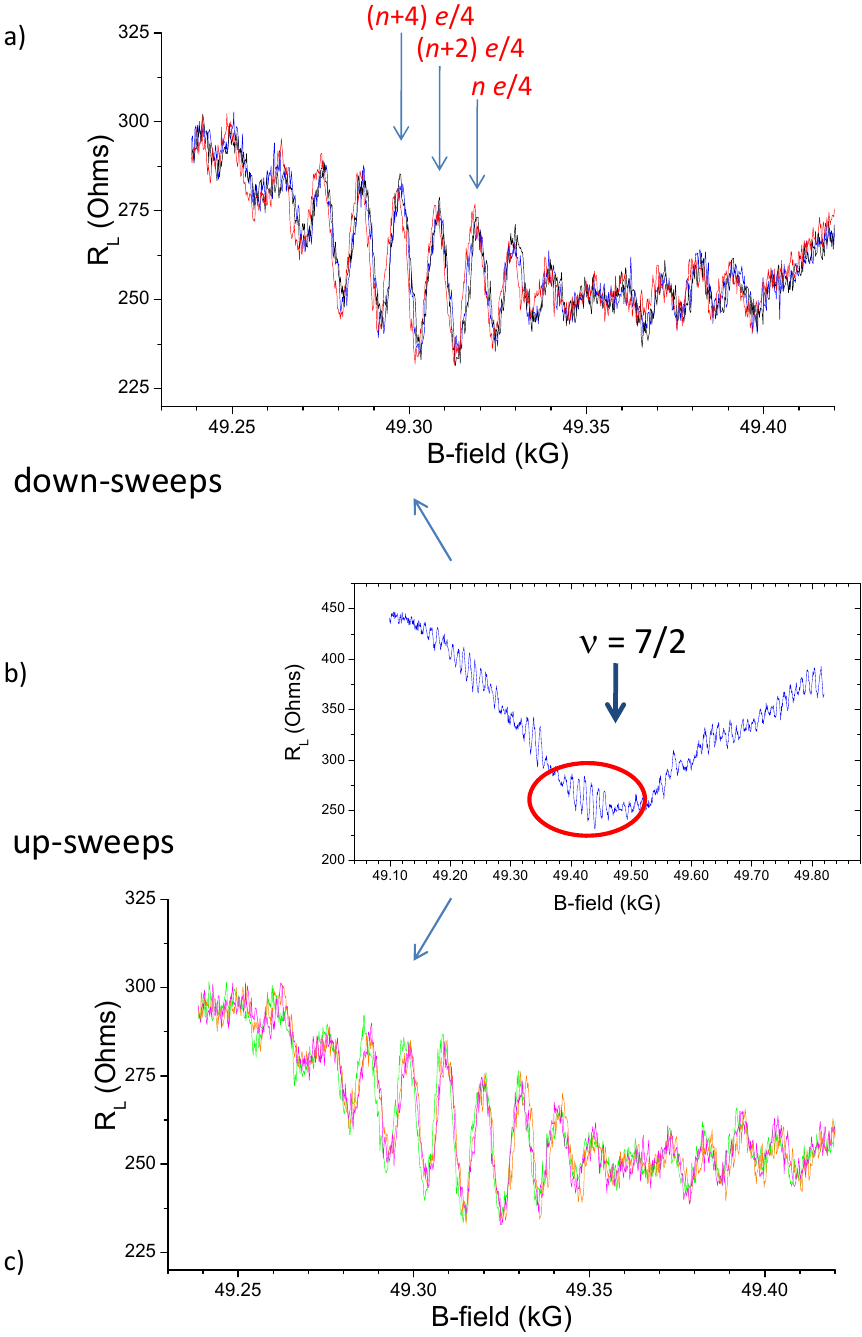}
\caption{Stability of interference oscillations at $\nu=7/2$.\\
(a) \& (c)  Top and bottom trace sets show interference oscillations in $R_L$ near $\nu=7/2$ for up and down magnetic sweeps; these sweeps together form Figure~\ref{fig:Hall_trace}(b).  The position of these oscillations in the 7/2 minimum is shown panel (b).   The two sets together are a continuous series of six sweeps (three in each displayed set).  Note the stability of the sweeps. A minor $B$-field hysteresis is apparent when comparing the data of Figure~\ref{fig:Hall_trace}(b) and these traces of isolated up and down sweeps.
The oscillation period corresponds to that of the $7f_0$ spectral weight shown in Figure~\ref{fig:7_halves_interf}, with the $f_0$ value determined above in Figure~\ref{fig:S5-2-1}(d). The spectral weight at $7f_0$ strongly points towards the non-Abelian even-odd effect due to charge-$e/4$ quasiparticles.
The set of six sweeps shown covers a time of 8.75 hours, 87.5 minutes for each directional sweep. The data demonstrate both a high level of stability for the time and magnetic field ranges over which the data were taken.
(Sample 6, preparation 16, $T=20$mK.)
}
\label{fig:S5-2-2}
\end{figure}

\begin{figure}[htb]
\includegraphics[width=\columnwidth]{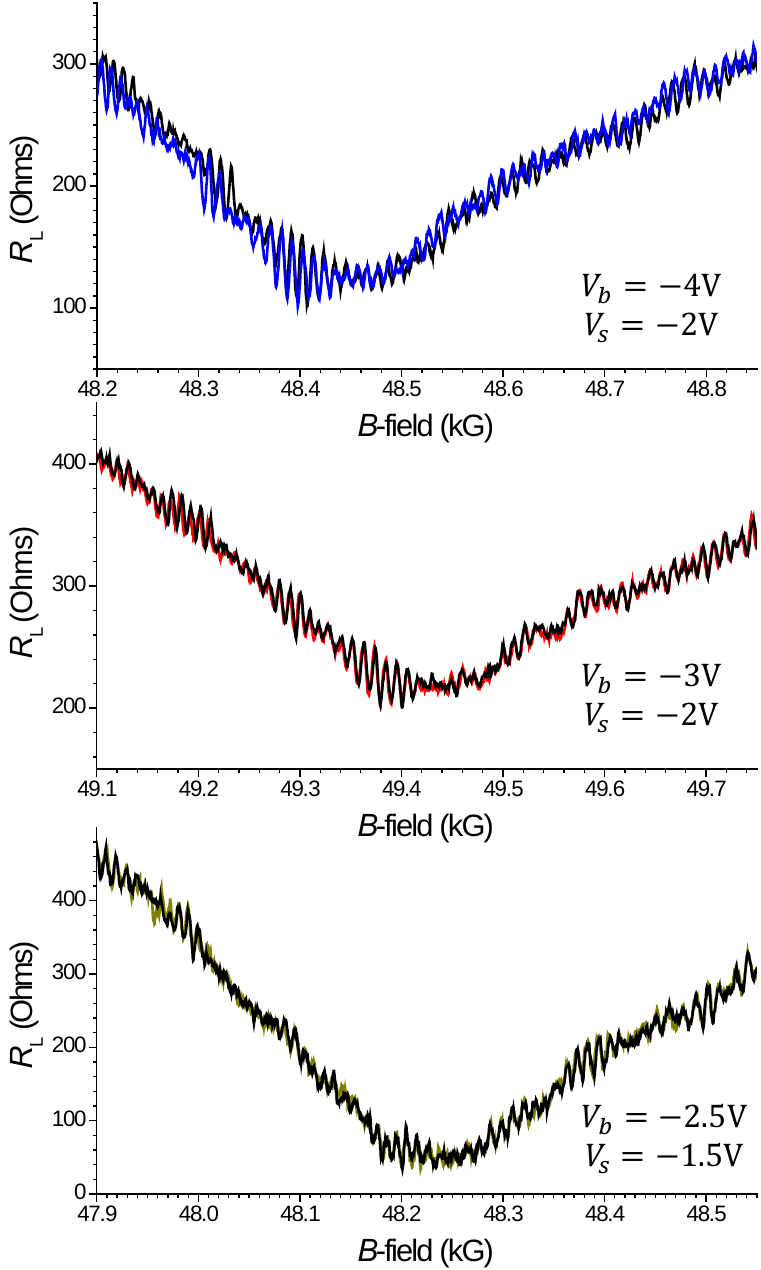}
\caption{Additional sets of interference oscillations at $\nu=7/2$ obtained for different preparations and with different voltage settings of both $V_s$ and $V_b$. Notice different values of the magnetic field corresponding to the $\nu=7/2$ resistance minima resulting from the density change in response to the different illumination histories and gating.  Top traces are from sample 6, preparation 18, middle traces are from sample 6, preparation 17, bottom traces are from sample 6, preparation 27 (see section~\ref{sec:S2c} for specific details); $T=20$mK in all cases.
}
\label{fig:add_osc1}
\end{figure}

\begin{figure}[htb]
\includegraphics[width=0.9\columnwidth]{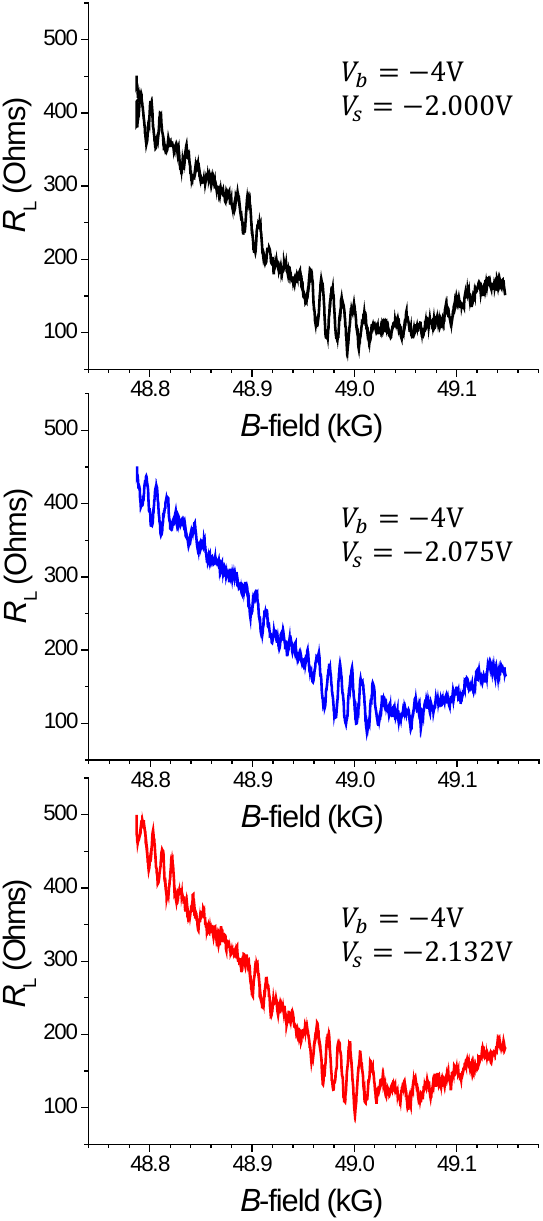}
\caption{Additional sets of interference oscillations at $\nu=7/2$ for different side gate voltage settings $V_s$, all with $V_b=-4.0$V. (Sample 6, preparation 21, $T=20$mK.)
}
\label{fig:add_osc2}
\end{figure}

Figure~\ref{fig:S5-2-2} breaks down the oscillation data of Figure~\ref{fig:Hall_trace}(b).
Figures~\ref{fig:add_osc1} and \ref{fig:add_osc2} demonstrate the generic nature of oscillations at $\nu=7/2$, which are observed over substantial gate voltage ranges and in multiple sample preparations. Fine tuning is not needed to achieve these prominent effects.

\onecolumngrid
% \twocolumngrid

\twocolumngrid

%\pagebreak
\clearpage

\onecolumngrid
\subsubsection{Supplements to Figure~\ref{fig:seventhirds}}
\label{sec:sipplement_fig6}

\begin{figure}[!h]
\begin{minipage}[c][\linewidth][t]{\linewidth}\centering
\includegraphics[width=\columnwidth]{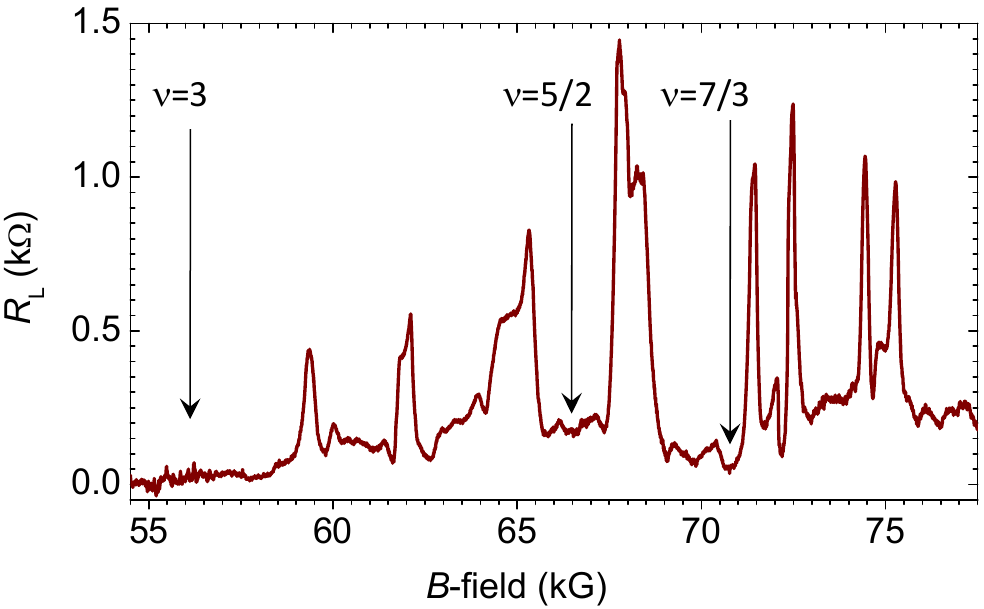}
\caption{Plot of the longitudinal resistance as a function of the magnetic field that shows the minima corresponding to QH states at $\nu = 7/3$, $5/2$ and $3$. Zooming into these minima and performing a Fourier transform of resistance oscillations in their vicinity results in the oscillations spectra shown in Figure~\ref{fig:seventhirds}. Sample~6, preparation~36, $T\sim 25$mK.
}
\label{fig:add_QH_trace}
\end{minipage}
\end{figure}

\begin{figure}[!b]
\begin{minipage}[c][\linewidth][t]{\linewidth}\centering
\includegraphics[width=0.95\columnwidth]{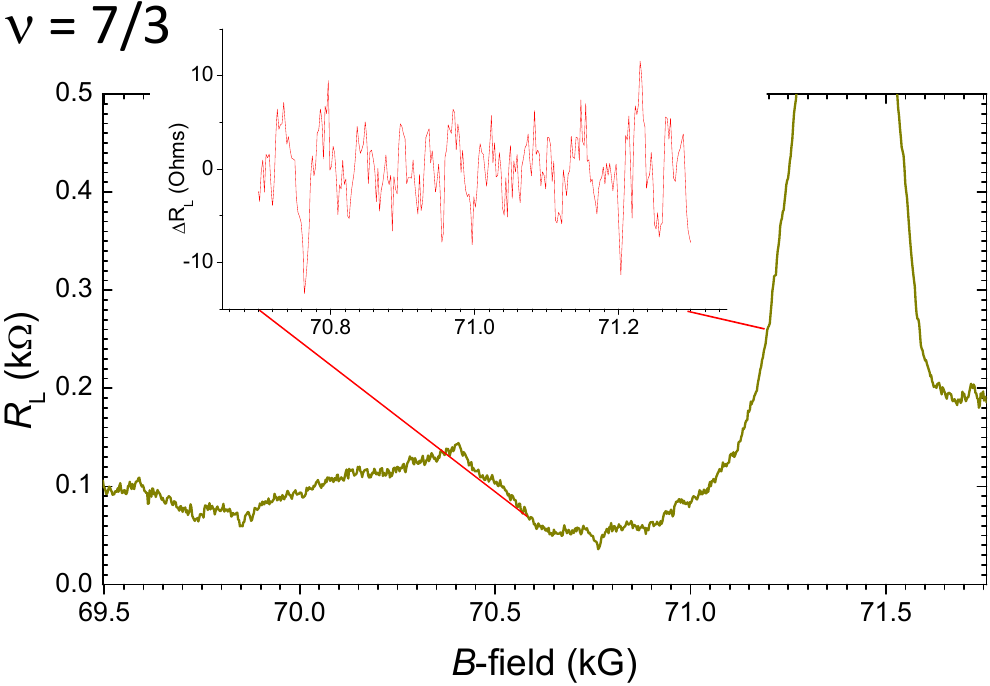}
\caption{Zooming into resistance oscillations in the vicinity of ${\nu = 7/3}$. The inset shows the same oscillations within the actual Fourier transform window used in obtaining the spectrum in Figure~\ref{fig:seventhirds} (after background subtraction). Sample~6, preparation~36, $T\sim 25$mK.
}
\label{fig:add_seven_thirds}
\end{minipage}
\end{figure}

\begin{figure}[!h]
\begin{minipage}[c][\linewidth][t]{\linewidth}\centering
\includegraphics[width=0.95\columnwidth]{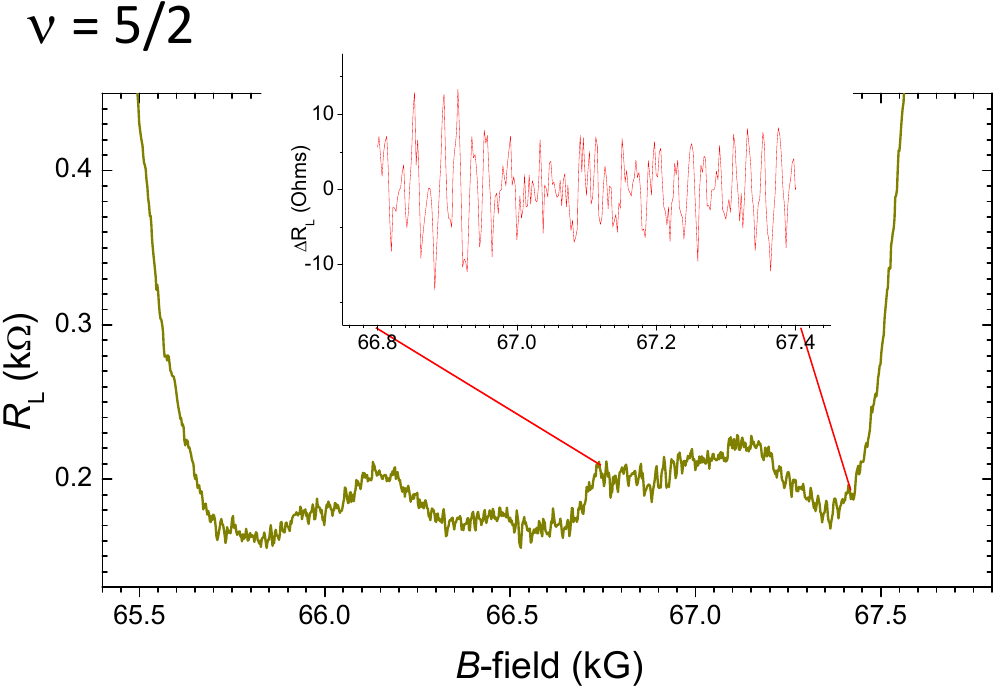}
\caption{Zooming into resistance oscillations in the vicinity of ${\nu = 5/2}$. The inset shows the same oscillations within the actual Fourier transform window used in obtaining the spectrum in Figure~\ref{fig:seventhirds} (after background subtraction). Sample~6, preparation~36, $T\sim 25$mK.
}
\label{fig:add_five_halves}
\end{minipage}
\end{figure}

\begin{figure}[!b]
\begin{minipage}[c][\linewidth][t]{\linewidth}\centering
\includegraphics[width=0.95\columnwidth]{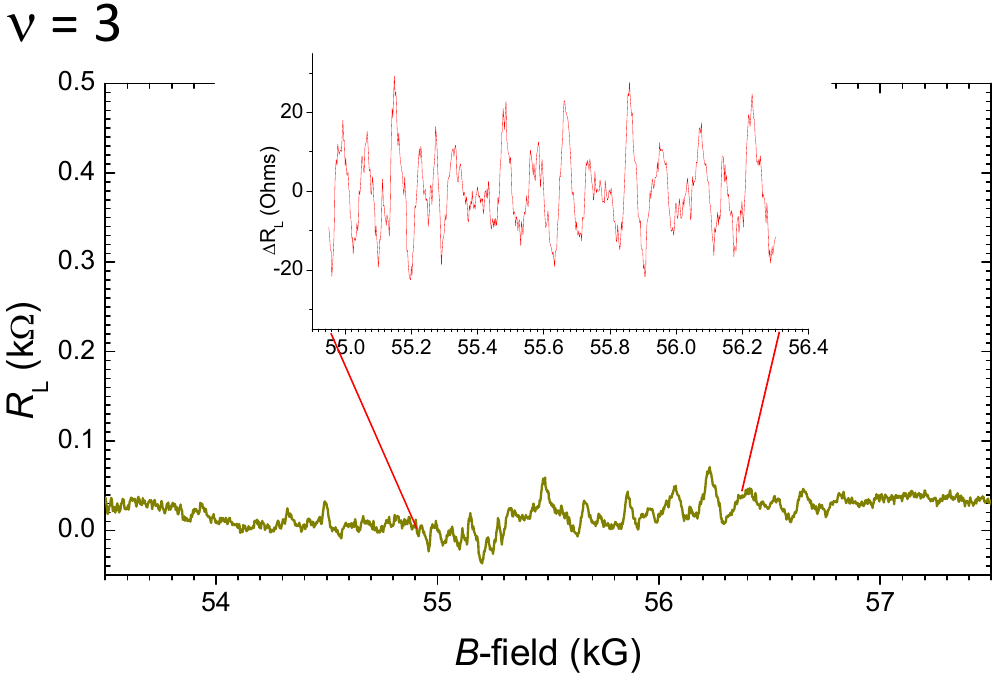}
\caption{Zooming into resistance oscillations in the vicinity of ${\nu = 3}$. The inset shows the same oscillations within the actual Fourier transform window used in obtaining the spectrum in Figure~\ref{fig:seventhirds} (after background subtraction). Sample~6, preparation~36, $T\sim 25$mK.
}
\label{fig:add_nu_three}
\end{minipage}
\end{figure}

In this section we present the actual transport data pertaining to the oscillation spectra shown in  Figure~\ref{fig:seventhirds}. Figure~\ref{fig:add_QH_trace} shows the overall Hall trace showing minima at $\nu = 7/3$, $5/2$ and $3$. Figures~\ref{fig:add_seven_thirds}, \ref{fig:add_five_halves} and\ref{fig:add_nu_three} show blow-ups of resistance oscillations near these minima. The spectra shown in Figure~\ref{fig:seventhirds} are a result of Fourier transforming these oscillations (after background subtraction -- see the main text).

\vspace{1cm}
\twocolumngrid

\vspace{3.8cm}

%\pagebreak
\clearpage
\twocolumngrid

\subsubsection{Supplements to Figure~\ref{fig:7_halves_interf}}
\label{sec:S5-4}

Figure~\ref{fig:7_halves_interf} of the main text demonstrates the frequency spectrum for the interference oscillations observed in the vicinity of the $R_L$ minimum at $\nu=7/2$, with the observed spectral features corresponding to $f_0$, $1.5f_0$, $3f_0$, and $(7\pm 1.5)f_0$. As described in the main text, these peak values are consistent with the frequency values of the expected interference of non-Abelian $e/4$ and Abelian $e/2$ quasiparticles at $\nu=7/2$. In this section we supplement the results of Figure~\ref{fig:7_halves_interf} with similar data for other samples and preparations.

\begin{figure}[htb]
\includegraphics[width=0.6\textwidth]{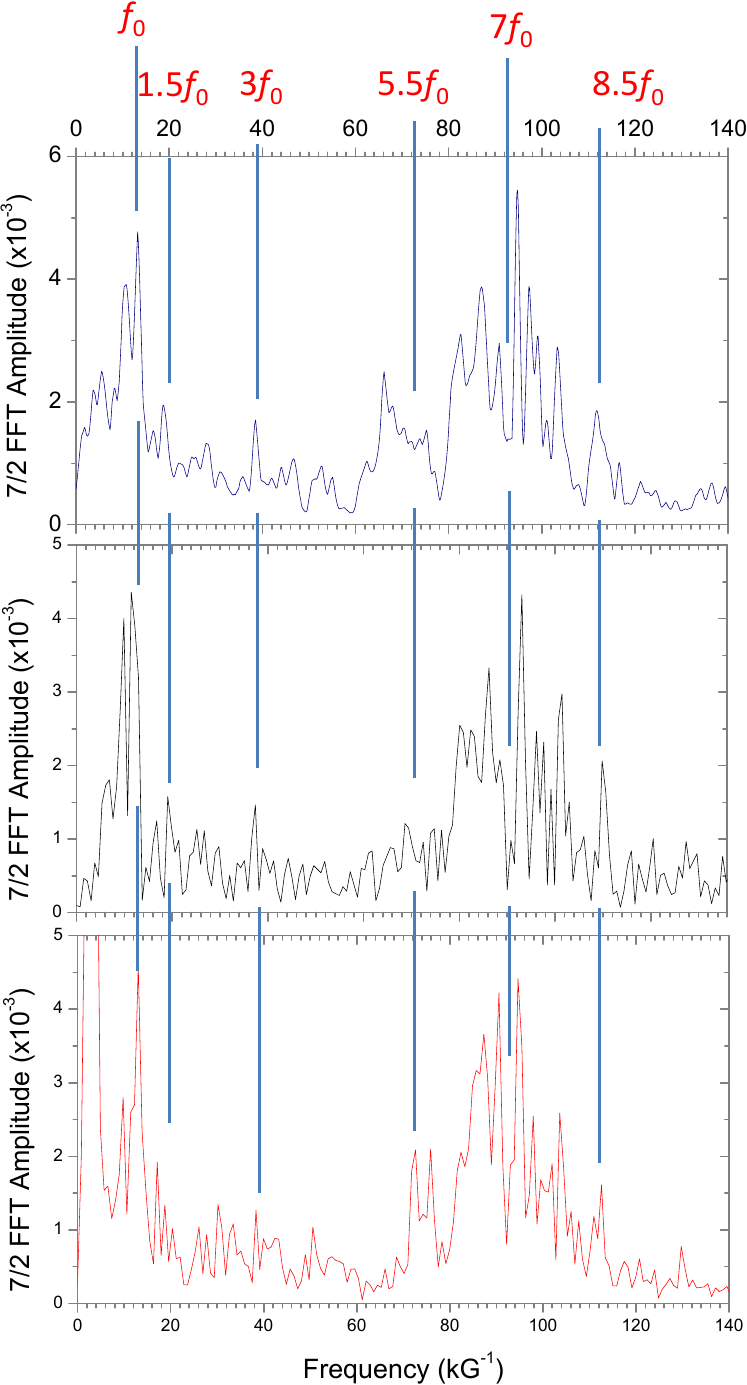}
\caption{The three panels show oscillation power spectra at $\nu=7/2$ from three different sample preparations, same sample. The vertical blue lines mark the frequencies $f_0$, $1.5f_0$, $3f_0$, $5.5f_0$, $7f_0$ and $8.5f_0$. Note that the horizontal axes have been slightly adjusted so that the peak corresponding to the reference frequency $f_0$ is aligned for all three plots.  Sample 6, preparation~16 (top), the average of preparations~6--8 (middle), and preparation~10 (bottom); $T=20mK$.}
\label{fig:S5-4-1}
\end{figure}

Figure~\ref{fig:S5-4-1} demonstrates common features of three $\nu=7/2$ power spectra. In all three the spectral peak at $f_0$ frequency is prominent but the most important dominant feature is the complex centered at $7f_0$. Within these dominant complexes, three specific features are indicated: minima at $7f_0$ and side peaks at $5.5 f_0$ and $8.5f_0$ (i.e. $(7\pm 1.5) f_0$).  Irrespective of the details, the complex centered at $7f_0$ is strongly indicative of the non-Abelian even-odd effect due to charge-$e/4$ qusiparticles; no Abelian interference can result at such high-frequency oscillations. The minimum at $7f_0$ occurs due to a small frequency modulation of this process.  Empirically the source of this modulation is the minor peak at roughly $0.4f_0$ present in all three traces; this feature is consistent with the overall $R_L$ minimum formation at $\nu=7/2$, a large period property due to the e/4 population -- see Figures~\ref{fig:7_halves_interf} and \ref{fig:S5-5-2}.

The spectral complex centered at $7f_0$ may also display side peaks at $(7\pm 1.5) f_0$, which, to different degrees, may be discerned in all three panels.  These peaks can be attributed to modulation of the even-odd effect at $7f_0$ by the interference of $e/4$ edge quaiparticles around $e/2$ bulk quasiparticles.

\begin{figure}[htb]
\includegraphics[width=0.67\textwidth]{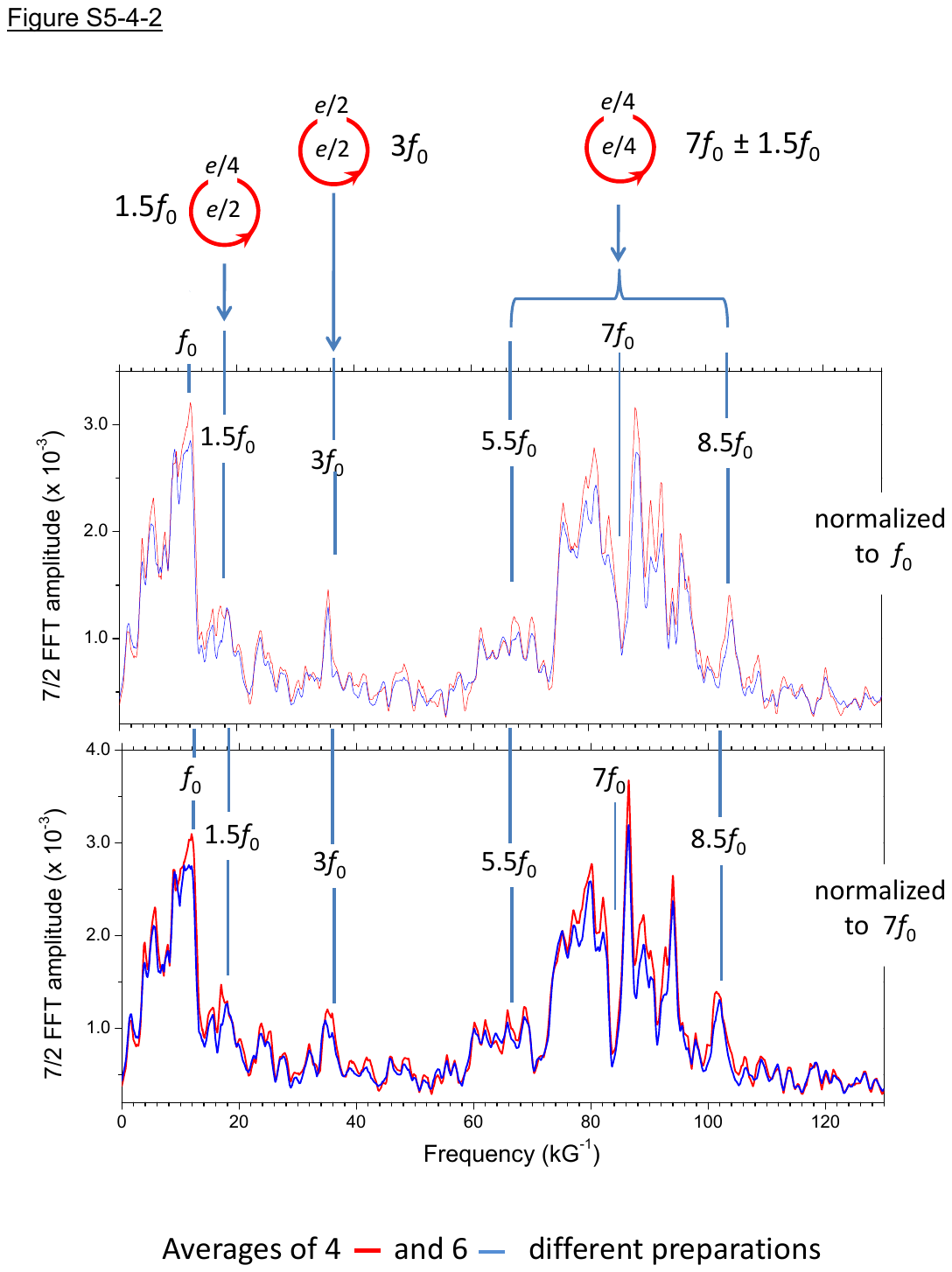}
\caption{Averages of spectra from four (red) and six (blue) different sample preparations at $\nu=7/2$ normalized to two different frequency points. The different spectra are normalized in frequency to have either the same $f_0$ positions (top panel) or the same $7f_0$ position (bottom panel).   Note the measured peaks in agreement with frequencies marked  $f_0$, $1.5f_0$, $3f_0$, $5.5f_0$ and $8.5f_0$, and the minimum marked at $7f_0$.  Sample 6, averages of preparations 8, 10, 13-16; $T=20mK$.}
\label{fig:S5-4-2}
\end{figure}

Figure~\ref{fig:S5-4-2} displays two sets of oscillation power spectra averages at $\nu=7/2$, each normalized to $f_0$ or $7f_0$.

\begin{figure}[htb]
\includegraphics[width=0.67\textwidth]{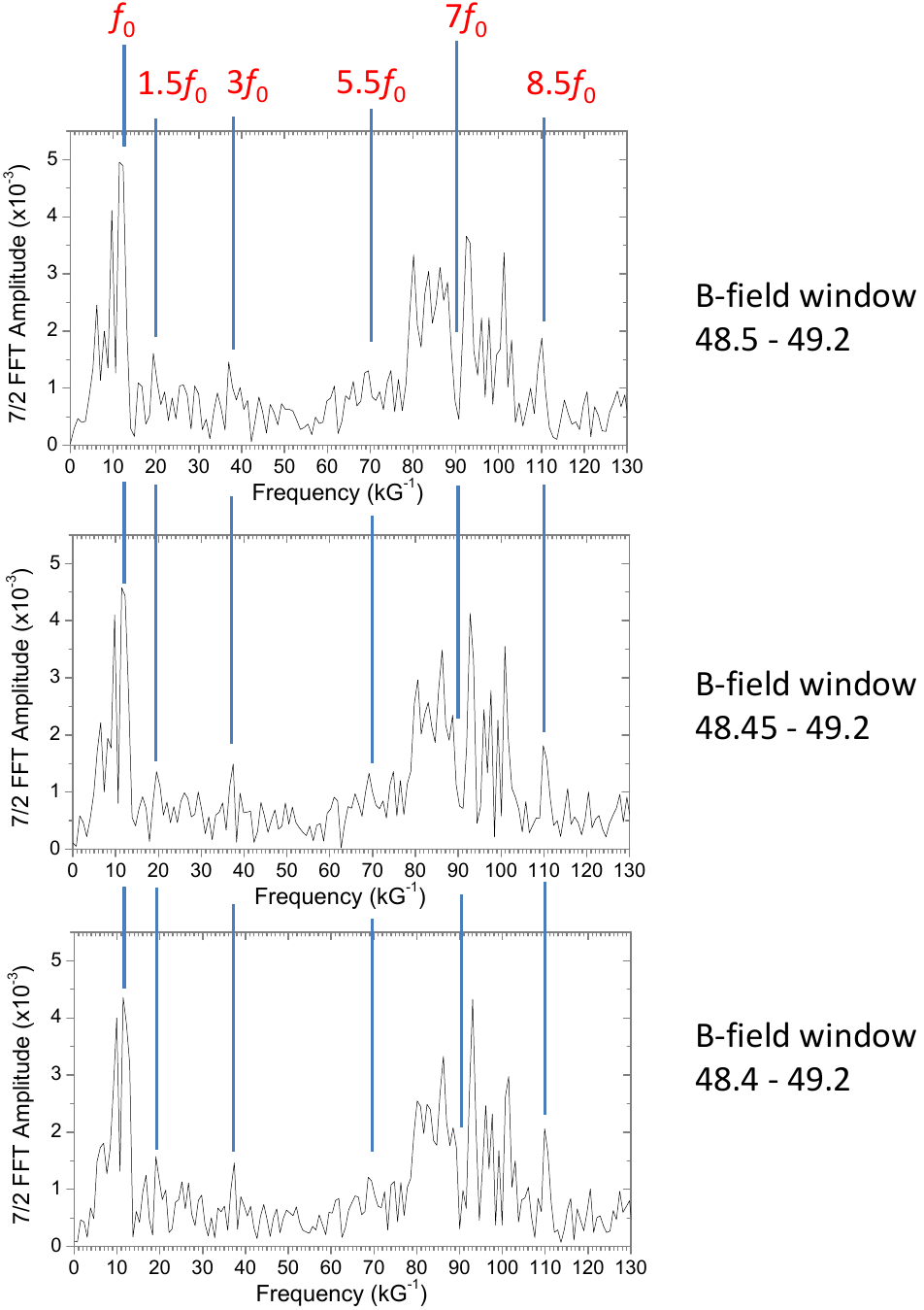}
\caption{The three power spectra presented here are from the same sample, preparation, and $B$-field sweep trace of $R_L$, but use different FFT windows. Note that the spectral features at $f_0$, $1.5f_0$, $3f_0$, $5.5f_0$, $7f_0$ and $8.5f_0$ are expressed to slightly different degrees. Sample 6, preparation 14, $T=20mK$.
}
\label{fig:S5-4-4}
\end{figure}

Meantime Figure~\ref{fig:S5-4-4} presents $\nu=7/2$ power spectra taken from the same device, same sweep, but different magnetic field windows for the FFTs.

\begin{figure}[htb]
\includegraphics[width=0.67\textwidth]{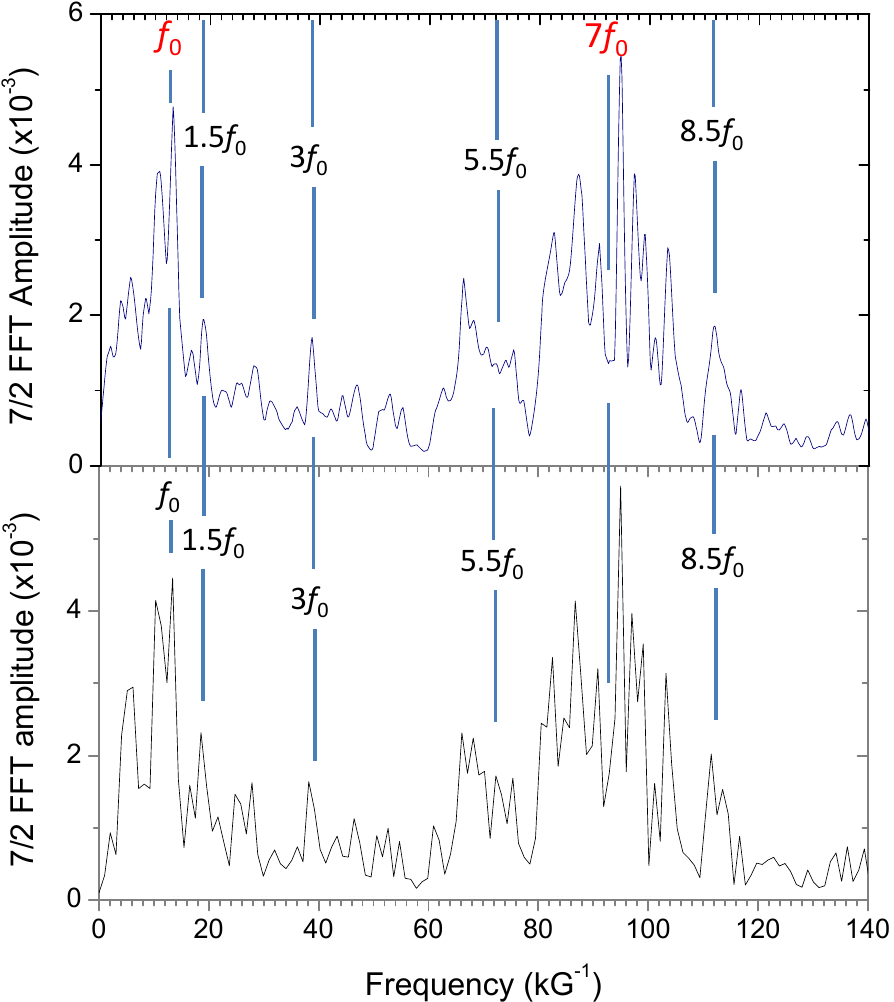}
\caption{. Single vs. averaged power spectra at $\nu=7/2$. All power spectra here are FFTs of the same magnetic field sweep using the same sample and preparation: bottom panel is a single trace, the top an average of four traces using different FFT windows.  Note that averaging reduces spurious peaks such as that at $\sim 26 \text{kG}^{-1}$ . Sample 6, preparation 15, $T=20mK$.
}
\label{fig:S5-4-3}
\end{figure}

Figure~\ref{fig:S5-4-3} presents comparison of a single FFT of the $\nu=7/2$ interference oscillations to the average of four FFTs  using different magnetic field windows.

\begin{figure}[htb]
\includegraphics[width=0.67\textwidth]{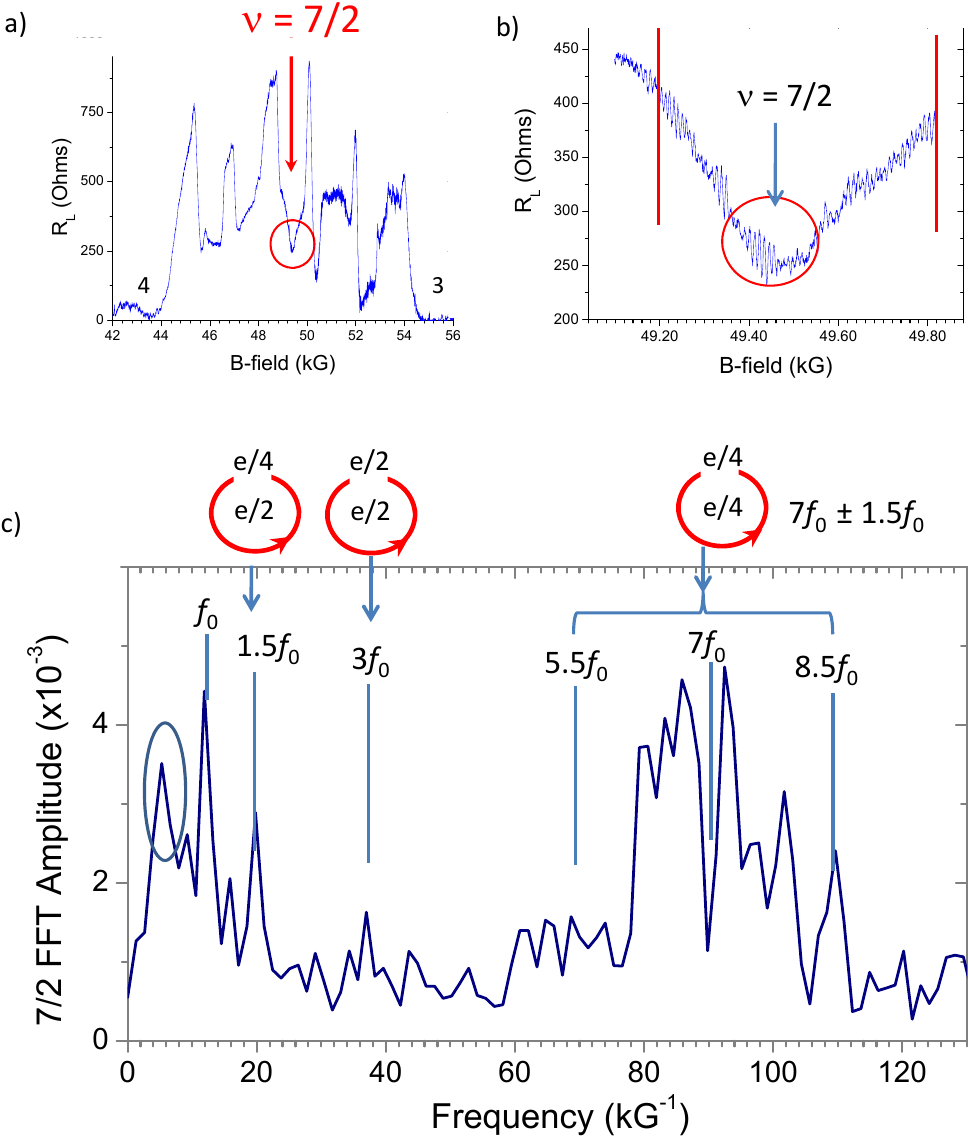}
\caption{(a)-(b) A blow-up of a QH trace $R_L(B)$. (c) The oscillation power spectrum averaged over two preparations. Frequencies $f_0$, $1.5f_0$, $3f_0$, $5.5f_0$, $7f_0$ and $8.5f_0$ expected for various interference processes are marked. The FFT window is marked in panel (b).
%Note that the  f0 vertical line is consistent with the   f0  value measured at integer filling.
Sample 6, preparations 8 and 14, $T=20mK$.
}
\label{fig:S5-4-5}
\end{figure}

Finally, Figure~\ref{fig:S5-4-5} shows the $\nu=7/2$ power spectrum averaged over two preparations of the same sample.

We provide these figures in order to drive home the main message: while the low-frequency features of the spectrum (including how well certain features attributed to low-frequency modulation are resolved around the main peaks) depend on such niceties as the background subtraction and the choice of the FFT window, the most remarkable and most physically significant feature, namely the amount of spectral weight in the vicinity of $7f_0$ is insensitive to these details.

\onecolumngrid

\twocolumngrid

\onecolumngrid

\twocolumngrid

\onecolumngrid

\twocolumngrid

\onecolumngrid

\twocolumngrid

\onecolumngrid

\twocolumngrid

\clearpage
\twocolumngrid
\subsubsection{Supplements to Figures~\ref{fig:fermionic_parity} and~\ref{fig:5_halves_interf}}
\label{sec:S5-6}

Figure~\ref{fig:fermionic_parity} of the main text demonstrates the temporal stability of interference oscillations at $\nu=7/2$.  In this section we present similar data for the oscillations at $\nu=5/2$, showing both their stability and occasional $\pi$ phase jumps.
These data are followed by a supplementary discussion of the  $\pi$ phase jumps and estimates of the $e/4$ quasiparticle density  over the $B$-field range corresponding to the $\nu=7/2$ minimum.

\begin{figure}[htb]
\includegraphics[width=0.67\textwidth]{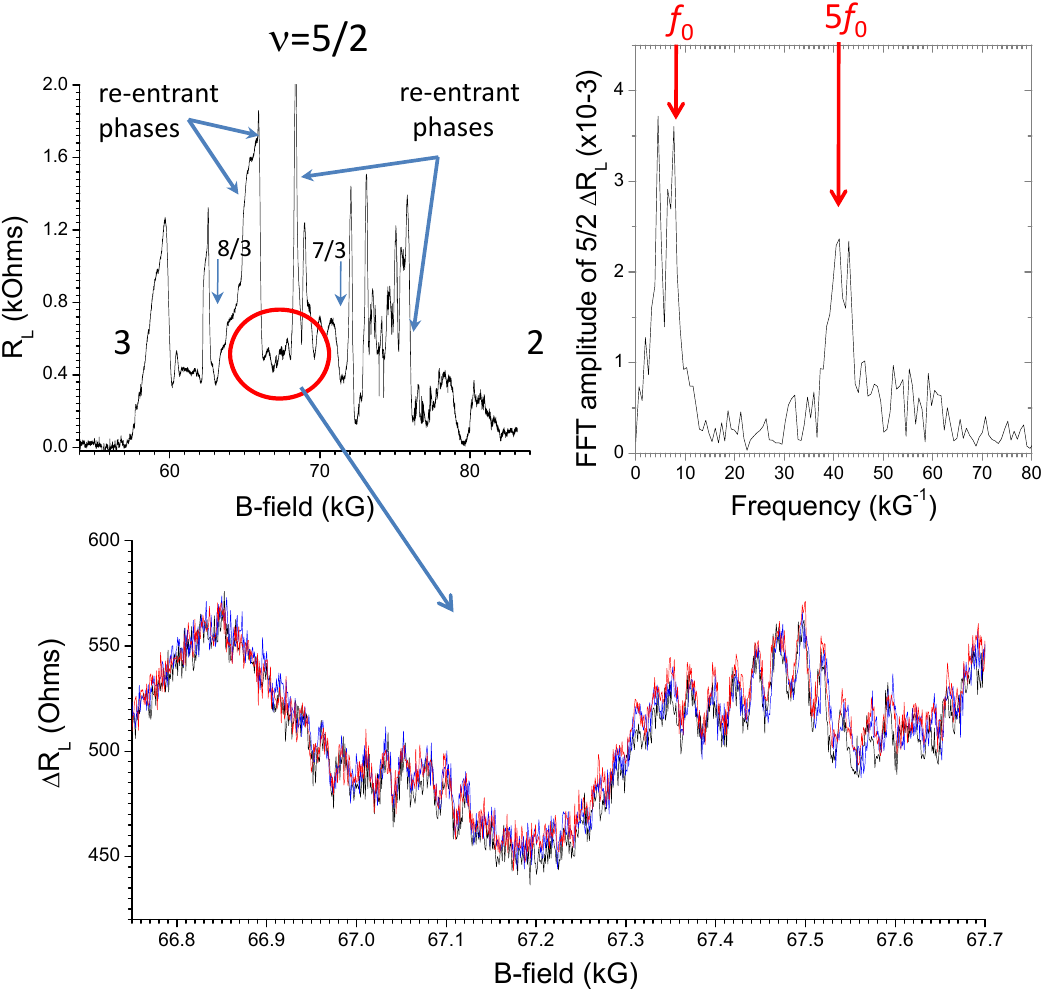}
\caption{Stability and power spectrum of interference oscillations at $\nu=5/2$.\\
Top traces are $R_L$ from filling factor $\nu=3$ to 2 centered at $\nu=5/2$ (circled area), and an FFT of a trace such as that shown in the lower panel of the figure.\\
The colored lower trace set is three $B$-field sweeps in a single direction showing overlap of the resistance oscillations.  Here the predominant oscillation period is that of $\sim 5f_0$ shown in the power spectrum.  Each trace is 60 minutes, so total data collection time is 6 hours (the opposite sweep direction is not shown).\\
Sample 6, preparation 18, $T\sim 20mK$.
}
\label{fig:S5-6-1}
\end{figure}

	As shown in Figure~\ref{fig:5_halves_interf} of the main text (and reproduced below as Figure~\ref{fig:S5-6-1} for ease of reference), similarly to the $\nu=7/2$ state, the $\nu=5/2$ state in the ultra-high purity Al material also demonstrates stable oscillations that can be attributed to the non-Abelian even-odd effect. Minor density shifts at filling factors below $\nu=3$ prevent direct comparison of these oscillations with the measurements at $\nu=7/2$. However, the principal spectral weight observed for $\nu=5/2$ is consistent with frequency $5f_0$ , with the $f_0$ values measured at the high field side of filling factor $\nu=3$ and at $\nu=2$, i.e. on the same side of the density shift as the observed $\nu=5/2$ state.
	
\begin{figure}[htb]
\includegraphics[width=0.65\textwidth]{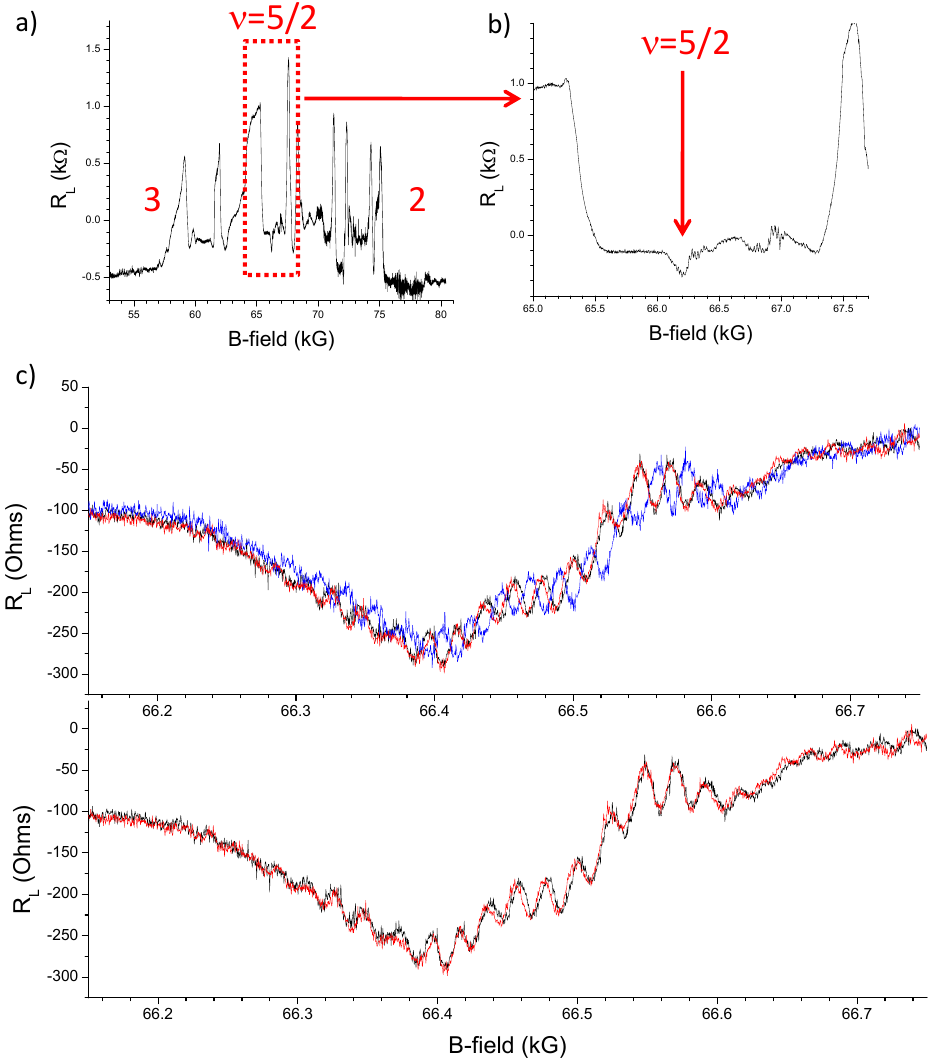}
\caption{Temporal stability of oscillations at $\nu=5/2$ and observed phase jumps:\\
Panels (a) and (b) show overall $R_L$ measurement between $\nu=2$ and $\nu=3$ and then focus at the region around $\nu=$5/2. The bottom part of panel (c) shows two consecutive up-sweeps in $B$-field, with close overlap. These two traces are reproduced in the top panel of (c) but also shown is the next consecutive up-sweep, demonstrating a $\pi$ phase jump over most of the oscillations.
Sample 6, preparation 23, $T\sim 20mK$.
}
\label{fig:S5-6-2}
\end{figure}

As with the 7/2 data, the data presented in Figures~\ref{fig:S5-6-1} and \ref{fig:S5-6-2} below demonstrates reproducibility and stability of the oscillations associated with the non-Abelian even-odd effect and indicate that the fermion parity remains stable over long time and magnetic field range near $\nu=5/2$. When the phase jump occurs, it is close to $\pi$ (see Figure~\ref{fig:S5-6-2}) as would be expected for fluctuations of either the fermion parity or the parity of enclosed charge-$e/4$ quasiparticles.
Similarly to what has been seen at $\nu=7/2$ (see Figure~\ref{fig:fermionic_parity}), the $\pi$  jumps are also observed at  $\nu=5/2$.
The relative prevalence of these phase flips at 5/2 versus 7/2 is an actively pursued issue.

\onecolumngrid

\clearpage
\twocolumngrid

A question to be addressed is what determines neutral fermion tunneling resulting in changes in the fermion parity, particularly in the $B$-field sweeps where the gate voltages remain fixed and the only changing parameter is the magnetic field.  As one moves away from the center of the plateau at $\nu=7/2$ or $\nu=5/2$, the number of quasiparticles or quasiholes increases; we can estimate their density and in so doing check plausibility of our model.  We can use the data in
Figure~\ref{fig:7_halves_interf} to assess the $e/4$ number and the quasiparticle density.  In Figure~\ref{fig:7_halves_interf}(b) the magnetic field interval of roughly 1kG spans the width of the 7/2 minimum.  Figure~\ref{fig:7_halves_interf}(c) (as well as \ref{fig:Hall_trace}(b)) shows a period of $\sim 12.5$G corresponding to oscillations at $7f_0$, i.e. 12.5G corresponds to changing the number of $e/4$ quasiparticles by 2.  By moving from plateau center to its edge we create about $500\text{G} /(12.5\text{G} /2) = 80$ charge-$e/4$ quasiparticles/quasiholes.  The period of $f_0$ oscillations is $\sim 88$G, resulting in an active area of the interferometer of $41\,\text{G}\,\upmu\text{m}^2/88\,\text{G} \sim 0.46\,\upmu\text{m}^2$.   Therefore the maximum density of the $e/4$ quasiparticles in the active area $A$ of the interferometer is $80/0.46\,\upmu\text{m}^2 \sim  170\,\upmu\text{m}^{-2}$.   This puts the separation between the quasiparticles at $\sim 0.07\,\upmu\text{m}=70$nm at the maximum density.   The magnetic length at 5~Tesla is about 12nm, so that the quasiparticle minimum separation is greater than 5 magnetic lengths.  From Figure~\ref{fig:7_halves_interf}(c), at about 48.7kG or $\sim 100$G from the $\nu=7/2$ minimum, 16 charge-$e/4$ quasiparticles should be present in area $A$ from counting the even-odd oscillations, with the quasiparticle density of $16/(0.46\,\upmu\text{m}^2) \sim 35 \,\upmu\text{m}^{-2}$, producing a separation of  about  $0.17\,\upmu\text{m} = 170$nm,  or about  14 magnetic lengths.  These separation numbers do not contradict the picture of an ensemble of charge-$e/4$ quasiparticles that can sustain a stable fermion parity over a range of $B$-field around $\nu=7/2$: the model would clearly be invalid if separation had been found to be near or less than the magnetic length over a substantial part of the $\nu=7/2$ minimum.  As is, the higher density at the boundaries of the $\nu=7/2$ minima away from the plateau center may contribute to fermion parity flips, and this must be examined experimentally.

\onecolumngrid

\clearpage

\twocolumngrid

%\clearpage

\subsubsection{Supplements to Figure~\ref{fig:power_spectra}}
\label{sec:S5-3}

Figure~\ref{fig:power_spectra} of the main text demonstrates the frequency spectrum for the interference oscillations observed in the vicinity of the $R_L$ minimum at $\nu=5/2$:  this spectrum showed four predominant peaks at $f_0$, $2f_0$, $4f_0$ and $6f_0$, with $f_0$ being the frequency corresponding to AB of the electron measured at integer filling factors.
As has been explained, these peak values are consistent with the frequency values expected for the interference of non-Abelian charge-$e/4$ and Abelian charge-$e/2$ quasiparticles at $\nu=5/2$.  This result serves as a metric for examining interference in a system containing both non-Abelian charge-$e/4$ and Abelian charge-$e/2$ quasiparticles, which we then use to examine the interference at $\nu=7/2$.  In this section we supplement the results shown in Figure~\ref{fig:power_spectra} for $\nu=5/2$ with data in other samples and preparations at the same filling fraction, and in one case a later measurement in the same device and preparation.

\onecolumngrid

\begin{figure}[htb]
\includegraphics[width=0.67\textwidth]{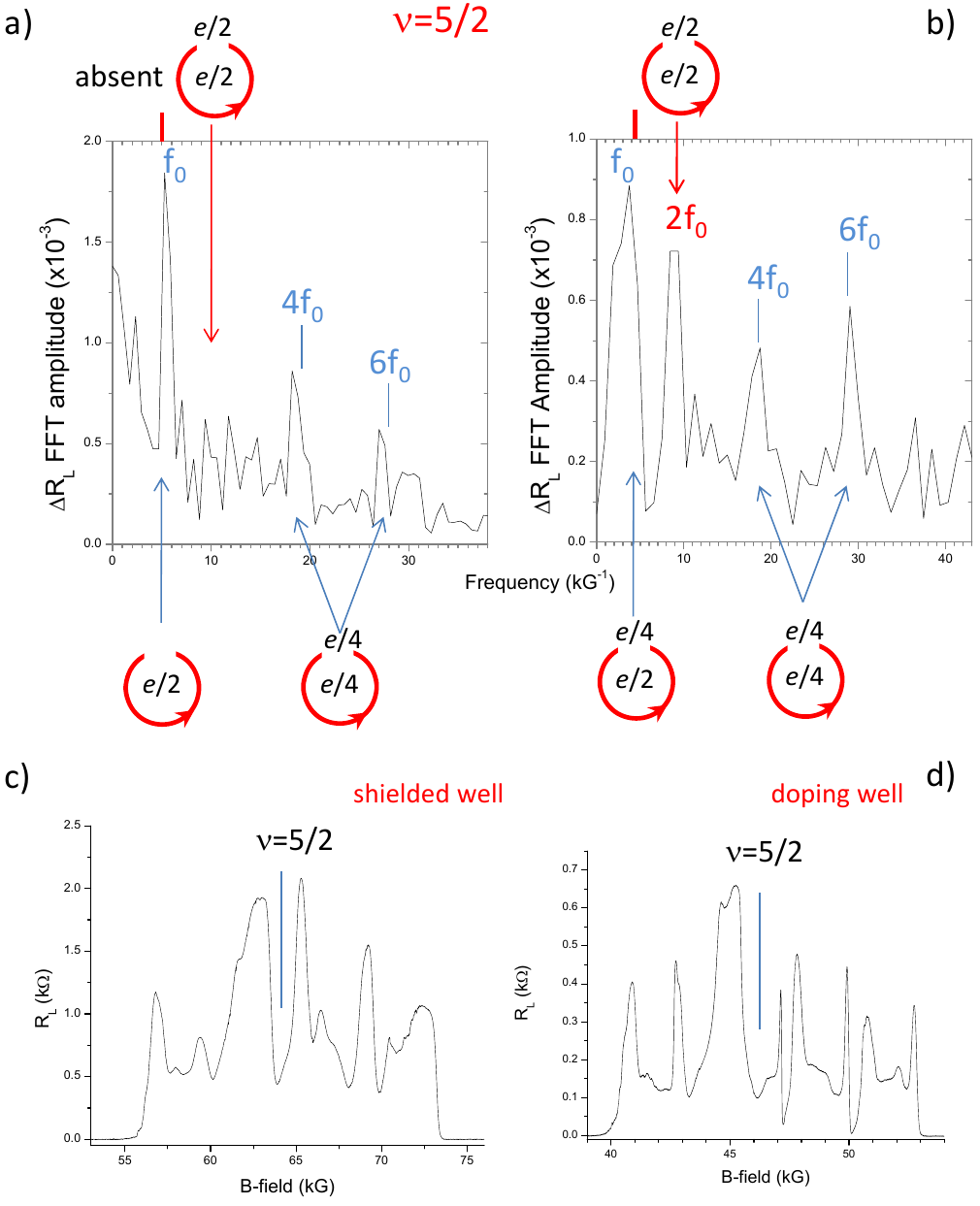}
\caption{Dominant peaks in the AB oscillation spectrum at $\nu=5/2$ in multiple samples.\\
(a) \& (b)  Power spectra of $R_L(B)$ near $\nu=5/2$ in samples 3 and 4 (top panels) fabricated from different heterostructures, using different sample illuminations and gate voltages (preparations). The interferometer is as shown in Figure~\ref{fig:S1-1}, or Figure~\ref{fig:interferometer} schematic and top electron micrograph.
The spectra are obtained from the $R_L$ data by first subtracting the background and then applying the FFT.
}
\label{fig:S5-3-1}
\end{figure}

\clearpage
\twocolumngrid

\begin{figure}[htb]
\includegraphics[width=\columnwidth]{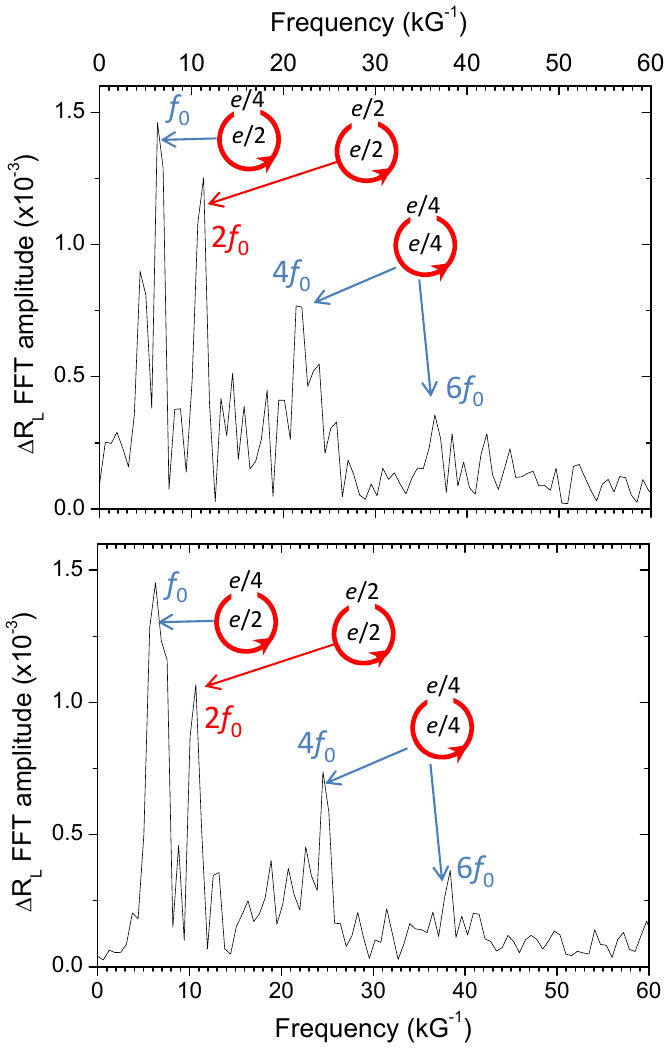}
\caption{Comparison of the oscillation spectra at $\nu=5/2$ from the same sample and device. The two spectra shown here correspond to two different magnetic field sweeps in the same device recorded days apart.  The bottom trace is the data of Figure~\ref{fig:power_spectra}(b) and the top trace is the data of Figure~\ref{fig:2f0_peak}(b).
Sample 2, preparation 2, $T=20mK$.
}
\label{fig:S5-3-2}
\end{figure}

\begin{figure}[htb]
\includegraphics[width=\columnwidth]{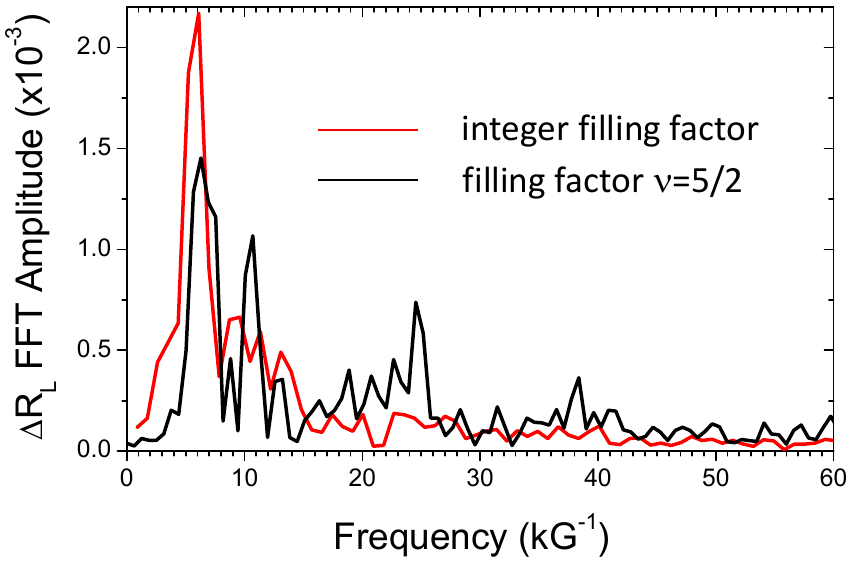}
\caption{Comparison of oscillation spectra  at an integer filling and at $\nu=5/2$.
This spectra have been obtained using the same sample, device, and interferometer gate voltages.  Note that the $\nu=5/2$ spectrum is the same as that of Figure~\ref{fig:power_spectra}(b) of the main text and Figure~\ref{fig:S5-3-2} above.
Sample 2, preparation 2, $T=20mK$.
}
\label{fig:S5-3-3}
\end{figure}

%\clearpage

\clearpage
\twocolumngrid
\subsubsection{Supplements to Figure~\ref{fig:sup_peak_fit}: Prevalence and assignment of spectral peaks at $f_0$,  $2f_0$,  $4f_0$ and  $6f_0$}
\label{sec:S5-7a}

\begin{figure}[h!tb]
\centering
    \includegraphics[width=0.67\columnwidth]{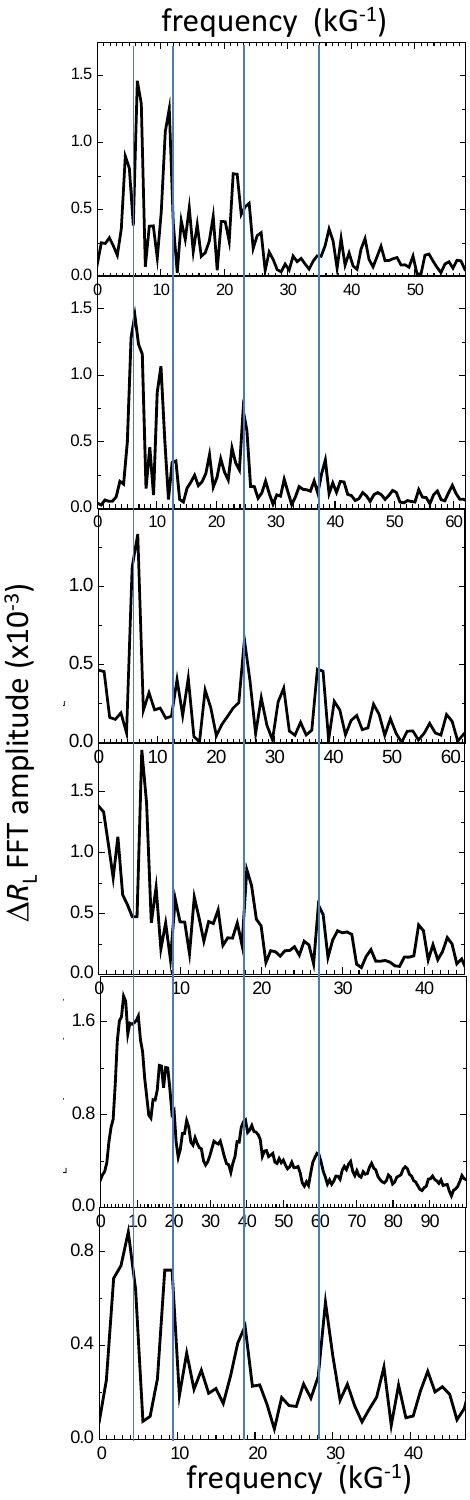}
    \caption{FFT spectra corresponding to the six data sets used in Figure~\ref{fig:sup_peak_fit} and also presented elsewhere in the  paper:  from top to bottom, (i) Figure~\ref{fig:2f0_peak}(b); (ii) Figure~\ref{fig:power_spectra}(b); (iii) Figure~\ref{fig:2f0_peak}(a); (iv) Figure~\ref{fig:S5-3-1}(a); (v) Figure~\ref{fig:S5-7-3} (smoothed); (vi) Figure~\ref{fig:S5-3-1}(b). The blue lines correspond to the multiples of $\tilde{f}_0$ for each sample, with the horisontal range adjusted to $10 \tilde{f}_0$ for each spectrum. Their precise location is not crucial for the purpose of peak identification; here they serve as mere guides for the eye and could be replaced by the multiples of $f_0$ (the frequency of oscillations at an integer filling fraction).}
    \label{fig:peak_positions}
\end{figure}

In this section we revisit the full oscillation spectrum observed at $\nu=5/2$ filling while focusing on the identification of the spectral peaks with various quasiparticle braids.
%Three important issues addressed are: 1) overall prevalence and assignment of the $\nu=5/2$ spectral peaks at $f_0$, $2f_0$, $4f_0$ and $6f_0$ , or $5f_0$; 2) Supplementary material on diminishment of the  $2f_0$ peak as interferometer backscattering is reduced, and 3)  the $f_0$ peak and potential contributions from background electron interference.
An important issue one must address first is the selection of spectral features designated as peaks for the purpose of this analysis. Given the noise invariably present in the measured data and the additional uncertainty introduced by the particulars of the Fourier transform (e.g. its window), formulating a clear objective criterion for such a selection is anything but straightforward. To illustrate this point, in Figure~\ref{fig:peak_positions} we present the FFT spectra used for the identification procedure outlined in Figure~\ref{fig:sup_peak_fit} of the main text.
The blue vertical lines here are rough guides for the eyes that have been used for selecting the most prominent spectral features near $f_0$, $2f_0$, $4f_0$ and $6f_0$. As one can clearly see, this procedure is not entirely satisfactory: in data sets 3 and 4 we identify no peaks in the vicinity of $2f_0$ while identifying such peaks near $6f_0$. A reasonable objection may be raised that some of the features seen between $f_0$ and $4f_0$ have their amplitude comparable to that near $6f_0$. The rationale for discarding the former while keeping the latter is empirical: throughout this entire study we have observed that the amplitude of all spectral features (including noise) predominantly decreases with frequency. While we have no clear explanation for why this is the case, we use this empirical fact to weigh in favor of higher frequency features while discarding features of similar amplitude at lower frequencies as likely noise.
%Power Spectra Data As In \Textcolor{Red}{Fig. 7B}

\begin{figure}[hbt]
\centering
    \includegraphics[width=0.95\columnwidth]{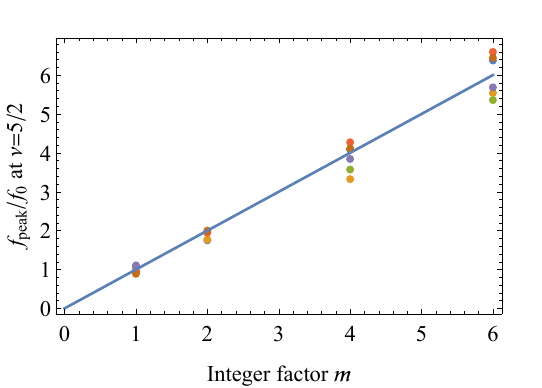}
    \includegraphics[width=0.95\columnwidth]{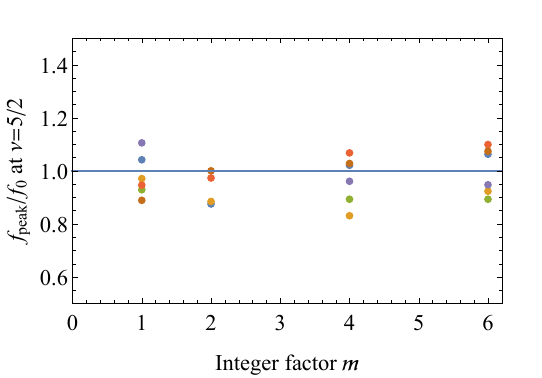}
    \caption{(a) Identification of FFT peaks at $\nu=5/2$ using highest values of observed spectral features. Frequencies corresponding to predominant spectral peaks observed at $\nu=5/2$ are divided by the value of $f_0$ obtained at $\nu=3$ for each sample and plotted as a function of integer multiples of $f_0$ with which they have been identified (samples~1--5, 2 preparations for sample~2 -- see traces in Fig.~\ref{fig:2f0_peak}(a,b), $T \sim 20$mK). Two of the samples did not exhibit peaks at or near $2f_0$.
    (b) Same data rescaled by the assigned factor $m$ in order to eliminate possible effects of errors increasing with $m$ due to a mismatch between $f_0$ measured at integer and fractional states.}
    \label{fig:peak_identification}
\end{figure}

Once the spectral features have been identified as peaks, we can proceed with their further analysis. In Figure~\ref{fig:sup_peak_fit} of the main text these
peak positions obtained for five different samples and multiple preparations have been normalized by the frequency $\tilde{f_0}$ corresponding to the coefficient of their linear fit have been then plotted versus expected integer factors of $m=1$, 2, 4, and 6. In Figure~\ref{fig:peak_identification} below we present the same data normalized by reference frequency $f_0$ determined by the oscillations at a nearby integer filling instead. The data is presented in two different plots with the second one showing the observed peak frequencies divided by both the reference frequency and the integer with which we identify the peak in order to eliminate the effect of ``fanning out'' of errors due to a potential mismatch between $f_0$ measured at the integer ($\nu=3$) and fractional ($\nu=5/2$) fillings. Within reasonable error margins (not exceeding 20\%), the points accumulate near a straight line, which validates our identification of these peaks with a set of quasiparticle braids:
 $e/4$ braiding $e/2$  ($f_0$),  $e/2$ braiding $e/2$ ($2f_0$), the non-Abelian even-odd effect due to charge-$e/4$ quasiparticles modulated by their braiding of bulk $e/2$ ($4f_0$ and $6f_0$).

\begin{figure}[htb]
\centering
    \includegraphics[width=0.95\columnwidth]{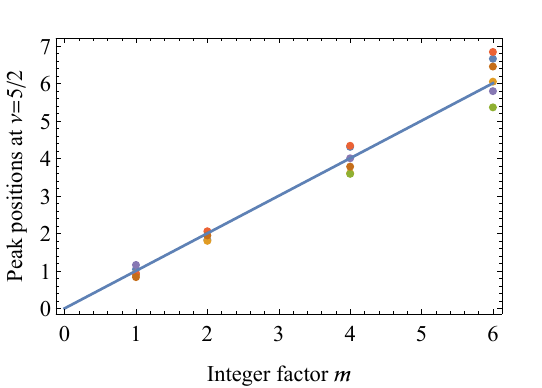}
    \includegraphics[width=0.95\columnwidth]{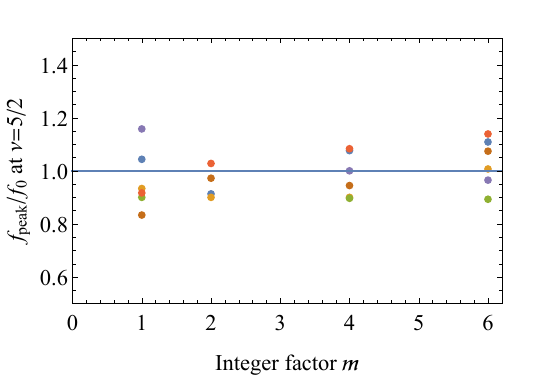}
    \caption{Identification of FFT peaks at $\nu=5/2$ using midpoints between frequencies corresponding to each peak's half-value instead of locations of peak values used in Figure~\ref{fig:peak_identification}. (a) Frequencies corresponding to midpoints of predominant spectral peaks observed at $\nu=5/2$ are divided by the value of $f_0$ obtained similarly at $\nu=3$ for each sample as a function of integer multiples of $f_0$ with which they have been identified (samples~1--5, 2 preparations for sample~2 -- see traces in Fig.~\ref{fig:2f0_peak}(a,b), $T \sim 20$mK). Two of the samples did not exhibit peaks at or near $2f_0$.
    (b) Same data rescaled by the assigned factor $m$ in order to eliminate possible effects of errors increasing with $m$ due to a mismatch between $f_0$ measured at integer and fractional states.}
    \label{fig:sup_peak_identification}
\end{figure}

Because the spectral data used for this identification is somewhat noisy, we supplement Figure~\ref{fig:peak_identification} with a similar figure where the peak location has been identified by the average of the two frequencies corresponding to each peak's half-value rather than the highest value. The result is presented in Figure~\ref{fig:sup_peak_identification}; it shows no significant difference with Figure~\ref{fig:peak_identification}.
The data behind Figures~\ref{fig:peak_identification} and ~\ref{fig:sup_peak_identification} are shown in part in Figures~\ref{fig:power_spectra} and \ref{fig:2f0_peak} and Sections~\ref{sec:S5-3} and ~\ref{sec:S5-7c}.

In previous studies \cite{Willett2013b} using generally smaller interferometer devices, the peaks at $4f_0$ and $6f_0$ were in some instances not resolved, resulting in single spectral peak at  $5f_0$. Observation of distinct $4f_0$ and $6f_0$ peaks versus a single peak  $5f_0$ likely results from better stability of fermion parity on experimentally-relevant timescales, as described above in the introductory section. This in turn could be related to the device parameters such as size, as is discussed below and in Section~\ref{sec:S2} of Supplementary Materials. The $B$-field window over which the FFT is applied can also limit this resolution, so that a small window as caused by impinging integer filling re-entrant phases could limit resolution of the modulated peaks.
Past measurements \cite{Willett2013b} in similar heterostructures and over a range of interferometer sizes focused on demonstrating the high frequency oscillations, occurring at $(5\pm1)f_0$, that are attributable to the non-Abelian properties of $e/4$ quasiparticles.  In those measurements, the lower frequency features at $f_0$ and $2f_0$ were filtered from the data to emphasize the discovered high frequency oscillations at $(5\pm1)f_0$, and also to minimize potential higher harmonic noise in an FFT.  By not employing that filtering, using generally larger interference devices, further minimizing potential negative impact of device fabrication on the quality of the electron gas, and most importantly improving heterostructure sample quality, we were able to examine the full frequency range of the expected oscillations.

\clearpage

\subsubsection{Supplements to Figure~\ref{fig:interf_model}}
\label{sec:S5-5}
%section S5-5:  Figure 5 supplements:

As has been explained in the main text, measured oscillation spectra at $\nu=7/2$ and $\nu=5/2$ are consistent with a simple picture of the high frequency even-odd oscillations that are modulated by lower frequency oscillations due to the interference of $e/4$ quasiparticles around Abelian quasiparticles, which occurs whenever the total number of bulk $e/4$ quasiparticles inside the interferometer is even. Depending on the stability of the overall Abelian anyonic charge -- the fusion channel -- the high frequency peaks may split into two peaks as a result of this modulation, generating spectral features at $(5\pm 1)f_0$ at $\nu=5/2$ and $(7\pm 1.5)f_0$ at $\nu=7/2$.

\onecolumngrid
%\clearpage
\begin{center}
\begin{figure}[b!]
\includegraphics[width=0.67\textwidth]{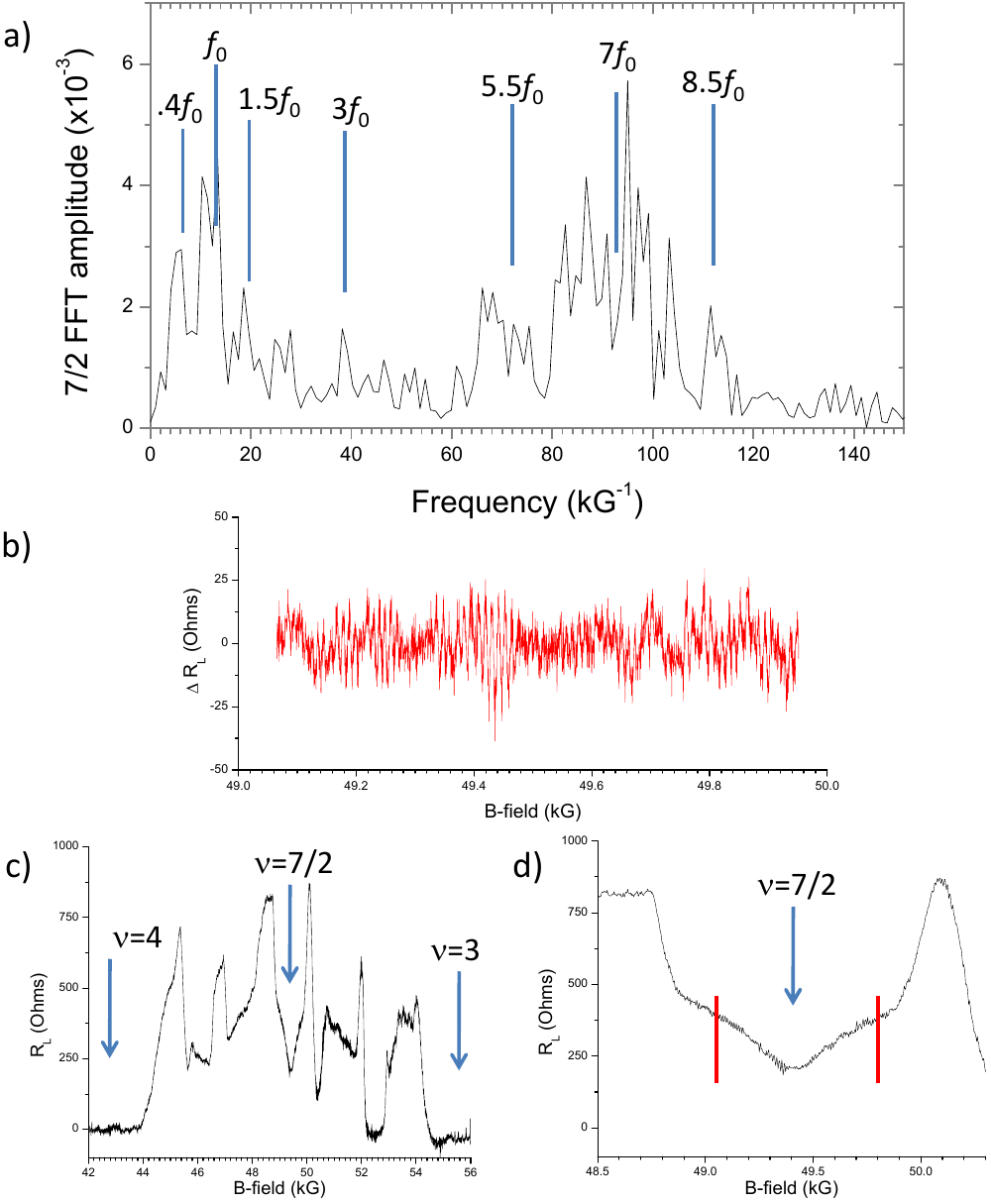}
\caption{The overall oscillation spectrum observed at $\nu=7/2$ is reproduced in panel (a) with the source data for $R_L(B)$ shown in panels (c)--(d). Panel (b) shows $R_L(B)$ around the $\nu=7/2$ minimum with the background subtracted to demonstrate the series of oscillations that are Fourier-transformed to produce the power spectrum in (a).
Sample 6, preparation 15, $T=20mK$.
}
\label{fig:S5-5-1}
\end{figure}
\end{center}

\twocolumngrid

However, the Abelian interference is not the only source of modulation of the high-frequency even-odd processes.
Artefacts such as the dependence of the oscillation amplitude on the precise filling factor in the vicinity of the minimum can result in additional satellite spectral features at high frequency. E.g., one feature that appears to be seen at $\nu=7/2$ is a spectral peak at around $0.4f_0$ -- see Figures~\ref{fig:7_halves_interf}(e) and \ref{fig:interf_model}(b) of the main text as well as Figure~\ref{fig:S5-5-1} (the fact that it is not seen in the $\nu=5/2$ data seemingly rules out some systematic error as its source). This peak appears in the new heterostructure/device samples and may be attributable to the proximity of the re-entrant phases to the observed $\nu=7/2$ resistance minimum. A proximity to a competing phase can affect the quasiparticle coherence time and hence the oscillation amplitude.
In order to see how this additional modulation would affect the oscillation spectrum, we model the longitudinal resistance as
\begin{multline}
\Delta R_L=\left[0.25\cos\left(2\pi B(0.4f_0 )\right)+0.75\cos\left(2\pi B(1.5f_0)\right)\right]\\
\times\cos\left(2\pi B(7f_0)\right)
\label{eq:supplement_modulation}
\end{multline}

This equation captures both sources of modulation, the Abelian interference at $1.5 f_0$ and the empirical contribution at $0.4 f_0$, with coefficients, but not frequencies, adjusted to match the measured data.
This expression is Fourier-transformed using the same FFT window that has been used in processing the actual experimental data and the resulting spectrum is plotted in Figure~\ref{fig:interf_model}(d). It shows good agreement with the measured spectrum above it, in panel (b): both the minimum at $7f_0$ and the ``descendent'' peaks at $5.5f_0$ and $8.5f_0$ are well reproduced given the coefficients employed in the model.

\begin{figure}[htb]
\includegraphics[width=0.95\columnwidth]{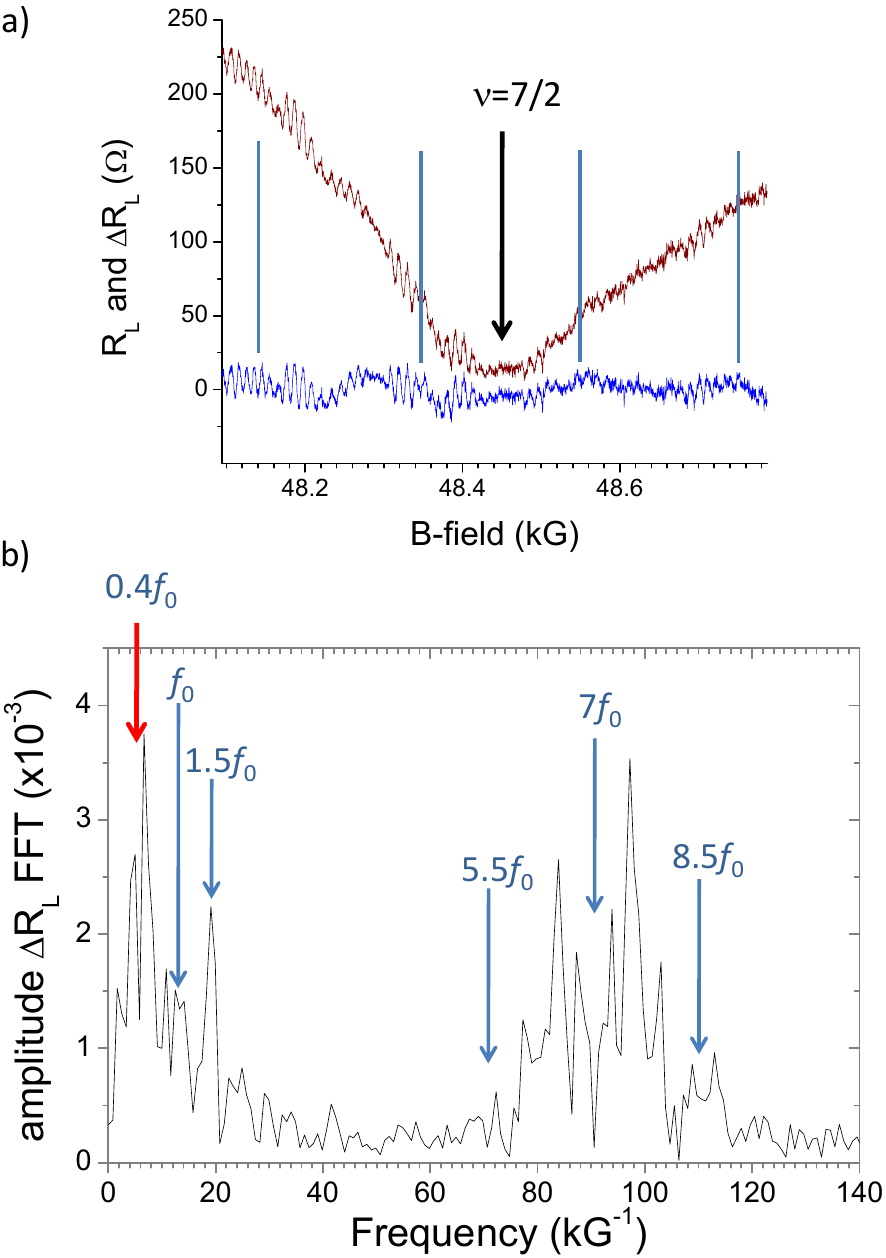}
\caption{Interference oscillations in a sample that shows a pronounced spectral feature $0.4f_0$:\\
(a) $R_L$ and $\Delta R_L$ around filling factor $\nu=7/2$.  Vertical marks indicate the periodicity corresponding roughly to frequency $0.4f_0$.\\
(b)  FFT of the full extent of $\Delta R_L$ displayed in (a); marked are the frequencies of  $0.4f_0$, $f_0$, $1.5f_0$, $5.5f_0$, $7f_0$ and $8.5f_0$, with $7f_0$ being a reference frequency.
Sample 6, preparation 23, $T\sim 20mK$.
}
\label{fig:S5-5-2}
\end{figure}

The source of the $0.4f_0$ peak in the $\nu=7/2$ spectra can be localized to the form of the  minimum in  $R_L$ which is distinct in the new higher purity Al heterostructures.  This large period oscillation can be exposed with a subtraction from  $R_L$ of a large $B$-field range moving average of   $R_L$ with the process results displayed in Figure~\ref{fig:S5-5-2}: the moving average here uses a window of  about 175G.  Shown in the top panel is the  $R_L$  trace around $\nu=7/2$ and below it is that  $R_L$ with such a background subtracted. Marked are vertical lines with a period that corresponds roughly to the $0.4f_0$ frequency. This periodicity is consistent with the overall minimum form in $R_L$ but more clear in the $\Delta R_L$ .
With this large $B$-field range background subtraction the $0.4f_0$ feature is more pronounced in the FFT of the full extent of $\Delta R_L$ displayed in panel (a).  Figure~\ref{fig:S5-5-2}(b) shows this FFT with well a developed  peak at $1.5f_0$, a minimum at $7f_0$, but with the majority of the spectral weight around that frequency, and a pronounced feature at $\sim 0.4 f_0$.

%\clearpage

\clearpage
\subsubsection{Supplements to Figure~\ref{fig:2f0_peak}: the  $2f_0$ peak and its dependence on backscattering at constrictions}
\label{sec:S5-7b}

\begin{figure}[b!th]
\includegraphics[width=0.9\columnwidth]{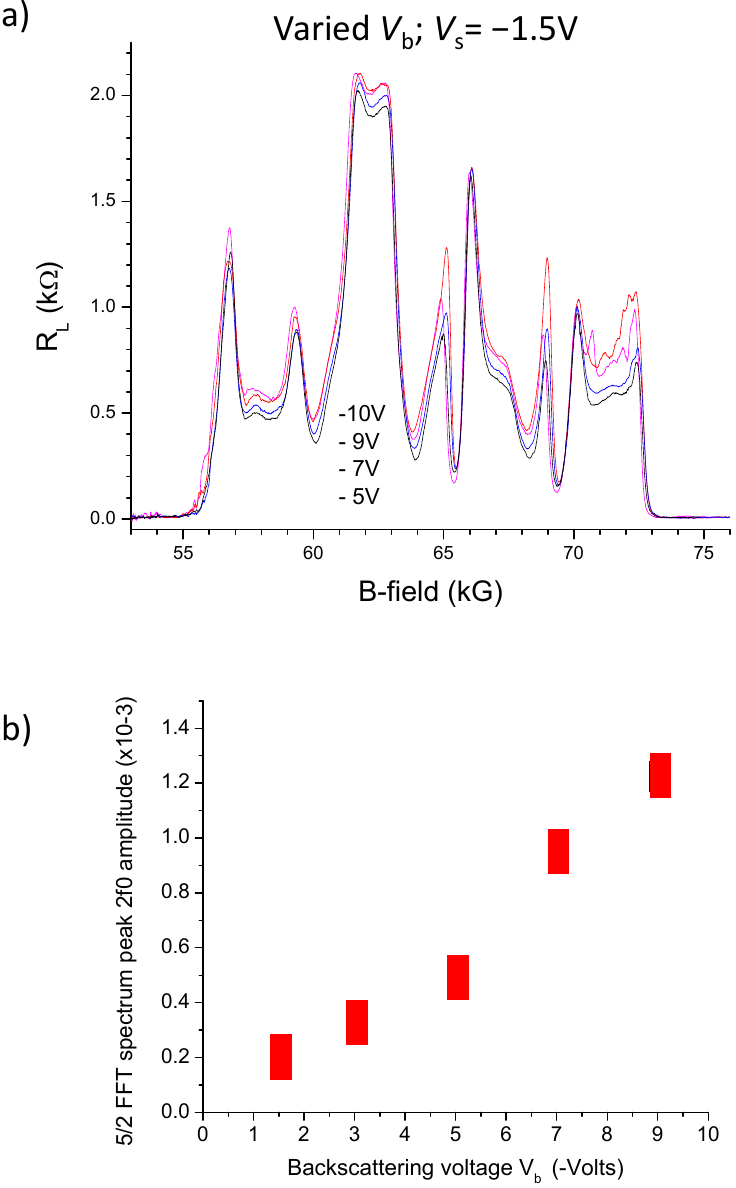}
\caption{(a) $R_L$ transport through sample 2 of Figure~\ref{fig:power_spectra}.  The progression of $V_b = -10, -9, -7, -5$V in the traces is red, pink, blue, black respectively.  The $\nu=5/2$ minimum is at 64kG.  Preparations 4-7, $T\sim 20$mK.\\
The amplitude of the $2f_0$ peak at $\nu=5/2$ versus gate voltage $V_b$, demonstrating increase in amplitude with smaller gate separation.  This result is similar to the finding in Figure~\ref{fig:2f0_peak} where $2f_0$ is normalized by the respective $f_0$ amplitude values.
}
\label{fig:S5-7-1}
\end{figure}

The application of larger voltages to the $V_b$ gates of the interferometers can preserve the overall transport spectrum in $R_L$ between filling factors $\nu=2$ and 3 for some samples, device designs, and preparations as shown in Figure~ \ref{fig:S5-7-1}. This is the same sample and device as that in Figure~\ref{fig:power_spectra}, and the overall spectrum displays only a minor increase in the resistance minima between filling factors $\nu=2$ and 3 as the backscattering voltage $V_b$ is increased.

Also displayed is a plot of the amplitude of the $\nu=5/2$ FFT spectrum peak at  $2f_0$ versus backscattering voltage $V_b$, showing that as the backscattering gate separation distance is made smaller (more negative $V_b$  ), the amplitude of the  $2f_0$ spectral peak increases. This dependence of amplitude on gate voltage is as expected for a mechanism where $e/2$ charges encircle area $A$ of the interferometer, necessarily backscattering at the gate constrictions where $V_b$  is applied, but with tunneling amplitude relatively more dependent on the gate separation than for the $e/4$ charges. The interference of $e/4$ charges should contribute to spectral peaks at $f_0$,  $4f_0$,  $5f_0$, and  $6f_0$, but not  $2f_0$.

\begin{figure}[b!ht]
\includegraphics[width=\columnwidth]{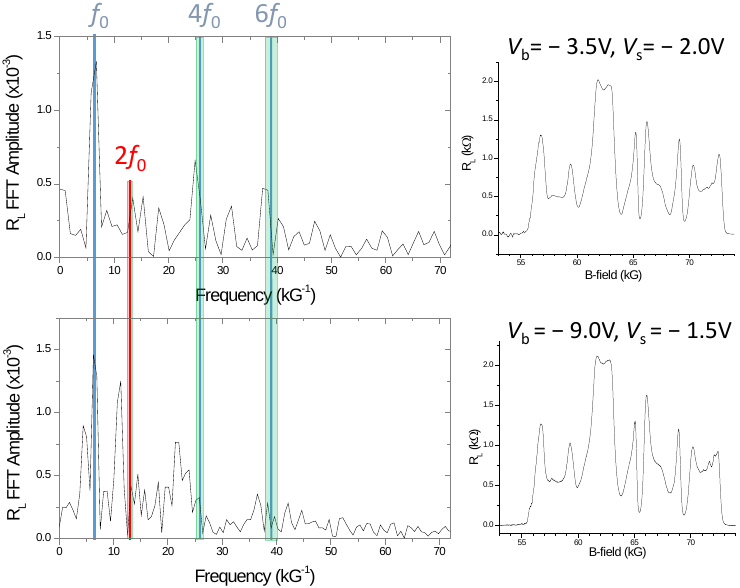}
\caption{Further data on the influence of backscattering on the $2f_0$ peak at $\nu=5/2$ filling factor. Reproduction of $\nu=5/2$ power spectra from Figure~\ref{fig:2f0_peak} with their respective $R_L$ transport traces to the right. Shaded areas indicate error margins due to potential errors in identifying $f_0$, which increases for its multiples.  Note the similarity of the transport traces for the two different gate set values of $V_b = -3.5$V, $V_s = -2.0$V versus $V_b = -9.0$,  $V_s  = -1.5$V.  Sample 2, preparations 1 \& 2, $T\sim 20$mK.
}
\label{fig:S5-7-2}
\end{figure}

\clearpage
\subsubsection{Origins of the $f_0$ peak, its temperature dependence and potential contributions from background electron interference.}
\label{sec:S5-7c}

The oscillation frequency of $f_0$ can result from either charge-$e/4$ quasiparticle interference around bulk charge-$e/2$ excitations or from the ordinary electron interference. Consequently, mere observation of the $f_0$ peak does not by itself establish $e/4 – e/2$ interference; for a proper identification of this peak it is necessary to examine its temperature dependence.  As the temperature is increased, such a peak should disappear or diminish in amplitude simultaneously with the loss of the $\nu=5/2$ resistance minimum in transport if the $e/4$  interference is contributing to the peak.  On the other hand, the AB interference of electrons could survive a disappearance of the $\nu=5/2$ state and persist at higher temperatures.  At the very least, the temperature range over which the $e/4$-quasiparticle contribution to the $f_0$ peak changes should be similar to the temperature range over which the $\nu=5/2$ state is observed.

\begin{figure}[b!]
\includegraphics[width=\columnwidth]{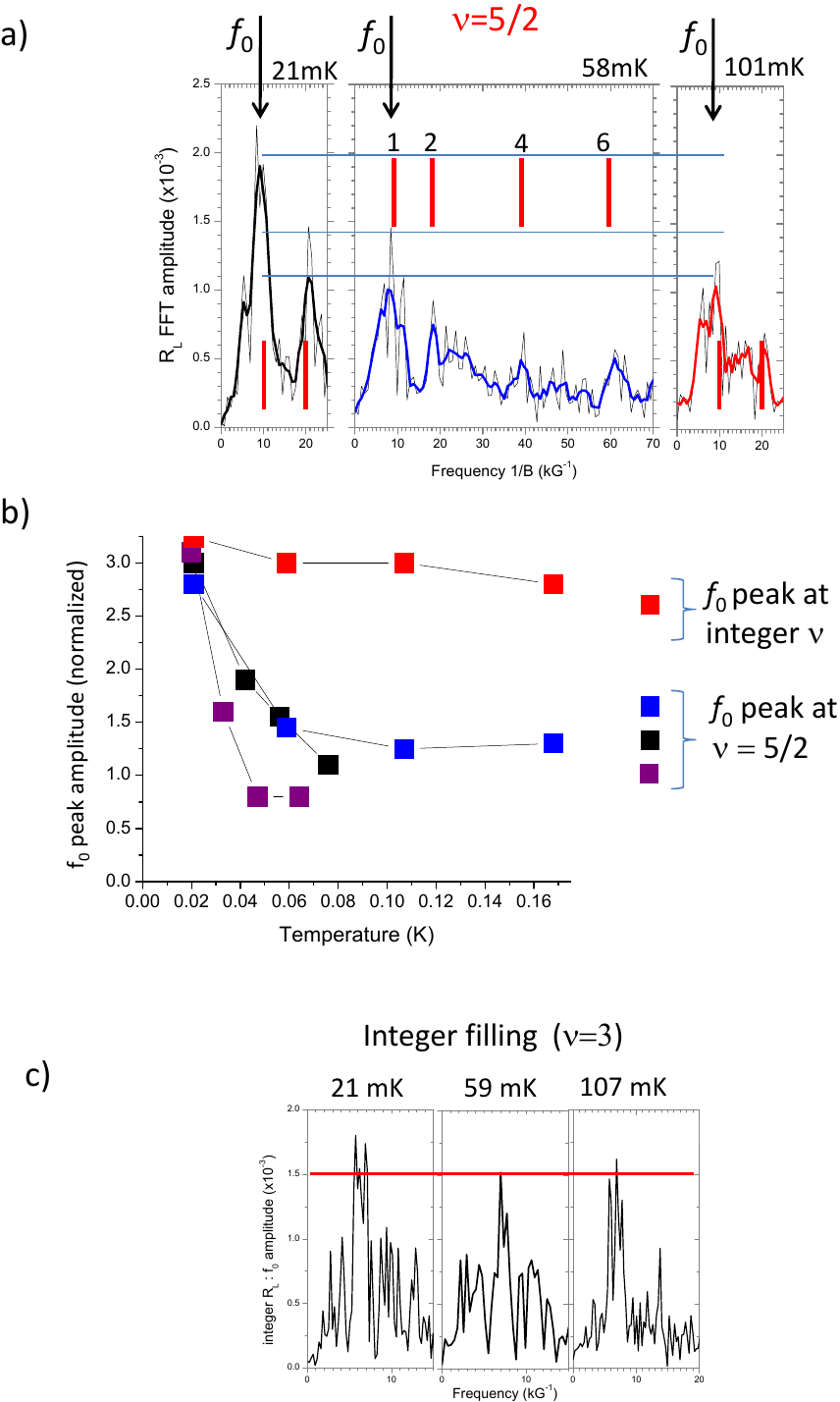}
\caption{Temperature dependence of the $f_0$ spectral peak at $\nu=5/2$.\\
(a) In order to demonstrate the temperature dependence of the spectral peak at $f_0$, the top panel shows the full spectrum of $\nu=5/2$ interference oscillations at $T=58$mK and the $f_0$ peaks at $T=21$mK and 101mK in sample 5. Note rather poor resolution of spectral features at 2, 4 and $6f_0$ in the middle plot due to relatively hight temperature $T=58$mK.\\
(b) Amplitude of the $f_0$ peak for these three temperatures (and a higher temperature measurement) versus temperature for this sample (blue data points).  Also shown are $\nu=5/2$ data for two other samples, demonstrating a substantial decrease in the $f_0$ peak amplitude as the temperature is increased from $\sim 20$mK to $\sim 80m$K.\\
(c) The spectral peak at $f_0$ for three temperatures at integer filling $\nu=3$ in sample 5.   The peak values are plotted in panel (b) for comparison to $\nu=5/2$: note the relative temperature independence of the integer peak over this temperature range in contrast to the $\nu=5/2$ data.
}
\label{fig:S5-7-3}
\end{figure}

The observed temperature dependence of the oscillation power spectra for $\nu=5/2$ and integer Hall states over a range of low temperatures is consistent with this picture. These data are displayed in the Figures~\ref{fig:S5-7-3} and \ref{fig:S5-7-4} below.  It is shown there that the amplitude of the $f_0$ peak in the power spectra of the integer quantum Hall states does not change significantly with increasing temperature (up to greater than 100mK) whereas over the same temperature range the $f_0$ peak at $\nu=5/2$ decreases in amplitude as the temperature is increased.   This decrease of the amplitude is commensurate with changes in the longitudinal resistance: the amplitude of the $f_0$ peak decreases with the rise of the measured value of $R_L$ at $\nu=5/2$, both saturating in their values at temperatures above roughly 80mK. These properties are consistent with the idea that the $f_0$ peak contains contributions from both $e/4$-quasiparticle and background electron interference at $\nu=5/2$.  These electron contributions can arise from degrading the fractional quantum Hall state by both venturing far from the center of the plateau during $B$-field sweeps and by temperature increase. Given the overall spectral features of the interference oscillations and their temperature dependence, these data indicate that several processes of different origin are likely to contribute to the $f_0$ peak at  $\nu=5/2$, in general agreement with our model. Specifically, the data in Figure~\ref{fig:S5-7-3} indicate that at $\nu=5/2$, the $f_0$ peak has both a contribution from the quasiparticles associated with the $\nu=5/2$ state and a background electron contribution. At lowest temperatures, the $f_0$ spectral peak at $\nu=5/2$ has a component whose temperature sensitivity roughly tracks the temperature behavior of the minimum value of resistance $R_L$ -- see Figure~\ref{fig:S5-7-4}.  As temperature is increased and the $\nu=5/2$ state is effectively destroyed, so is the contribution from its charge-$e/4$ quasiparticles whereas the background electronic contribution is still present. This results in the leveling of the temperature dependence of the $f_0$ spectral peak; at these higher temperatures measurements of the $f_0$ peak at integer filling factors shows its relative temperature independence.

\begin{figure}[h!]
\includegraphics[width=\columnwidth]{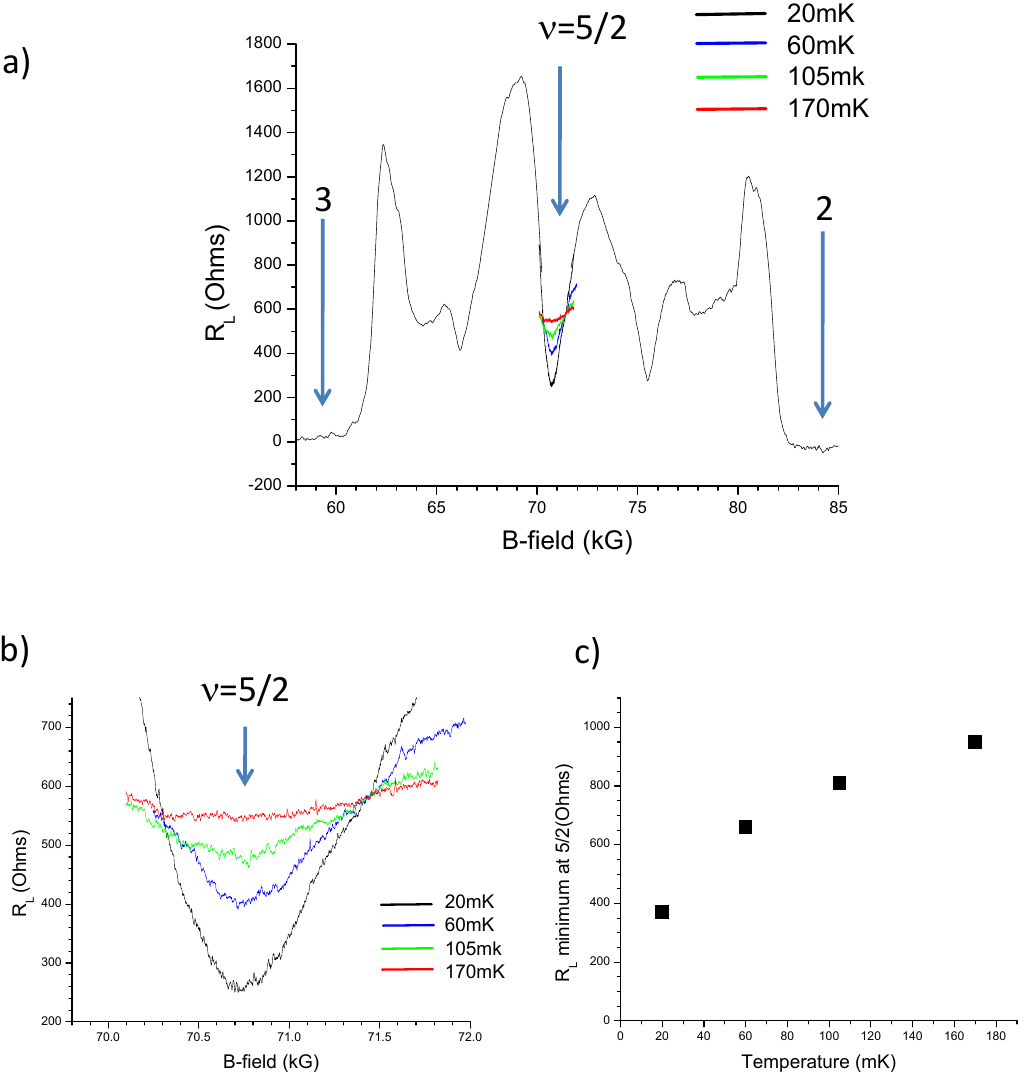}
\caption{Temperature dependence of the $R_L$ minimum at $\nu=5/2$ in the sample in Figure~\ref{fig:S5-7-3}.
The total change of $R_L$ with increasing temperature is roughly exhausted by the temperature of 100mK, and is similar to the saturation shown in the amplitude of the $f_0$ peak at $\nu=5/2$ in Figure~\ref{fig:S5-7-3}(b)\&(c).\\
(a) Trace of the longitudinal resistance $R_L$ through the interferometer at base temperature, with $R_L$ measurements immediately around $\nu=5/2$ at higher temperatures superimposed.\\
(b) A blowup of these traces in the vicinity of $\nu=5/2$.\\
(c) The minimum value of $R_L$ plotted for different temperatures; note the similar leveling at high temperatures of both minimum resistance and in the $f_0$ peak at $\nu=5/2$  shown in Figure~\ref{fig:S5-7-3}(b).
}
\label{fig:S5-7-4}
\end{figure}

%\clearpage
\subsection{Spectral dependence on Fermionic parity stability}
\label{sec:S6}
\setcounter{table}{0}
\setcounter{figure}{0}

Expression of different frequencies in the interference oscillation spectrum at $\nu=5/2$ and 7/2 is dependent upon both the tunneling probability for the $e/4$ and $e/2$ edge excitations as they encircle the active area of the interferometer, but, crucially, it is also dependent upon the stability of the fermion parity for the fusion channels of the charge-$e/4$ quasiparticles.

While potential contributions of charge-$e/2$ quasiparticle interference and electron interference has been already addressed in Sections~\ref{sec:S5-7b} and \ref{sec:S5-7c}, here we focus on the
stability of the fermion parity as the determining factor in the expression and potential splitting of the high-frequency even-odd oscillation peak. The fermion parity stability is presently assumed to depend on the sample and device parameters, but the exact nature of these factors has not been tested.  To describe the effect of this stability on the experimental measurements one must compare the fermion parity autocorrelation time to the time needed to measure the various interference periodicities in the resistance at $\nu=5/2$ and 7/2 due to the interference of both $e/4$ and e/2. To observe or discern the fundamental non-Abelian braid of $e/4$ encircling e/4, the change of magnetic field in a $B$-sweep should at the very least be sufficient to result in the flux change of $\Phi_0/5$ or $\Phi_0/7$. To discern the convolution of this non-Abelian braid and the $e/4$ braid of e/2, the fermion parity must remain stable for a longer time, sufficient to change the flux by  $\Phi_0$ or  $2\Phi_0/3$, respectively for $\nu=5/2$ and 7/2. Therefore, the relevant metric is given by the magnetic field sweep times for those periods.  Implicit in this assessment of the fermion parity stability is that the assumption that the time scale associated with the change of magnetic field is the same as determines the rate at which the number of quasiparticles changes inside the interferometer during the the magnetic field sweep.

As described in the theory sections (Section~\ref{sec:S1} and the review of non-Abelian interferometry in the main text), in order to see the frequencies corresponding to the interference of $e/4$ around $e/2$, i.e. $f_0$ at $\nu=5/2$ or $1.5f_0$ at $\nu=7/2$,  the fermion parity should remain stable during time required to change the flux by (at least) $\Phi_0$ at $\nu=5/2$ and  $2\Phi_0/3$ at $\nu=7/2$.  In addition, these processes will split the high-frequency peaks corresponding to the non-Abelian even-odd effect by the same amounts, producing spectral peaks at $4f_0$ and $6f_0$ for $\nu=5/2$, and $5.5f_0$ and $8.5f_0$ for $\nu=7/2$.  But the fermion parity stability can be of shorter time duration; if it is shorter than the time periodicity of the even-odd effect (measurement time corresponding to $\Delta\Phi=\Phi_0/5$ or $\Phi_0/7$), these peaks will be washed out completely.  In the intermediate case, i.e. when the parity is stable on the time scale of $\Delta\Phi=\Phi_0/5$ or $\Phi_0/7$, but not on the scale of $\Delta\Phi=\Phi_0$ or $2\Phi_0/3$, the splitting of the high-frequency peaks would not occur and the spectral peaks would appear at $5f_0$ (5/2) and $7f_0$ (7/2).  Also, for such intermediate duration of fermion parity stability, the enhanced $f_0$ for $\nu=5/2$ and the $1.5f_0$ for $\nu=7/2$ should not occur.

The details of the spectrum can be further complicated if the stability varies over the range of the $\nu=5/2$ or $\nu=7/2$ $B$-field sweep.  An example to consider is where the stability is long around the center of the $\nu=5/2$ or $\nu=7/2$ minimum ($R_L$) or plateau ($R_D$), but shortens away from the center.  This scenario could lead to a spectrum for $\nu=5/2$ that would include the $f_0$, $4f_0$, $6f_0$ peaks, but additionally the $5f_0$: the shortened stability time away from the plateau center contributes the $5f_0$, whereas the $f_0$, $4f_0$ and $6f_0$ peaks are due to the long stability time.  In this picture one could conjecture that the $e/4$ quasiparticle/quasihole density is increased away from $\nu=5/2$ and this may increase the tunneling amplitude of the neutral fermion mode between the edge and the bulk.  As for $\nu=7/2$, the spectrum would contain, $1.5f_0$, $5.5f_0$, $7f_0$, and $8.5f_0$, noting that the background electron interference peak at $f_0$ should also be present.

These possible contributions to the spectra, again all originating from the non-Abelian nature of the $e/4$ and the Abelian contribution of the fused $e/4$ charges (i.e. charge-$e/2$ quasiparticles), have sets of spectra at $\nu=5/2$ and 7/2 defined by the fermionic stability duration:  these are tabulated below, followed by examples of power spectra at $\nu=5/2$ and 7/2 that represent some of these possibilities.  For $\nu=5/2$ filling, examples are shown of spectra with a single $5f_0$ peak, the intermediate fermionic stability time, and also spectra with 4 and $6f_0$ peaks, longer fermionic stability time.  At $\nu=7/2$ filling examples are shown of a single  $7f_0$ peak and no discernible side peaks at 5.5 and  $8.5f_0$, again consistent with a fermionic stability time only long enough to resolve the rapid non-Abelian oscillation, and another example showing spectral features around $7f_0$ with resolved $8.5f_0$ peak, not a discernible $5.5f_0$, and a weak $1.5f_0$ peak, consistent with a longer fermionic stability time.  An example of fully resolved features (1.5, 5.5 and $8.5f_0$) consistent with a fermionic stability time longer than the $1.5f_0$ measurement time is shown in Figure~\ref{fig:7_halves_interf} of the main text.

\begin{table}[htb]
  \centering
  \includegraphics[width=0.95\columnwidth]{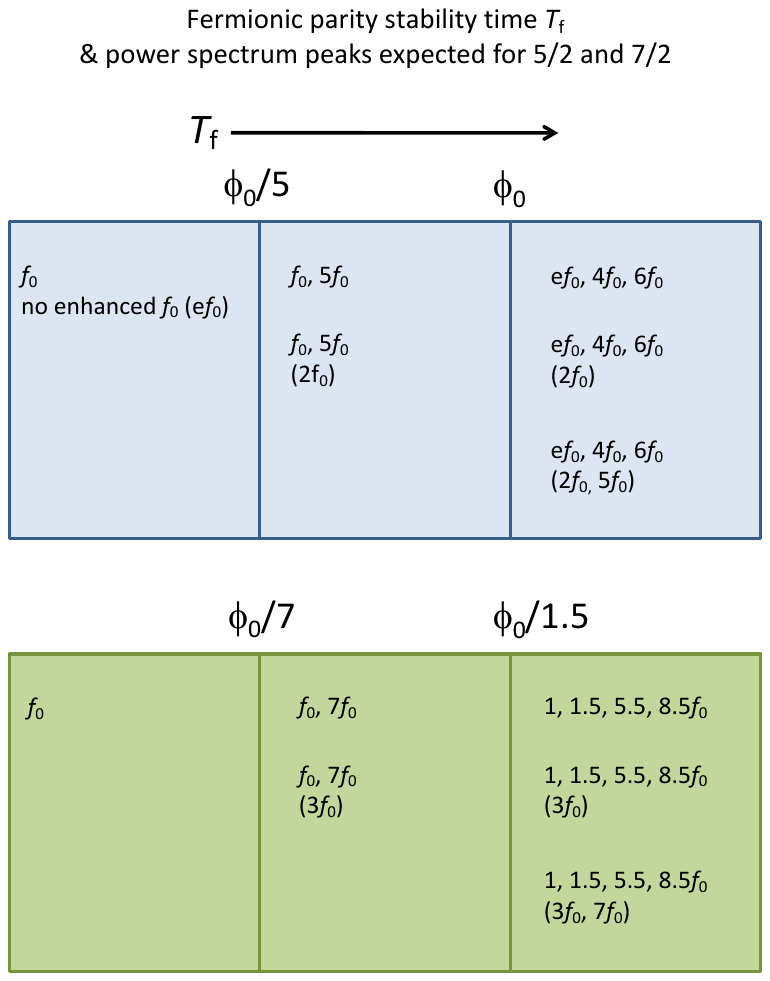}
  \caption{ Expected/possible power spectrum peaks for filling factors $\nu=5/2$ and 7/2 given a fermionic stability time $T_\text{f}$.  The determining time factors are where the stability time $T_\text{f}$ is with respect to the measurement time, and the metrics for this time are two points:   even/odd parity change with B-field sweep, $e/4$ braiding $e/4$  ($5/\Phi_0$ or $7/\Phi_0$) measurement time, and measurement time for the $e/4$ braid of $e/2$ ($1/\Phi_0$ or $3/2\Phi_0$). For example, if the stability time is less than the time to measure a full period of number parity change due to $B$-field sweep, no resistance oscillation will be observed, except for a background electron interference at $f_0$ (we refer to the enhanced $f_0$ peak due to contributions from electrons plus $e/4$ braiding $e/2$ at $\nu=5/2$ as e$f_0$ in the table).   If the stability time is only long enough for a measurement of number parity change ($5/\Phi_0$ or $7/\Phi_0$), that period will be measured but not the modulation or splitting of that peak. If the stability time is longer than the $e/4$ braid $e/2$ measurement time, then the full spectrum of peaks should be observed.  Note that the $e/2$ braid of $e/2$ can be added to each set independently of the fermionic stability time.}
  \label{tab:S6-T1}
\end{table}

\clearpage
\onecolumngrid

\begin{figure}[htb]
\includegraphics[width=0.67\textwidth]{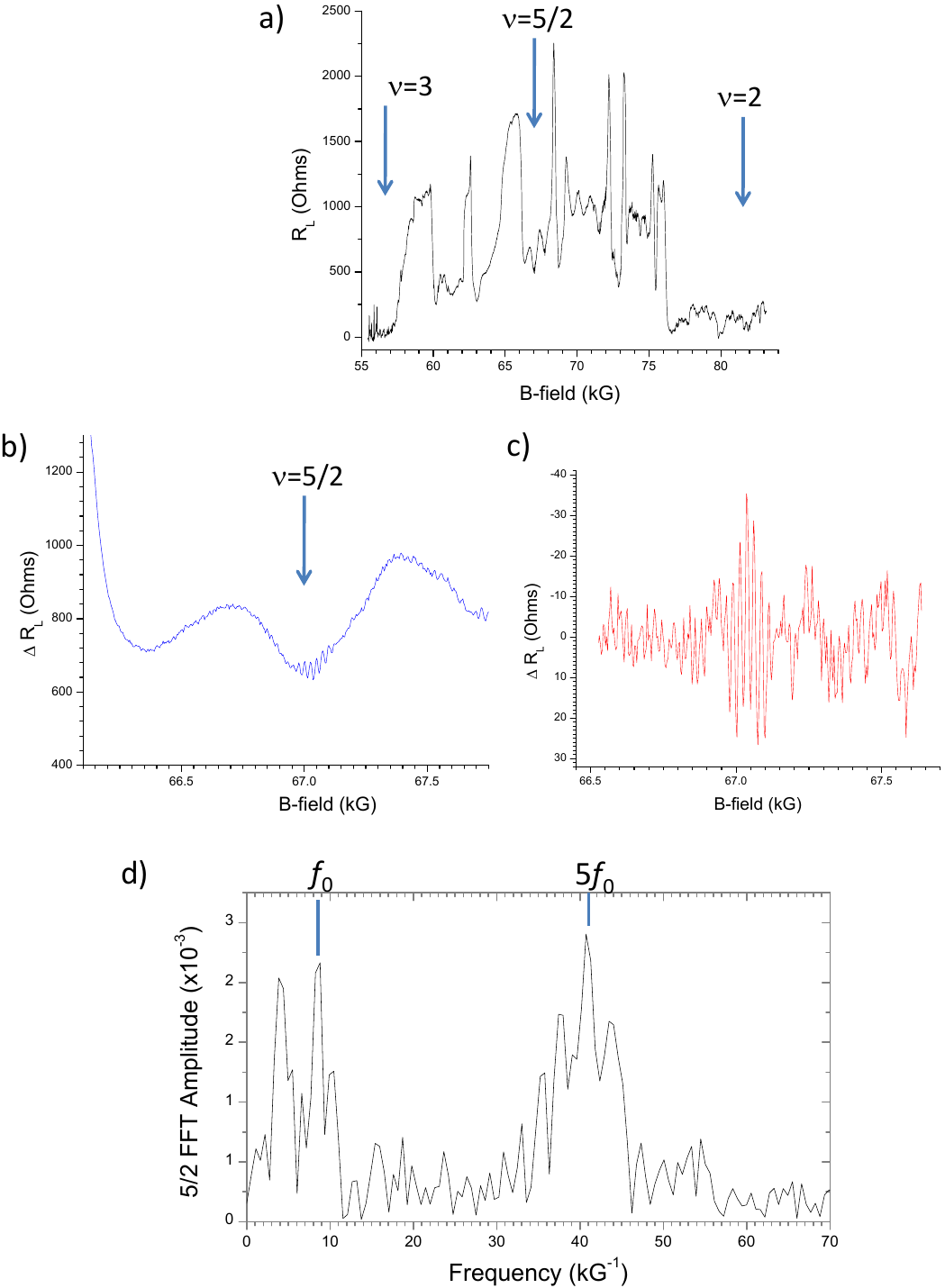}
\caption{ $\nu=5/2$ power spectrum showing only $5f_0$ and $f_0$ peaks, consistent with a fermionic stability time less than the time to measure $e/4$ braiding $e/2$ modulating the non-Abelian braid.  $R_L$ through the interferometer (a), focus of $R_L$ near $\nu=5/2$ (b), and background resistance subtracted from $R_L$ at and near $\nu=5/2$, $\Delta R_L$ versus magnetic field, (c), and (d) power spectrum taken using the data of (c). Sample 6, preparation 3,  $T\sim 20$mK,  magnetic field sweep rate is 28 Gauss/min over the range of (c).
}
\label{fig:S6-1}
\end{figure}

\begin{figure}[htb]
\includegraphics[width=0.67\textwidth]{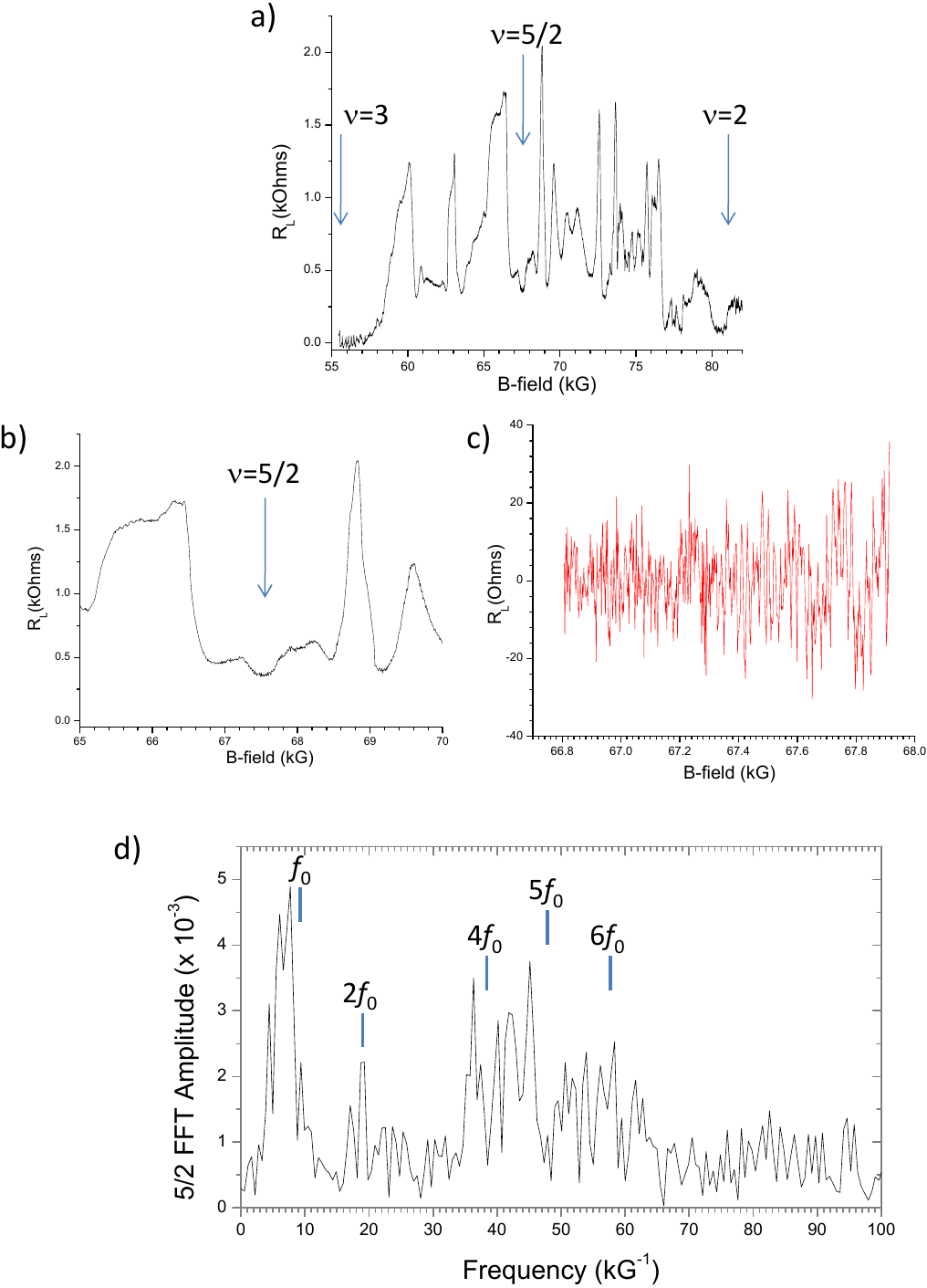}
\caption{Another $\nu=5/2$ power spectrum showing coarse $4f_0$, $6f_0$ peaks, and the $2f_0$ peak.  This device has wider backscattering gates that may lend to the relatively poorly resolved 4 and $6f_0$ peaks, yet present a distinct $5f_0$ minimum. This complex of $5f_0$ minimum and even coarse 4 and $6f_0$ peaks suggests a long enough fermionic stability time for convolution of the $e/4-e/2$ braid and the non-Abelian braid.   (a) through (d) as in Figure~\ref{fig:S6-1}. Sample 6, preparation 12, $T\sim 20$mK; (c) magnetic field sweep rate is 28 Gauss/min.
}
\label{fig:S6-2}
\end{figure}

\begin{figure}[htb]
\includegraphics[width=0.67\textwidth]{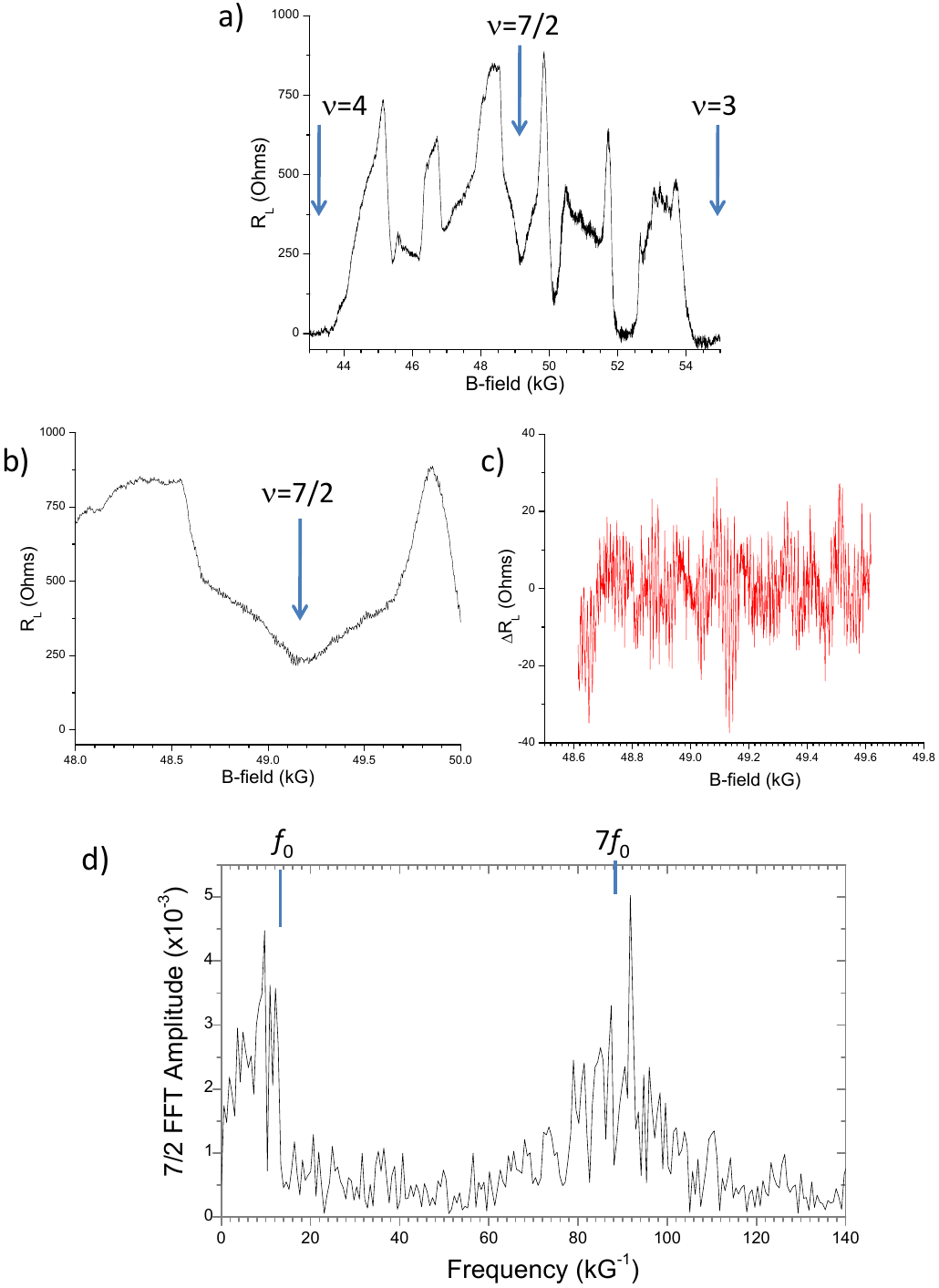}
\caption{$\nu=7/2$ power spectrum showing only $7f_0$ prominent peak, and presumed background electron $f_0$ peak.  Note the absence of the $1.5f_0$ peak even with $7f_0$.  This is consistent with a fermionic stability time less than the time to measure $e/4$ braiding $e/2$ modulating the non-Abelian braid.   As above, longitudinal magneto-resistance ($R_L$) through the interferometer (a), focus of $R_L$ near $\nu=7/2$ (b), background resistance subtracted from $R_L$ at and near $\nu=7/2$, $\Delta R_L$ versus magnetic field (c), and (d) power spectrum taken using the data of (c). Sample 6, preparation 6, $T\sim 20$mK, (C) magnetic field sweep rate is 28 Gauss/min.
}
\label{fig:S6-3}
\end{figure}

\begin{figure}[htb]
\includegraphics[width=0.67\textwidth]{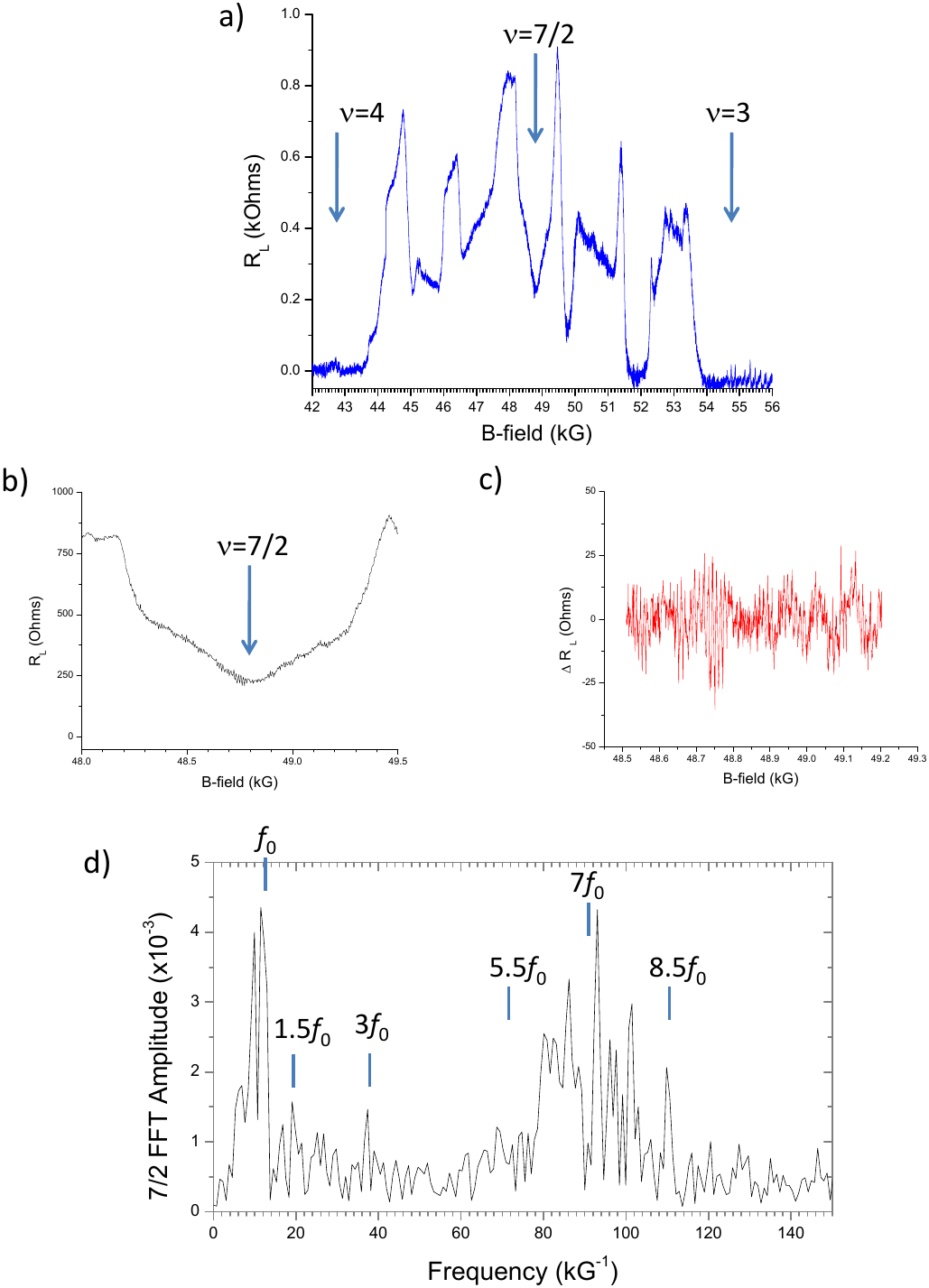}
\caption{$\nu=7/2$ power spectrum showing distinct minimum at $7f_0$ and distinct peaks at $1.5f_0$, $3f_0$, and $8.5f_0$. This sample preparation represents the model of fermionic stability long enough that the modulation of the $7f_0$ oscillation is measurable.  (a) through (d) as above. Sample 6, preparation 14, $T\sim 20$mK; (c)
magnetic field  sweep rate is 28 Gauss/min.
}
\label{fig:S6-4}
\end{figure}

\end{document}